%% file: etacrv.tex
\documentclass[apj]{emulateapj} 
\slugcomment{{\sc Accepted to ApJ:} November 9, 2015} 
\usepackage{amsmath}
\usepackage{amsfonts}
\usepackage{amssymb}
\usepackage{subfigure}
\usepackage[utf8]{inputenc}
\usepackage{ae,aeguill}
\usepackage{graphicx}
\usepackage{color}
\usepackage{epsfig}
\usepackage{fancyhdr} 
\usepackage{amssymb, amsmath}
\usepackage{longtable}
\usepackage[figuresright]{rotating} 
\usepackage{url}
\usepackage{textcomp}
\usepackage{multirow}
\usepackage{enumerate}
\usepackage{pdfpages}
\usepackage{natbib}
\usepackage{version}
\usepackage{placeins}
\usepackage[bookmarks=true,bookmarksnumbered=true,plainpages=false,breaklinks=true,colorlinks=true,urlcolor=blue,citecolor={black},linkcolor={black}]{hyperref}
\usepackage[english]{babel}
\newlength{\firstcol}      
\newlength{\secondcol} 
\newlength{\largecol}     
\include{commands}

\begin{document}

\begin{abstract}
Debris disks are signposts of analogues to small body populations of the Solar System, often however with much higher masses and dust production rates. The disk associated with the nearby star \etacrv\ is especially striking as it shows strong mid- and far-infrared excesses despite an age of $\sim$1.4\,Gyr.
We undertake to construct a consistent model of the system able to explain a diverse collection of spatial and spectral data. 
We analyze Keck Interferometer Nuller measurements and revisit {\sc Spitzer} and additional spectro-photometric data, as well as resolved {\sc Herschel} images to determine the dust spatial distribution in the inner exozodi and in the outer belt. 
We model in detail the two-component disk and the dust properties from the sub-AU scale to the outermost regions by fitting simultaneously all measurements against a large parameter space.
The properties of the cold belt are consistent with a collisional cascade in a reservoir of ice-free planetesimals at 133\,AU. It shows marginal evidence for asymmetries along the major axis. 
KIN enables us to establish that the warm dust consists in a ring that peaks between 0.2 and 0.8\,AU. 
To reconcile this location with the $\sim$400\,K dust temperature, very high albedo dust must be invoked and a distribution of forsterite grains starting from micron sizes satisfies this criterion while providing an excellent fit to the spectrum. 
We discuss additional constraints from the LBTI and near-infrared spectra, and we present predictions of what JWST can unveil about this unusual object and whether it can detect unseen planets. 
\end{abstract}

\bibliographystyle{aa}
\title{Models of the $\eta$ Corvi debris disk 

from the Keck Interferometer, {\sc Spitzer} and {\sc Herschel}}  
\author{J. Lebreton\altaffilmark{1,2}, C. Beichman\altaffilmark{1,2,3}, G. Bryden\altaffilmark{3}, D. Defr\`ere\altaffilmark{4}, \\B. Mennesson\altaffilmark{3}, \and R. Millan-Gabet\altaffilmark{1,2}, \and A. Boccaletti\altaffilmark{5}}
\altaffiltext{1}{Infrared Processing and Analysis Center, California Institute of Technology, Pasadena, CA 91125, USA}
\altaffiltext{2}{NASA Exoplanet Science Institute, California Institute of Technology, 770 S. Wilson Ave., Pasadena, CA 91125, USA}
\altaffiltext{3}{Jet Propulsion Laboratory, California Institute of Technology, 4800 Oak Grove Dr., Pasadena, CA 91107, USA}
\altaffiltext{4}{Department of Astronomy, University of Arizona, 993 N. Cherry Ave, Tucson, AZ, 85721, USA}
\altaffiltext{4}{LESIA, Observatoire de Paris, CNRS, University Pierre et Marie Curie Paris 6 and University Denis Diderot Paris 7, 5 place Jules Janssen, 92195 Meudon, France}
\email{lebretoj@gmail.com}

\maketitle

\section{Introduction}

The luminosity function of debris disks surrounding Main Sequence (MS) stars is a decreasing function of age, due to the progressive grinding-down of dust producing planetesimals \citep{2007ApJ...663..365W, 2008ApJ...673.1123L}. In this context, the nearby \citep[18.2 parsec,][]{Holmberg09} F2V star \etacrv\ (\object{HD\,109085, HIP\,61174}) is particularly unusual. 
Despite an estimated age of $\sim$1.4 Gyr \citep{Wyatt:2005fj,2012ApJ...747...93L}, $\eta$\,Corvi shows evidence for a strong infrared excess.

The debris disk exhibits two distinct dust populations. A cold Kuiper belt-like disk was first detected through its far-infrared excess with IRAS and later imaged in the sub-millimeter with SCUBA \citep{Wyatt:2005fj} and the {\sc Herschel} Space Observatory \citep{Matthews:2010qy}. The {\sc Herschel} images reveal an inclined dust belt at an orbital distance of {$\sim$150\,AU} clearly separable from a warm component in the inner stellar system  \citep{Duchene:2014yu}. The outer belt has a fractional luminosity ${L\dma{D}/L\dma{\star}\,\sim\,3\times10\uma{-5}}$ and remains undetected in scattered light images. 
The distinctive feature of the \etacrv\ Spectral Energy Distribution (SED) is the presence of strong excess emission with respect to the photosphere in the mid-infrared with a fractional luminosity ${L\dma{D}/L\dma{\star}\,\sim\,3\times10\uma{-4}}$ \citep[\textit{e.g.}][]{Bryden:2006eu,Beichman:2006lr}. Only few debris disk stars harbor such signatures of large amounts of warm material residing in their close environment, \textit{i.e.} exozodiacal disks (exozodis).
Besides, the {\sc{Spitzer}}/IRS spectrum of \etacrv\ \citep{Chen06} harbors strong spectral features that are rarely seen in MS circumstellar disks and are thought to trace collisionally very active systems \citep[\textit{e.g.} \object{HD\,69830},][]{2005ApJ...626.1061B}. 
Mid-infrared spectral features make it possible to address the dust mineralogy in detail \citep[e.g.][]{Olofsson:2012vn}.
\citet{2012ApJ...747...93L} {and \citet{Chen06}} successively produced elaborate models of the debris disk through a thorough analysis of its spectrum. The latter proposes that the warm dust consists of primitive cometary (ice- and carbon-rich) material in the Habitable Zone (HZ) in addition to impact-produced silicas. These findings yield the interpretation that the exozodi originates from the relatively recent collision of a Kuiper Belt object with a larger body in the HZ, possibly during a Late Heavy Bombardment-like event (LHB).
Conversely, no hot dust was detected around this star using near-infrared interferometry at the 2\% level \citep{2013A&A...555A.104A}.

{To date there are no unambiguous measurements of the spatial location of the inner dust, besides spectral models that are intrinsically degenerate between dust size and location.
Mid-infrared interferometry was performed at the VLTI by \citet{Smith:2009zr}, suggesting that the exozodi is concentrated at less than a few AU and that it is coaligned with the outer disk but more detailed measurements are needed.
One of the key goals of the present paper is to analyze thoroughly interferometric data from the Keck Interferometer Nuller (KIN) originally presented by \citet{2011ApJ...734...67M} and \citet{Mennesson:2014qy}. The warm {$\eta$\,Crv} exozodi was resolved both spatially and spectrally by the KIN, revealing the spatial distribution within the null pattern. We here interpret this data using detailed dust disk models in order to refine the exozodi location. In addition, new results for the Large Binocular Telescope Interferometer (LBTI) were recently presented in a companion paper \citep{Defrere:2015lr} and we here present supporting information about the models.}

Analyzing such an ensemble of data requires one to go beyond simple models and to carefully solve the radiative transfer in the dust disk for a large enough parameter space using spectral libraries while accounting for the  specific instrumental response. 
Although such detailed approaches are now routinely used to model cold debris disks, only few have been attempted for exozodis, which include Vega \citep{2011A&A...534A...5D}, \bp\ \citep{2012A&A...546L...9D} and Fomalhaut \citep{Lebreton:2013uq}.

In the present paper, we build a detailed model of \etacrv\ valid from its innermost regions (exozodiacal disk) to its outermost ones (cold belt) that is able to reproduce the SED from mid-infrared to millimeter wavelengths as well as spatial constraints from Keck mid-infrared interferometry and {\sc Herschel} far-infrared imaging.
We first introduce photosphere models in Section\,2.
In Section\,3, we present KIN interferometric data and the SED of the disk including in particular a revision of the {\sc{Spitzer}}/IRS spectrum, as well as our analysis of {\sc Herschel}/PACS images. In Section\,4 we draw first conclusions from the data and we detail our modelling strategy. Modelling results are presented in detail in Section\,5 for the inner disk and Section\,6 for the outer one.
{We discuss and analyze our findings in Section\,7. In particular we confront them to new LBTI observations and test their compatibility with scattered light spectra. Finally a summary and conclusions are presented in Section\,8}.

\begin{table}[tpb]\caption{Stellar properties}\label{tab:properties}
\begin{center}
\begin{tabular}{ccc}
\hline\hline 
RA & $12\degr32\arcmin04\arcsec$ \\
dec & -$16\degr 11\arcmin 46\arcsec$ \\
Distance	  &	18.2 pc$\uma{[1]}$\\
\hline
Spectral type  &	F2V$\uma{[1]}$ \\
v$\dma{\sin{i}}$   & $92\,\textrm{m/s}\uma{[2]}$\\
$\theta\dma{LD}$   & 0.819\,mas$\uma{[2]}$\\ 
$L\dma{\star}$ & $L\dma{\star} = 5.06\pm0.05\uma{[3]}$\\
$T\dma{eff}$	&  	6900\,K$\uma{[3]}$\\
\hline
$M\dma{V}$ (0.55\um) &	$4.305\pm0.009\uma{[1]}$  \\   
$M\dma{B}$	 (0.44\um) &  $4.598\pm0.014\uma{[1]}$   \\
$M\dma{Rc}$	 (0.64\um) & $4.014\pm0.014\uma{[1]}$   \\
\hline
$M\dma{Hp}$	 (0.402\um) &	$4.385\pm0.001\uma{[5]}$ \\
$M\dma{Bt}$	(0.420\um) &	$4.728\pm0.014\uma{[4]}$ \\
$M\dma{VT}$  (0.532\um) &  $4.338\pm0.009\uma{[4]}$ \\
$M\dma{2J}$  (1.235\um) & $3.609\pm0.250\uma{[6]}$\\
$M\dma{2H}$  (1.662\um) & $3.372\pm0.24\uma{[6]}$\\
$M\dma{2Ks}$  (2.159\um) & $3.372\pm0.302\uma{[6]}$\\
$M\dma{L}$  (3.6\um) & $3.51\pm0.05\uma{[7]}$ \\
$M\dma{L'}$ (3.8\um) & $3.54\pm0.05\uma{[7]}$ \\
$M\dma{M}$  (4.8\um) & $3.58\pm0.05\uma{[7]}$ \\
\hline
\end{tabular}
\end{center}
{{\sc Notes --} The table lists apparent magnitude. Data from: 
[1] Simbad, 
[2] \citet{2013A&A...555A.104A},
[3] This study,
[4] \citet{Hog:2000qf}, 
[5] \citet{Hipparcos},
[6] \citet{Cutri:2003zl} and
[7] \citet{Sylvester:1996eu}}.
\end{table}

\begin{figure*}[h!tpb]
\begin{center}
  \includegraphics[angle=0,width=0.8\textwidth,origin=bl]{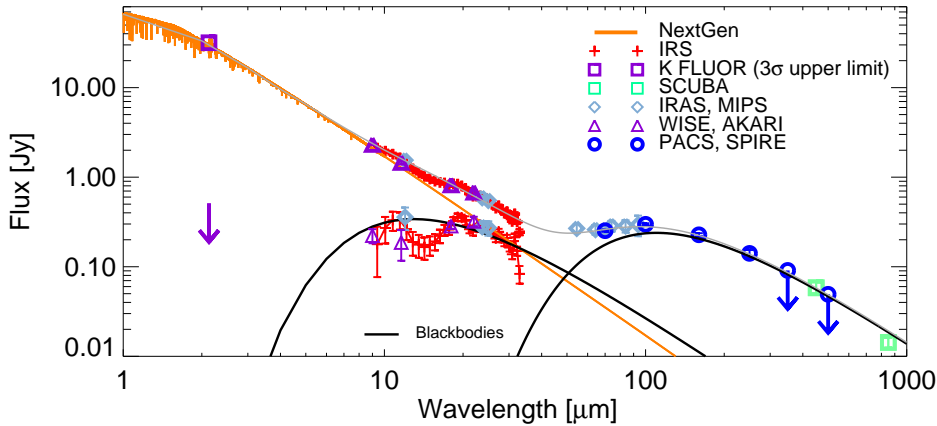}
  \caption{Spectral Energy Distribution of $\eta$ Crv. The SED includes spaceborn photometry from {\sc{Spitzer}}, Herschel, WISE, AKARI and sub-millimeter fluxes from SCUBA. The Nextgen photosphere model with T$\dma{eff}~=~6900\,\textrm{K}$ and $\log\,g = 4.5$ is overplotted and subtracted to the data to reveal the excess spectrum. Blackbody fits to the warm and cold parts of the SED are shown on an indicative basis.}\label{fig:sed}
  \end{center}
\end{figure*}

\newpage
\section{Stellar photosphere}\label{sec:photosphere}
Determining an accurate stellar spectrum is a critical step especially when trying to study faint excesses in the mid-infrared. 
\etacrv\ is usually referred to as an F2V star ($M = 1.52\,M\dma{\sun}$), which corresponds to an effective temperature T$\dma{eff} = 7000$ K and $\log\,g = 4.5$. However, the Vizier database lists various estimations that range from 6800\,K to 6900\,K and 4.06 to 4.22 respectively. 
We test a sample of tabulated synthetic spectra calculated with the NextGen model \citep{1999ApJ...512..377H} with T$\dma{eff} = 6800 \textrm{~or~} {7000\,\textrm{K}}$ and $\log\,g = 4.0 \textrm{~or~} 4.5$ assuming solar metallicity adequate for \etacrv. 
Two ${\textrm{T}\dma{eff} = 6900\,\textrm{K}}$ spectra are synthesized by averaging the above models for each $\log\,g$.
The models need to be scaled to the intrinsic magnitude of \etacrv\ which is equivalent to adjusting the stellar radius knowing the distance, or its luminosity. 

The scaling is obtained by fitting the high-resolution NextGen spectrum to flux measurements from Hipparcos in BT, VT and Hp bands \citep{Hog:2000qf,Hipparcos}, 2MASS in J, H and Ks bands \citep{Cutri:2003zl}, complemented by UKIRT in L, L' and M bands \citep{Sylvester:1996eu}, using appropriate filter profiles (Table\,\ref{tab:properties}). We test several subsets of the data: (1) all fluxes, (2) visible fluxes only, (3) all fluxes excluding the blue band, (4) all fluxes minus 2\% in the infrared (to correct for a possible excess).
A least-square fit to all measurements significantly favors the T$\dma{eff} = 7000$ K, $\log\,g = 4.5$ model although the agreement is not excellent at the shortest wavelengths. The star luminosity is $L\dma{\star} = 5.06\pm{0.05}\dma{\odot}$ for the various models and data subsets.
The value of $\log\,g$ has moderate impact on the spectrum, but the value of T$\dma{eff}$ is critical. Relaxing the constraint on the longer wavelengths yields colder spectra that best fit the blue channel, with less bright spectra. On the other hand T$\dma{eff} = 7000$ K improves the consistency with the near-infrared spectrum as noted by \citet{Duchene:2014yu}. In Section\,\ref{sec:moredust} we discuss the evidence for a 5\um\ excess proposed by \citet{2012ApJ...747...93L}, which has large uncertainties compared to optical photometry and cannot impact the fitting results.

Overall, depending on the spectrum assumed and the subset of wavelengths fitted, a standard deviation of 4\% (averaged over wavelengths) is observed in the final spectrum.
We use the T$\dma{eff}~=~6900\,\textrm{K}$, $\log\,g = 4.5$ spectrum with a scaling factor corresponding to a stellar luminosity $L\dma{\star} = 5.09\,M\dma{\odot}$.
The uncertainty on the stellar spectrum is propagated as a relative error term in the SED to account for photosphere-subtraction error, {with a strong impact in the mid-infrared part of the excess spectrum (in particular in the 3-8\um\ range, see Sec.\,\ref{sec:moredust} and Fig.\,\ref{fig:nearir_spectrum}.)}

\section{Observations and data reduction}
The full dataset that we use to constrain our models of the $\eta$\,Corvi system consists of:

(1) Nulls from the Keck Interferometer Nuller in the mid-infrared (Sec.\,\ref{sec:kin}),\\
(2) Broad-band photometry constituting the SED (Sec.\,\ref{sec:sed}) and 
(3) A higher resolution mid-infrared {\sc{Spitzer}}/IRS spectrum (Sec.\,\ref{sec:irs}),\\
(4) Herschel PACS resolved images in the far-infrared domain (Sec.\,\ref{sec:pacs}).

Each dataset is described hereafter.

\subsection{Keck Nuller data}\label{sec:kin}
\begin{figure*}[h!tbp]
\begin{center}
  \includegraphics[angle=0,width=0.65\textwidth,origin=bl]{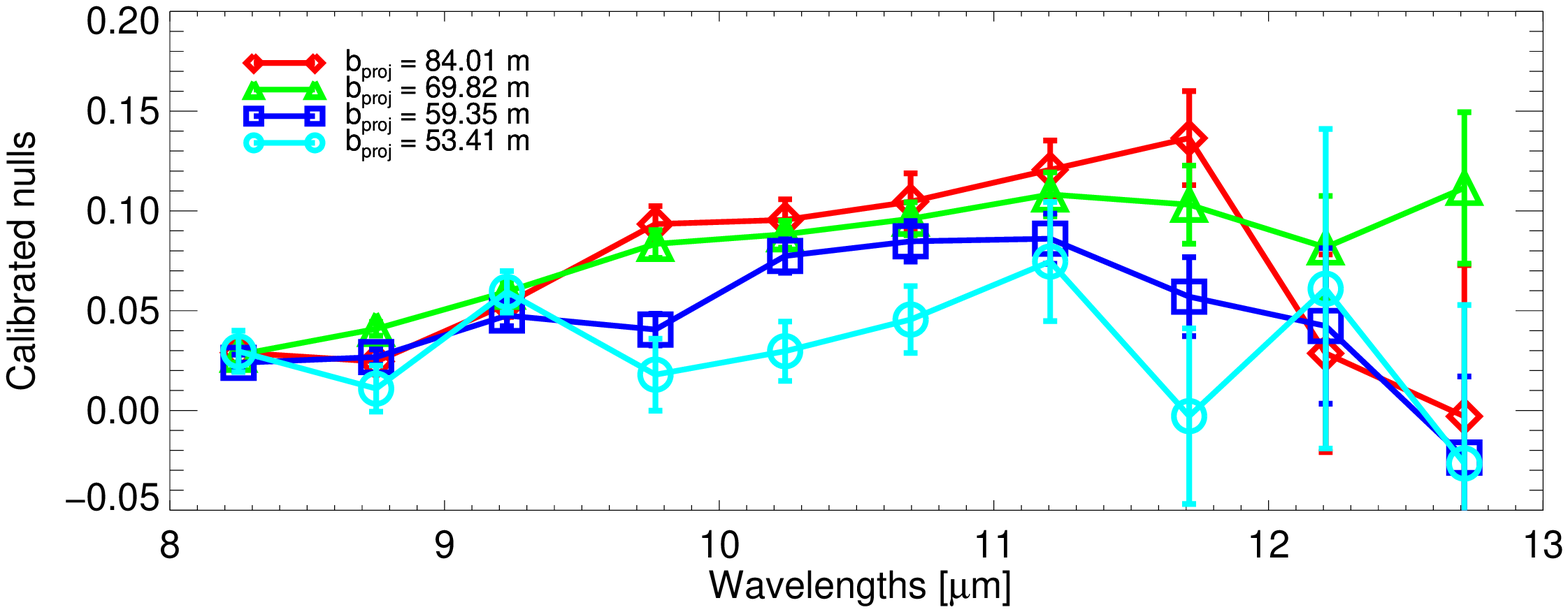}
    \includegraphics[angle=0,width=0.32\textwidth,origin=bl]{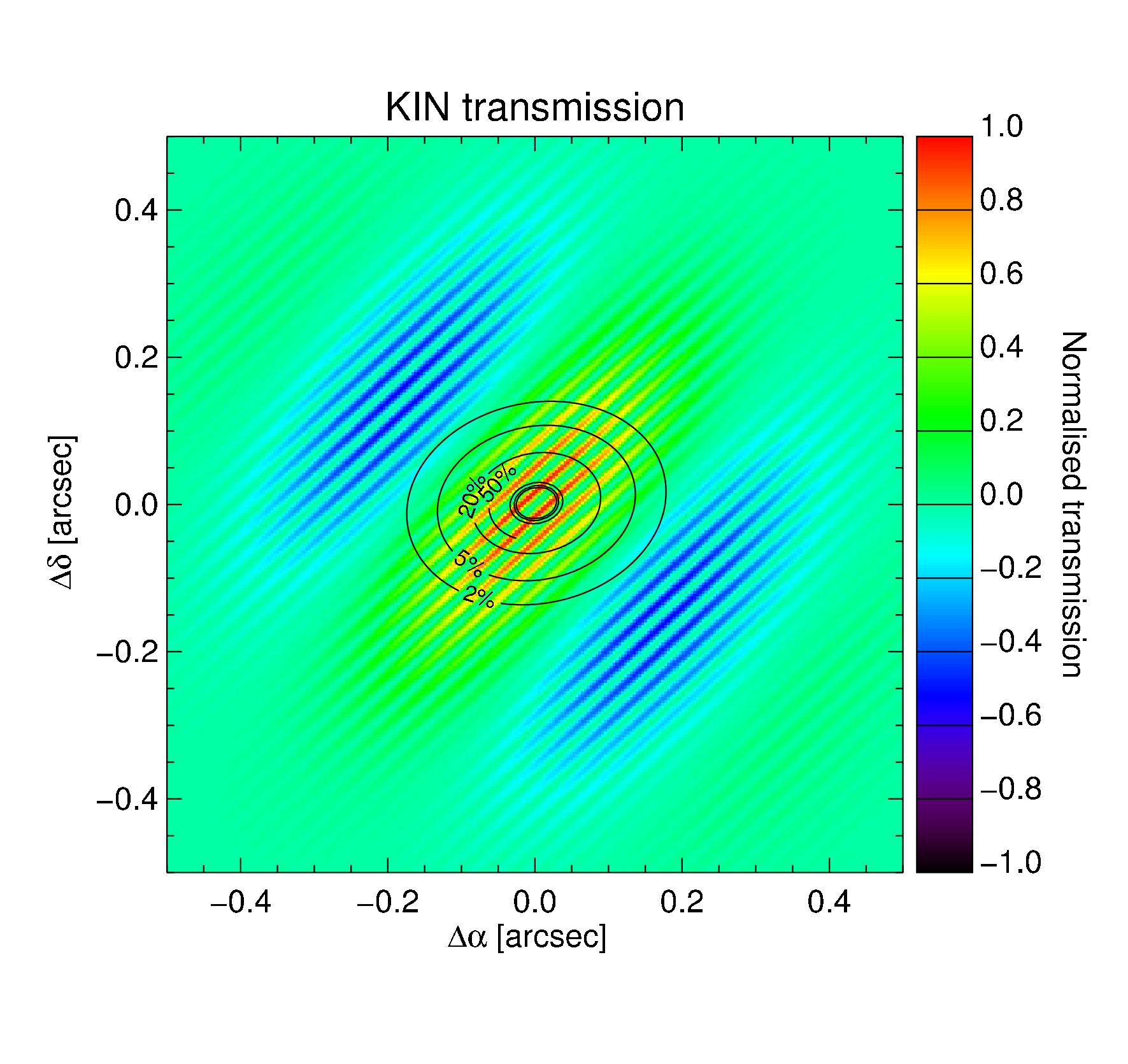}
        \includegraphics[angle=0,width=0.32\textwidth,origin=bl]{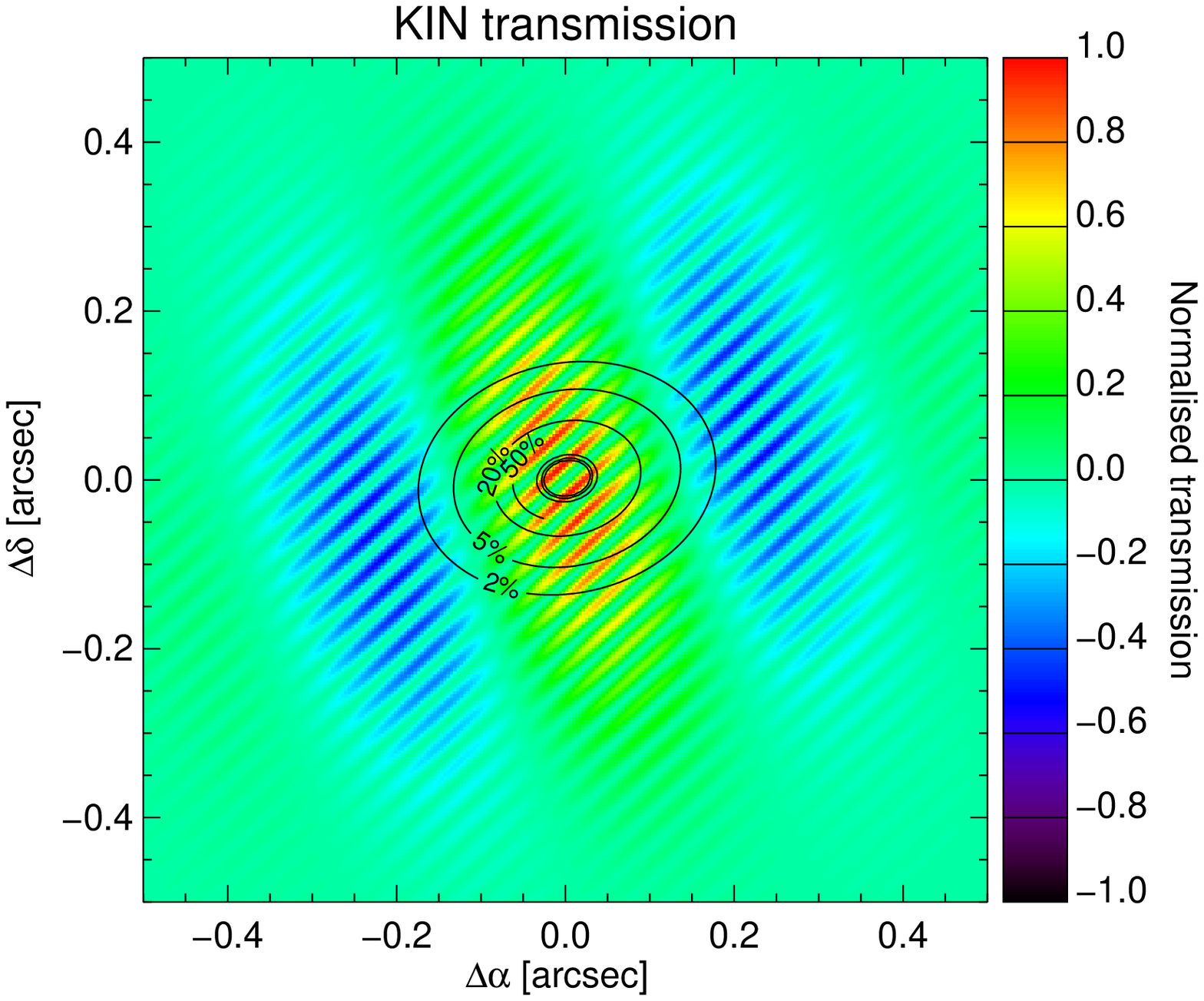}
            \includegraphics[angle=0,width=0.32\textwidth,origin=bl]{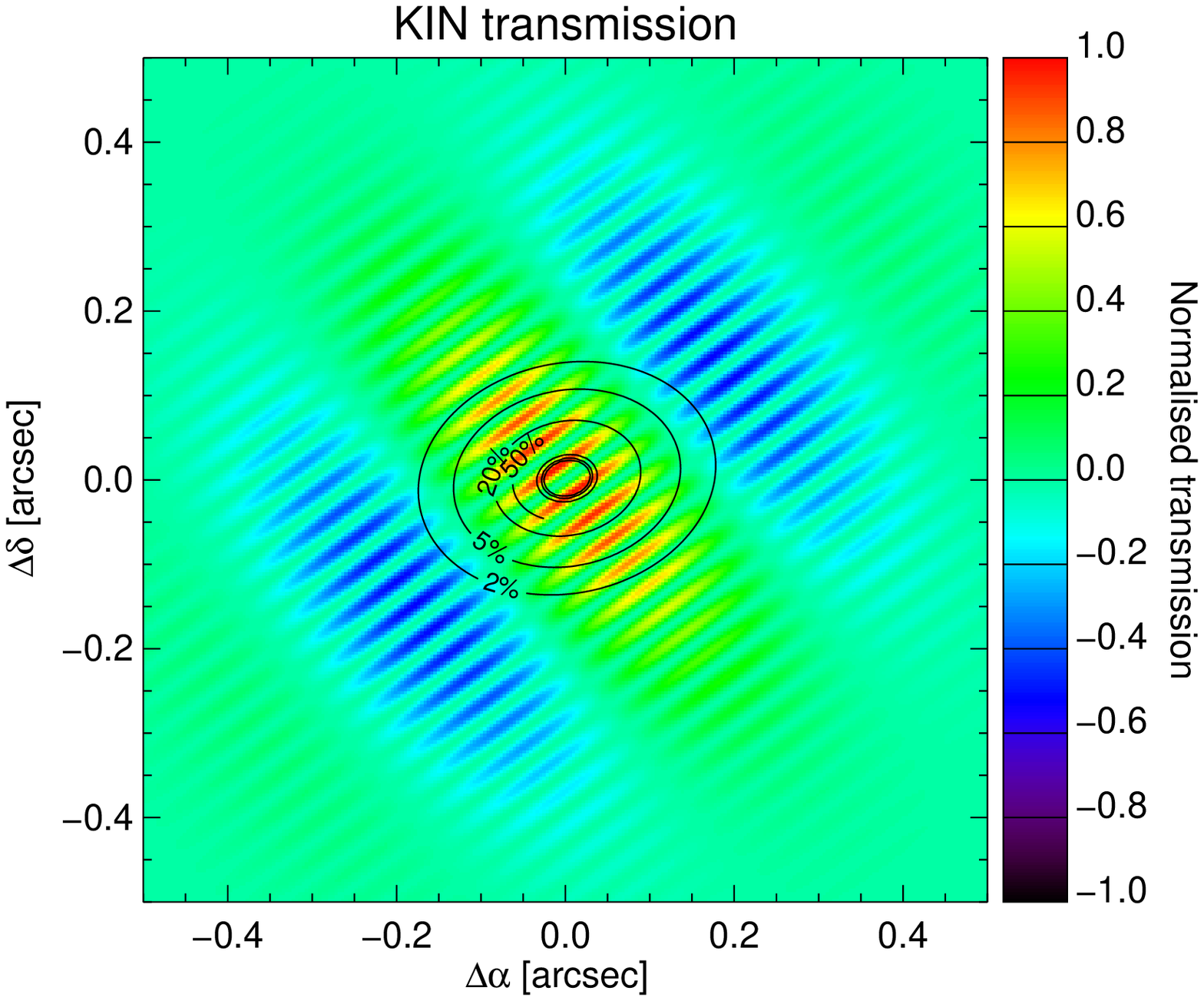}
                \includegraphics[angle=0,width=0.32\textwidth,origin=bl]{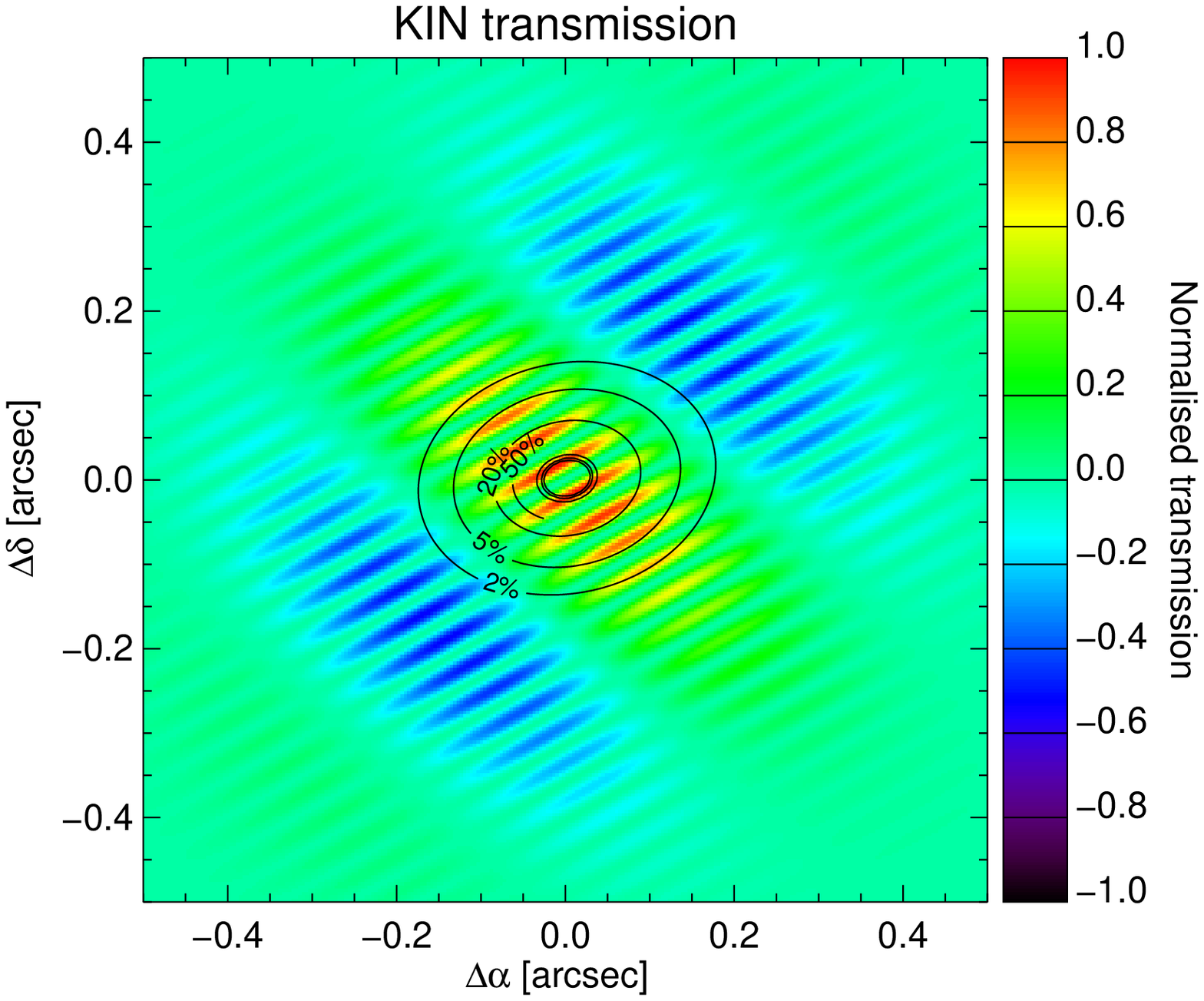}
  \caption{\textbf{Top-Left:} {Calibrated KIN nulls as a function of wavelength. The data include the stellar contribution, are corrected from the system nulls and are given with $1\sigma$ uncertainties \citep[formal error + external error, ][]{2011ApJ...734...67M}. Four epochs are shown corresponding to four different configurations of the interferometer, in particular in terms of baseline length $b\dma{proj}$. \textbf{Right and bottom:} KIN transmission maps at 10.25 $\mu m$ for the four epochs (Table\,\ref{tab:kin}) sorted in order of decreasing baseline and contour levels for a possible disk geometry that has an inner edge of 0.2\,AU and a $-1.5$ outer slope (Section\,\ref{sec:zodimodel}). The fringe pattern rotates with time and the size of the small fringes increases with decreasing projected baseline and with increasing wavelength.}} \label{fig:kin_data}
  \end{center}
\end{figure*}

$\eta$\,Crv was observed with the Keck Interferometer Nuller on April 17th and May 24th, 2008 (Table\,\ref{tab:kin}). The data were first presented by \citet{2011ApJ...734...67M} and re-reduced by \citet{Mennesson:2014qy}. 
{
Qualitatively the nulling technique consists in observing the source through a fringe pattern designed to cancel out the stellar contribution through destructive interference, while the circumstellar flux is transmitted through partially or fully constructive interference. 
Four beams are recombined by the KIN system: a pupil-splitting mirror divides the light gathered by each of the two telescopes into ``left'' and ``right'' beams. Interferometric nulling is achieved between the two Keck beams, and between the two right beams. 
The KIN transmission map is defined as the superposition of small fringes corresponding to the long baseline between the two Keck telescopes ($B \simeq 85$\,m) and large fringes corresponding to the short cross combiner baseline ($b \simeq 4$\,m), modulated by the transmission function of each telescope $T_L(\vec{\theta})$ and $T_R(\vec{\theta})$. It reads:}

\begin{eqnarray}
T(\vec{\theta}) = \sin^2{({\frac{\pi B \vec{\theta}}{\lambda}})}  \cos{({\frac{2\pi b \vec{\theta}}{\lambda}})}\sqrt{ T_R(\vec{\theta})T_L(\vec{\theta})}
\label{eq:nulldef}
\end{eqnarray}

{Simulated transmission maps are shown in Fig.\,\ref{fig:kin_data}.
The analytical expression of the astrophysical null can then be approximated as \citep[see][ for the complete expression]{Mennesson:2013mz}:}

\begin{eqnarray}
\begin{aligned}
N_{\rm ast}(\lambda) \simeq \frac{1}{F_*}  \int&  \left(I_d(\vec{\theta})+\left(\frac{\pi B {\theta}_*}{4\lambda}\right)^2\right) T(\vec{\theta}) d\vec{\theta}  
\end{aligned}
\label{eq:nulldef}
\end{eqnarray}

for a source composed of a partially resolved central star of diameter ${{\theta}_*\ll \lambda/B}$ and flux ${F_*}$ and an extended circumstellar disk of diameter ${{\theta}\gg \lambda/B}$ and sky brightness distribution $I_d(\vec{\theta})$.
The contribution from the central star to the observed null is very small ($\left({\pi B {\theta}_*}/{4\lambda}\right)^2\sim0.06\%$) given eta Crv's 0.8 mas photospheric diameter.
{In the case of an extended source, the measured null level is not only affected by the long baseline nulling pattern (fast oscillating squared sine term), but also by the cross fringe pattern (slowly oscillating cosine term).}

{The science data consists of the four astrophysical null measurements (hereafter ``nulls'') dispersed across the N band presented in Figure \,\ref{fig:kin_data}. }
{Because the projected long baseline of the interferometer changes as the sky rotates for the 4 epochs (1 on the first night, 3 on the second night), the nulls correspond to 4 different projected baselines ranging from 53 to 84\,m and covering as many spatial frequencies. The orientation of the telescopes with respect to the target also varies resulting in a fringe pattern that rotates with time.
Owing to an intermediary focal plane pinhole, the field of view is at maximum 450\,mas (FWHM), along the direction perpendicular to the left-right split. Thus at \etacrv's distance the emission from further than approximately 4\,AU, where the transmission is zero, cannot possibly contribute to the measured null. In fact, the transmission map shows that regions of positive transmission extend no further than 2\,AU in radius.
The first constructive peak of the nuller is at 12.3\,mas at 10 $\mu$m for the 84m baseline and 19.6 mas for the 53 meter baseline. 
Dust can be detected down to an inner working angle $\lambda/4B$ which correspond to 6.1 mas (0.11\,AU) for the 84m baseline and 9.7 mas (0.18\,AU) for the 53\,m baseline.}
The contour levels on Figure \,\ref{fig:kin_data} represent a possible disk geometry where most of the dust intercepts the first fringes. 
By construction, only part of the inner disk intercepts the constructive fringes. As the maps show, the nulls in the constructive fringes is approximately equal to half of the excess ratio given by the entire exozodi (the other half of the transmission map has zero transmission). The nulls are of the order of 10\% at 11$\mu$m while the Spitzer excess is about 20\%. This implies that they both arise from the same region \textit{i.e.} between $\sim$0.22-0.35\,AU and $\sim$2\,AU. Should the exozodi location be inside or outside these boundaries, it would not be detected by the interferometer. The exact value of the null for each baseline depends on the fraction of the dust that is intercepted by the fringes, offering a fine constraint on the disk geometry, which we will model in Section\,\ref{sec:zodimodel}.

{Complementary LBTI nulling data were later obtained by our team and we discuss them in Sec.\,\ref{sec:lbti}.}\\

%
\begin{table*}[h!tpb]
\begin{center}\caption{KIN observations and simulator setup}\label{tab:kin}
\begin{tabular}{ccccccc}
\hline\hline 
Date		    &	MJD	& Long baseline	(m) & Azimuth LB ($\degr$)	&	Azimuth SB ($\degr$) \\
4/17/2008/07:15:11	 & 54573.30221 & 84.01     &    42.09     &     50.61 \\
5/24/2008/07:27:41	 & 54610.31089 & 69.82     &    41.87     &     106.5 \\
5/24/2008/08:39:47	 & 54610.36096 & 59.35     &    35.39     &     129.0 \\
5/24/2008/09:21:26	 & 54610.38990 & 53.41     &    28.72     &     138.2 \\
\hline
\end{tabular}
\end{center}
\end{table*}

\subsection{Spectral Energy Distribution}\label{sec:sed}

Photometric data are available for \etacrv\ for a wide range of wavelengths. Table\,\ref{tab:photom} lists the SED measurements used in the present paper along with relevant references. The full SED is shown in Figure\,\ref{fig:sed}.
Data from IRAS, MIPS, AKARI, and WISE in the mid-infrared (${\lambda<35\,\mu m}$) constrain mainly the emission from the inner debris disk as a result of the different characteristic temperatures at play. Color corrections were made with significant impact on the IRAS fluxes only (8\%).
Far-infrared and millimeter data from IRAS (upper limits), MIPS, SCUBA and Herschel (${\lambda>50\,\mu m}$) trace mainly the emission from the outer disk. MIPS-SED data from 54 to 94 $\mu m$ data obtained after binning to reduce the resolution by a factor 5 and scaling to MIPS 70 $\mu m$ are also used (courtesy of Kate Su).

Herschel fluxes are particularly constraining as they provide sensitive measurements across the emission peak of the cold belt. We perform our own extraction of the photometry from PACS (70, 100 and 160 \um) in Section\,\ref{sec:pacs} and use the SPIRE photometric measurements (250, 350 and 500 \um) derived with PSF fitting by \citet{Duchene:2014yu}. Background contaminants were carefully removed except at 350 and 500 $\mu m$ where the limiting resolution did not allow to separate them; we use upper limits at these wavelengths.

At optical and mid-infrared wavelengths, we only considered observations performed with space observatories which are scattered by less than 2$\sigma$ while ground-based photometry used by previous authors \citep[\textit{e.g.}][]{Smith:2008fr} are typically higher by 3$\sigma$. 
WISE bands 1 and 2 are saturated thus they are not listed.
Optical magnitudes from Hipparcos and 2MASS are discussed in Sec.\,\ref{sec:photosphere} and listed in Table\,\ref{tab:properties}.
A limit on the near-infrared excess is derived from visibility measurements performed with the CHARA/FLUOR interferometer \citep{2013A&A...555A.104A}. The source was undetected implying that the disk flux is smaller than 2.0\% at the 3$\sigma$ confidence level. This value is converted to a flux given the photosphere level derived in Sec.\,\ref{sec:photosphere}.

\subsection{{\sc{Spitzer}}/IRS spectrum}\label{sec:irs}
\begin{figure*}[t!]
\begin{center}
  \includegraphics[angle=0,width=0.85\textwidth,origin=bl]{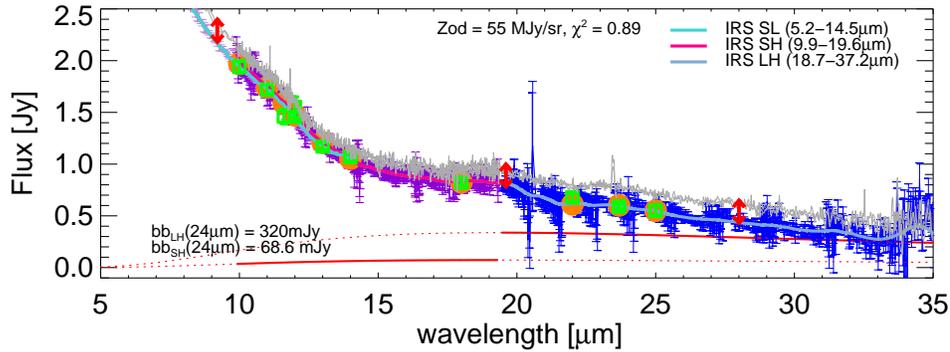}
  \caption{{\sc{Spitzer}} IRS spectrum and SED of $\eta$ Crv in the mid-infrared. The spectra are shown at the native spectral resolution of the three IRS modules with error bars (SL: light purple, SH: dark purple, LH: dark blue) and after applying a Gaussian smoothing filter (see legend). The SL spectrum was reduced with SPICE and includes background subtraction. The green squares are reference fluxes either extracted from the SL spectrum (3\% calibration uncertainty included) or are photometric measurements from MIPS, AKARI, IRAS and WISE. The orange circles are binned from the raw high-resolution spectra after adjusting the background level (``Zod'') until the best agreement with the reference fluxes is achieved. The background level is approximated by 265\,K blackbodies (red curves) adjusted to each slit area. The gray curve is the original \citet{Chen06} spectrum ($\pm 1 \sigma$) displayed for comparison, the difference is highlighted by the red arrows.}\label{fig:IRS}
  \end{center}
\end{figure*}

\begin{figure}[h!btp]
\begin{center}
  \includegraphics[angle=0,width=0.99\columnwidth,origin=bl]{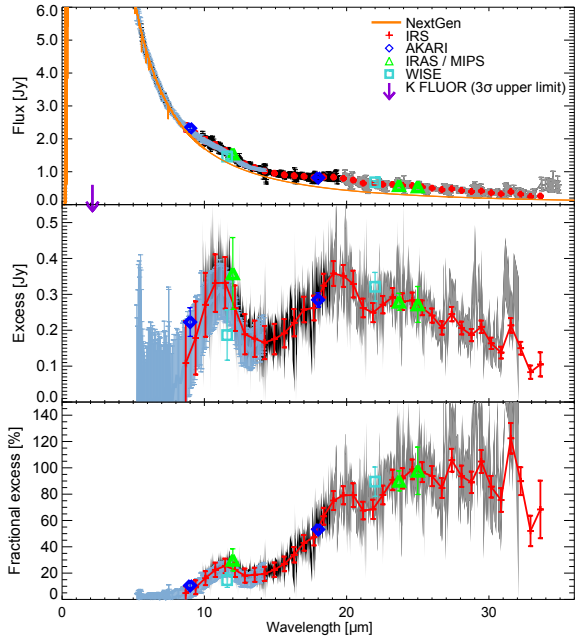}
  \caption{{{\sc{Spitzer}}}/IRS spectrum (top panel), excess spectrum (middle panel) and relative excess spectrum (lower panel) of $\eta$ Crv in the mid-infrared. The black and the grey curves are the high-resolution SH and LH spectra respectively. The blue curve is the SL spectrum. The error bars account for the statistical error and an additional 3\% calibration uncertainty. Measurements with a relative error higher than 15\% are excluded. The NextGen spectrum depicted in orange is subtracted in panel 2, and the spectrum is divided by it in panel 3. The red curves show the composite IRS spectrum resampled with 35 linearly-spaced points from 8 to 35 $\mu m$.}\label{fig:irs_data}
  \end{center}
\end{figure}

Eta Crv was observed by {\sc{Spitzer}}'s IRS spectrograph on 5 Jan 2004 as
part of the IRS instrument team's guaranteed time.
Both high spectral resolution modules (SH and LH) were utilized, 
covering wavelengths from 9.9 to 37.2 \um\ (9.9 to 19.3\um\ and 20.0 to 36.8\um\ respectively).
Data were also obtained with the short-wavelength low-resolution
module (SL1/2) ranging in wavelength from 5.2-14.5 \um,
partially overlapping with the high-resolution spectra.
Long-wavelength low-resolution spectra (LL1/2) were not taken.
The results have already been analyzed by \citet{Chen06};
we consider here the same data, drawn from the {\sc{Spitzer}} archive. 

{The IRS slit was moved according to a $2\times3$ dithering pattern. In order to assess possible slit loss, we extract spectra from the six measurements and fit Gaussians to the flux in several bands versus positional offset perpendicular to the slit. We verify that the central pairs of slit positions maximize the extracted flux. In 4 representative bands of the high resolution spectra, the Gaussian fit shows that the slit is centered on the star with a precision better than 0.3\arcsec\ and that any possible flux loss is smaller than 1\%. We finally use the single slit position that maximizes the flux which alleviates any possible off-axis pointing issue. 
We assess possible fringing issues using the {\sc irsfringe} package on all subsets of spectra as well as on the excess spectra. The tool detects no fringes, the amplitude of which was expected to be smaller than the error bars anyway.
Differential removal of the background was used when available. For the high-resolution spectra we develop a methodology to perform a fit to the background. This method is preferred over previous authors choice to apply a scaling factor to the spectrum to match photometric measurements.}

The overall calibration of the high-resolution spectra is less well known due to uncertainty in the level of background emission.
The problem arises from a lack of offset background measurements.
We set the amount of background subtraction by optimizing its level to match the SL spectrum and the relevant photometric measurements previously introduced (MIPS24, AKARI, WISE, IRAS25). Reference fluxes from the SL spectrum are extracted at 10, 11, 12, 13 and 14 \um\ using a direct interpolation and adding 3\% calibration uncertainty. 
Based on a generic infrared background calculator provided by the {\sc{Spitzer}} Science Center \citep[see][]{2003Icar..164..384R}, the background level (``Zod'') is between 23.9MJy/sr (medium-background) and 65.6MJy/sr (high-background) at 24 \um. We assume it varies with wavelength as a 265\,K blackbody, i.e. that it is dominated by zodiacal emission, and we integrate its emission over the SH and LH slit areas ($11.3\arcsec \times4.7\arcsec$ and $22.3\arcsec \times 11.1\arcsec$ respectively). 
After binning the high-resolution spectra at the reference wavelengths, we look for the smallest least-square fit to the reference fluxes by adjusting the parameter ``Zod''.
The result is shown in Fig.\,\ref{fig:IRS} and has a background level of 55MJy/sr ($\chi^2_r = 0.91$ with a bin width of 0.7\um ; we also test the convergence with different bin widths) appropriate to eta Crv's ecliptic latitude.
The subtracted flux is particularly large for the longest wavelength module (LH) whose collecting area is $\sim$5 times larger than the shorter module's.
We overplot a smoothed version of the combined high-res spectrum using a Gaussian filter for illustration, revealing that the final spectrum provides a good continuity between the three IRS modules.

Lastly, in order to improve the SNR, the high-resolution data have been binned to 35 linearly-spaced points between 10 and 34 $\mu m$ (Table\,\ref{tab:photom}). 
A calibration uncertainty of 3\% is quadratically added to the binned statistical error. 
Our resulting spectrum for \etacrv\ is qualitatively similar to the one originally published by \citet{Chen06} with variations due to upgrades to the data pipeline and differences in the background subtraction.

To evaluate the circumstellar excess, we subtract the photosphere model discussed in Sec.\,\ref{sec:photosphere} which adds up an error term of 4\% relative to the photosphere flux. Figure\,\ref{fig:irs_data} shows the total spectrum (disk+star), the excess spectrum (disk) and the relative excess (disk/star). We consider that the excess spectrum below $9\mu m$ is not significant ($\sim1\sigma$ limit) given the uncertainties on the photosphere. The relative excess is smaller than 25\% up to 15$\mu m$, where it rises steeply up to 25$\mu m$ with a notable dip between 20 and 23$\mu m$. Then it flattens up to 30$\mu m$, and possibly decreases although the data quality degrades. In the rest of the study, the IRS spectrum is only considered between 9 and 33\,\um.
Table\,\ref{tab:relative_excess} lists interpolated values of the relative excess (disk/star) for a selection of characteristic wavelengths. We underline that these are monochromatic and should be properly integrated to estimate the flux in a given filter. In this table we assume the photosphere is well determined and we neglect the uncertainty on its spectrum.

\begin{table}[h!tpb]
\caption{Relative excesses (disk/star) at characteristic wavelengths for the inner and the outer disk.}\label{tab:relative_excess}
\begin{center}\begin{tabular}{ccc}
\hline\hline 
Wavelength & \multicolumn{2}{c}{Relative excess ($F\dma{tot}-F\dma{*})/F\dma{*}$} \\
(\um) & Inner & Outer \\
\hline
8.0\um & $(4.42\pm5.10)\times10^{-2}$ & --\\
9.0\um & $(6.54\pm5.23)\times10^{-2}$ & --\\
11\um & $(24.1\pm5.47)\times10^{-2}$  & --\\
13\um & $(17.9\pm5.35)\times10^{-2}$  & --\\
15\um & $(22.9\pm5.47)\times10^{-2}$  & --\\
20\um & $(80.1\pm7.62)\times10^{-2}$  & --\\
25\um & $(98.3\pm7.53)\times10^{-2}$  & --\\
31\um & $(85.0\pm9.58)\times10^{-2}$	& $<2.2\times10^{-2}$ \\
\hline
55\um & <0.60	& $3.61\pm0.530$ \\
70\um & <1.15 & $6.19\pm0.19$ \\
100\um & <3.06 &$16.4\pm0.5$ \\
160\um &  -- &$33.6\pm1.5$ \\
250\um & -- &$52.2\pm4.8$ \\
850\um & -- & $64.2\pm7.2$ \\
\hline
\end{tabular}
\end{center}
{{\sc Notes --} {The listed values are always the total excess relative to the photosphere. For $\lambda<50\mu$m, the contribution for the outer disk is negligible, for $\lambda>50\mu$m, the contribution for the inner disk is negligible (within the photometric precision)}. Upper limits on the contribution from the outer disk at short wavelengths, and from the inner disk at long wavelengths are included when relevant.
The stellar spectrum is a NextGen model for 6900\,K and $log\,g=4.5$. The values are extracted from the corrected {\sc{Spitzer}}/IRS spectrum and are monochromatic below 31 \um. For longer wavelengths, the fluxes are extracted from the PACS, MIPS, and SCUBA SED. The error bars includes the statistical error and calibration error but omit the $4\%$ uncertainty on the stellar spectrum. The complete data set can be found in Table\,\ref{tab:photom} in flux units.}
\end{table}

{The new spectrum presents differences with respect to the \citet{Chen06} and \citet{2012ApJ...747...93L} ones.
We find an 11 um feature that is stronger that previously thought and a fainter excess upon 21\,$\mu$m.
This is a consequence of our more accurate photosphere model. 
Furthermore, unlike previous authors we did not use a multiplicative factor to scale the raw data because it seemed to us not to be physically motivated. Subtracting an additive term (the background) instead naturally yields a different slope for the global spectrum. 
We highlight that the previous IRS spectrum is compatible at 1$\sigma$ with our more conservative error bars and we consider this as evidence that it is intrinsically impossible to measure the excess spectrum better than a few percent of the photosphere level.}

\subsection{Herschel images and photometry}\label{sec:pacs}

\begin{figure*}[h!btp] 
\begin{center}
  \includegraphics[angle=0,width=0.49\textwidth,origin=bl]{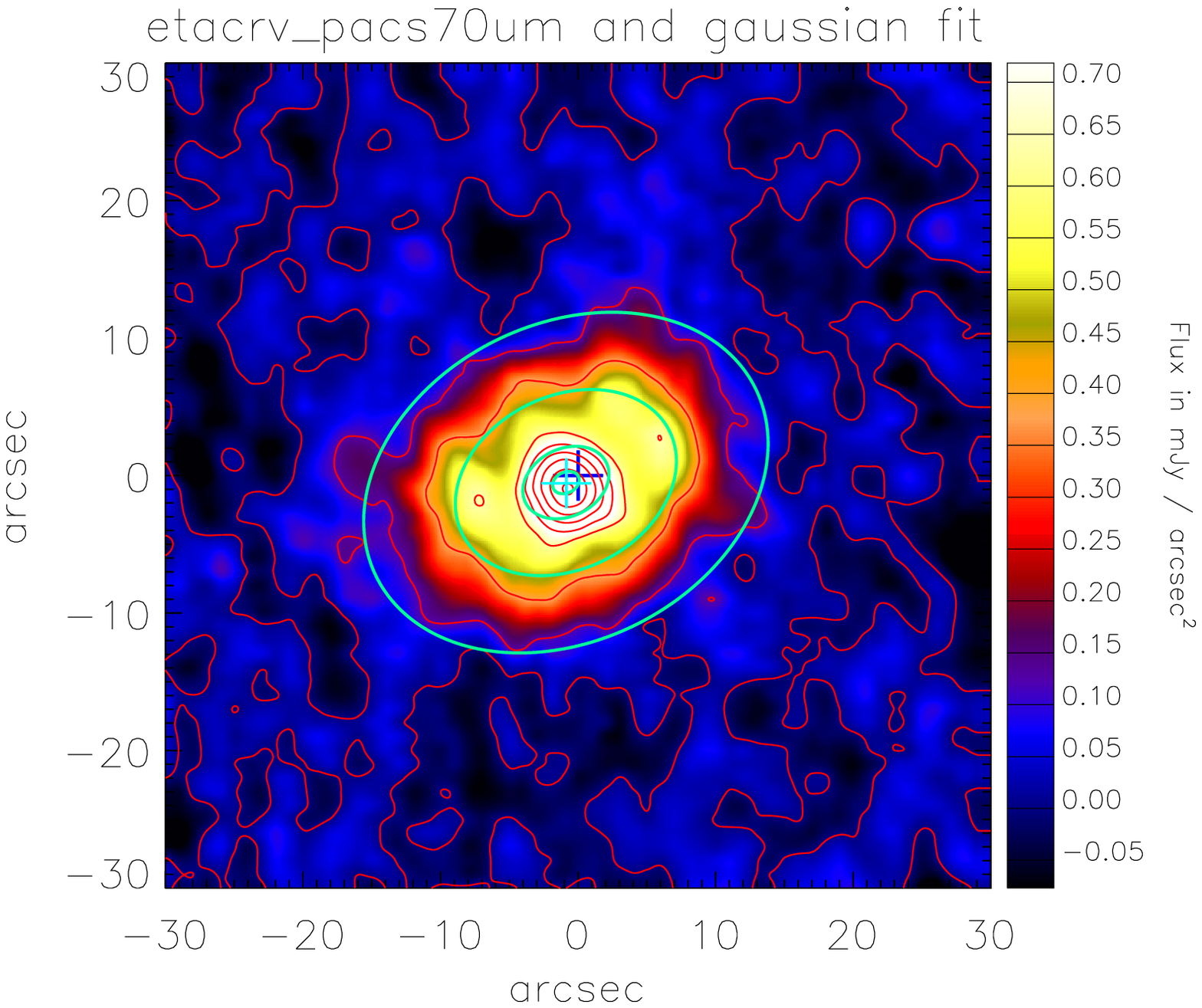}
    \includegraphics[angle=0,width=0.49\textwidth,origin=bl]{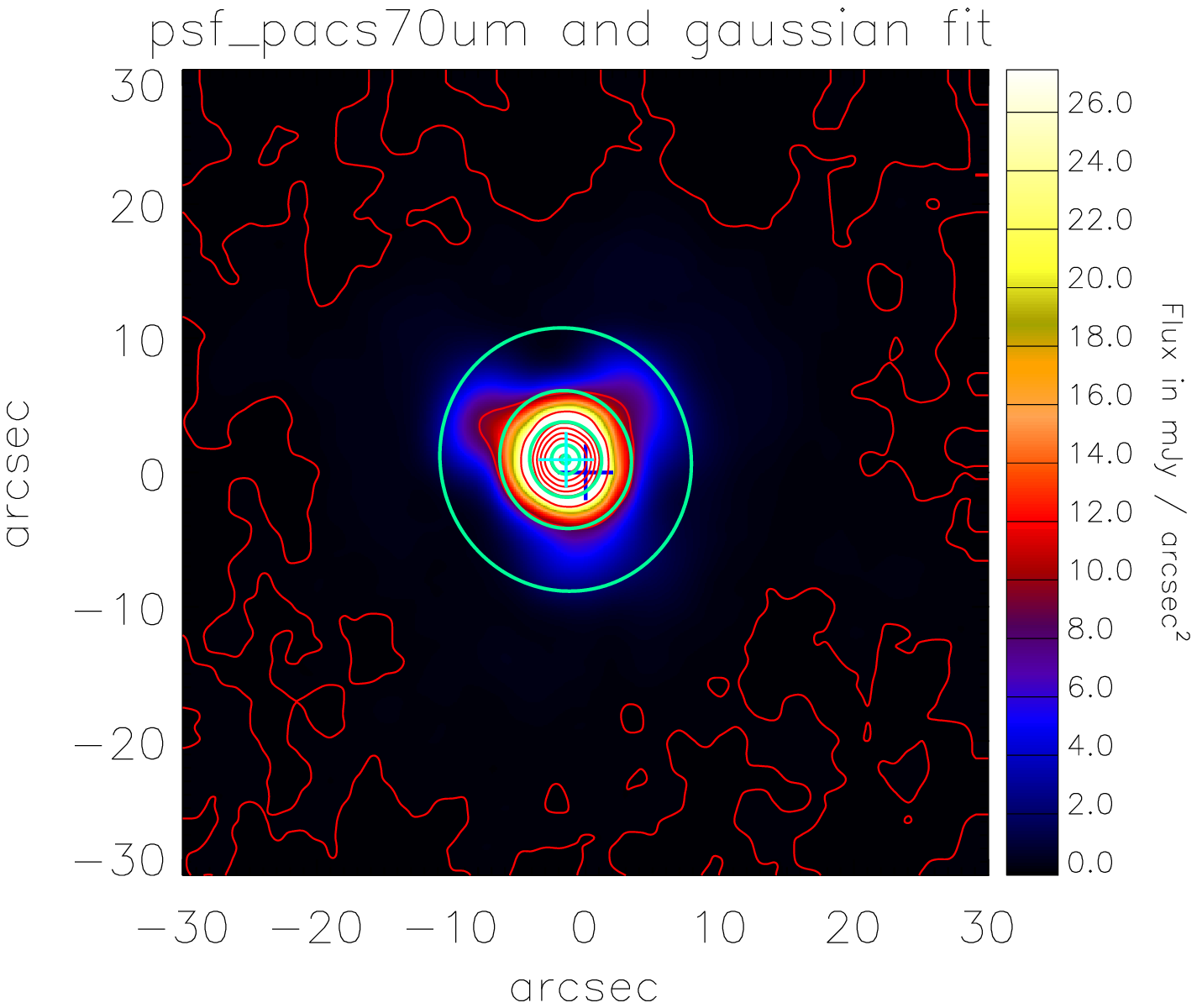}
    \includegraphics[angle=0,width=0.49\textwidth,origin=bl]{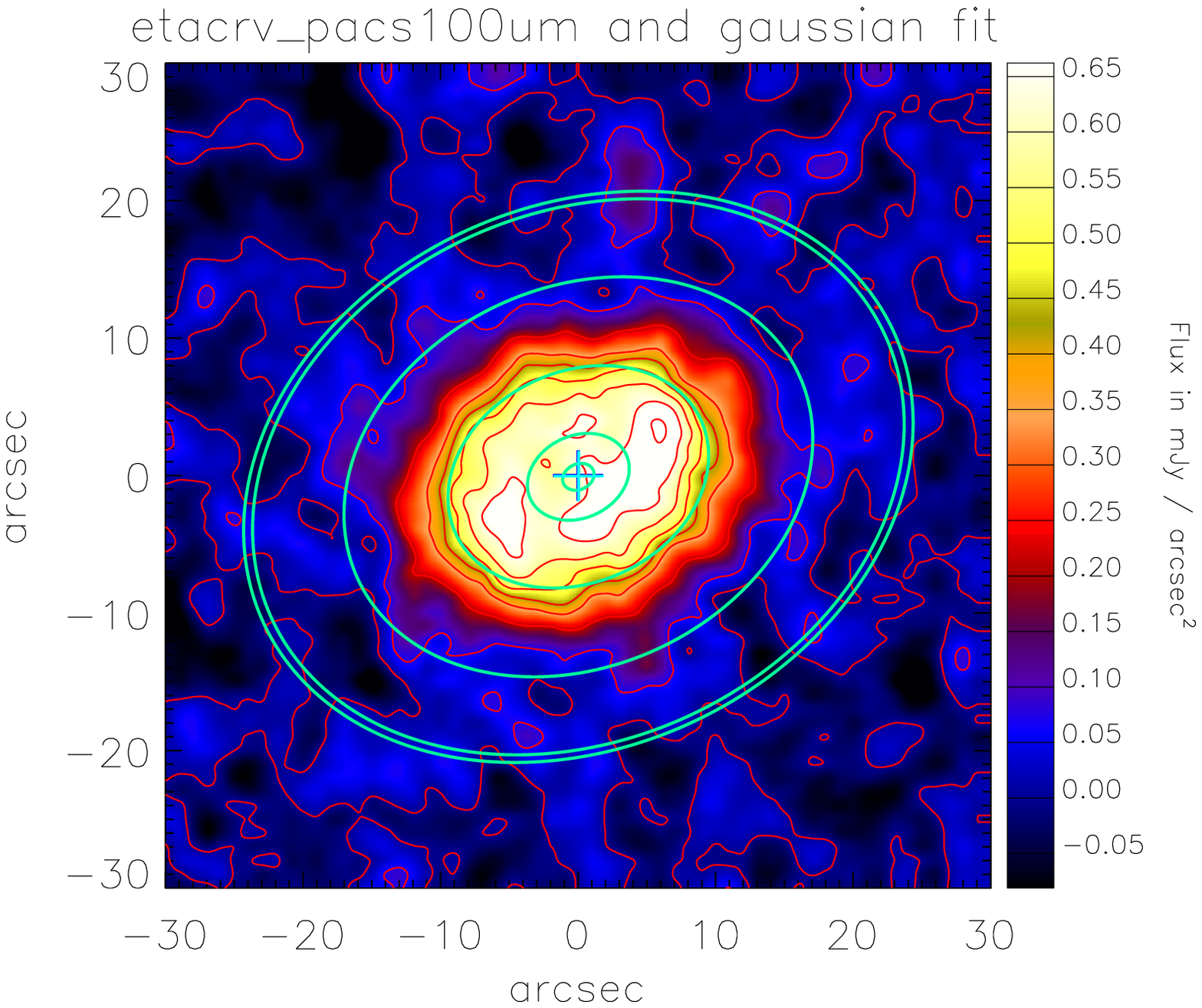}
        \includegraphics[angle=0,width=0.49\textwidth,origin=bl]{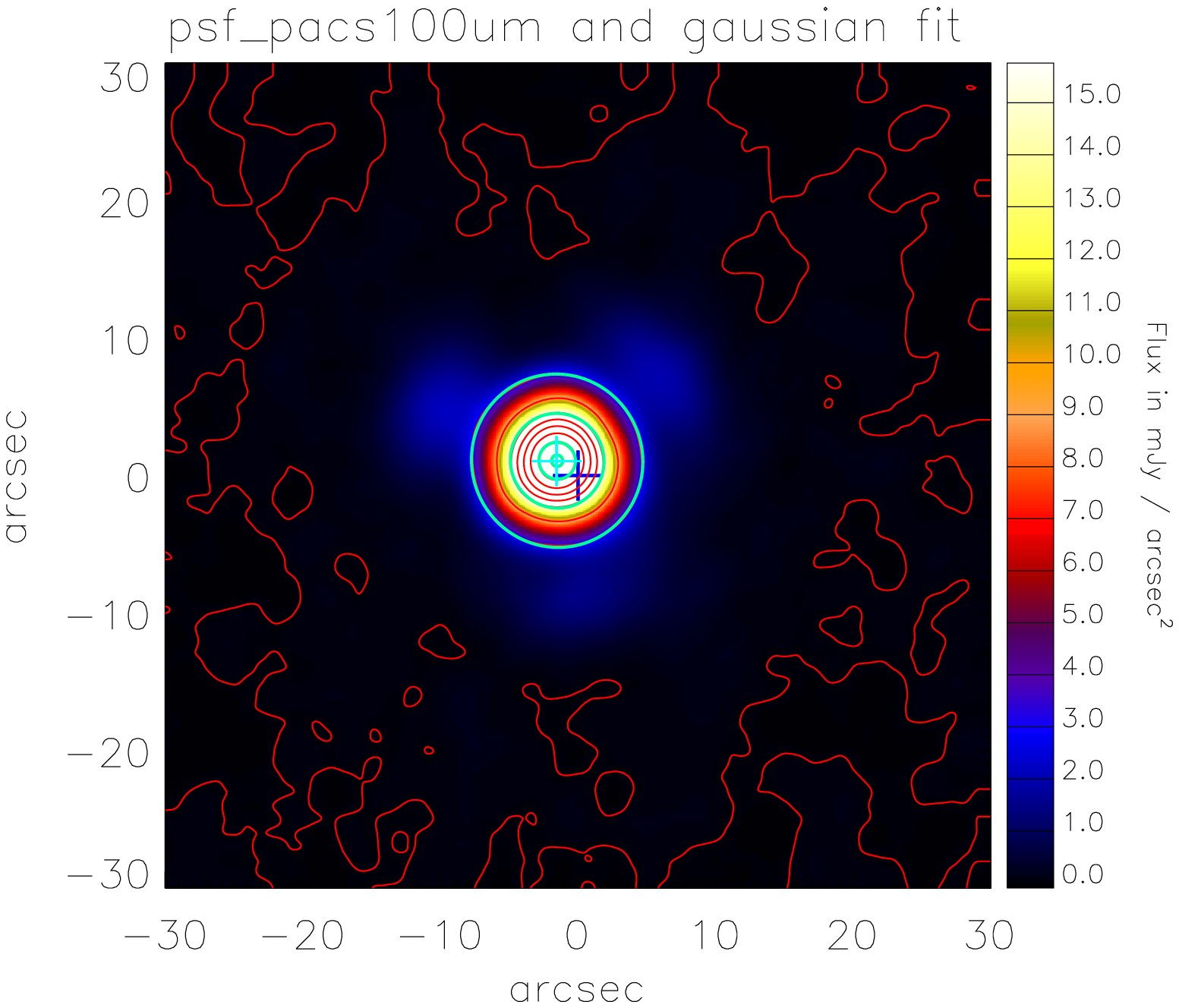}
      \includegraphics[angle=0,width=0.49\textwidth,origin=bl]{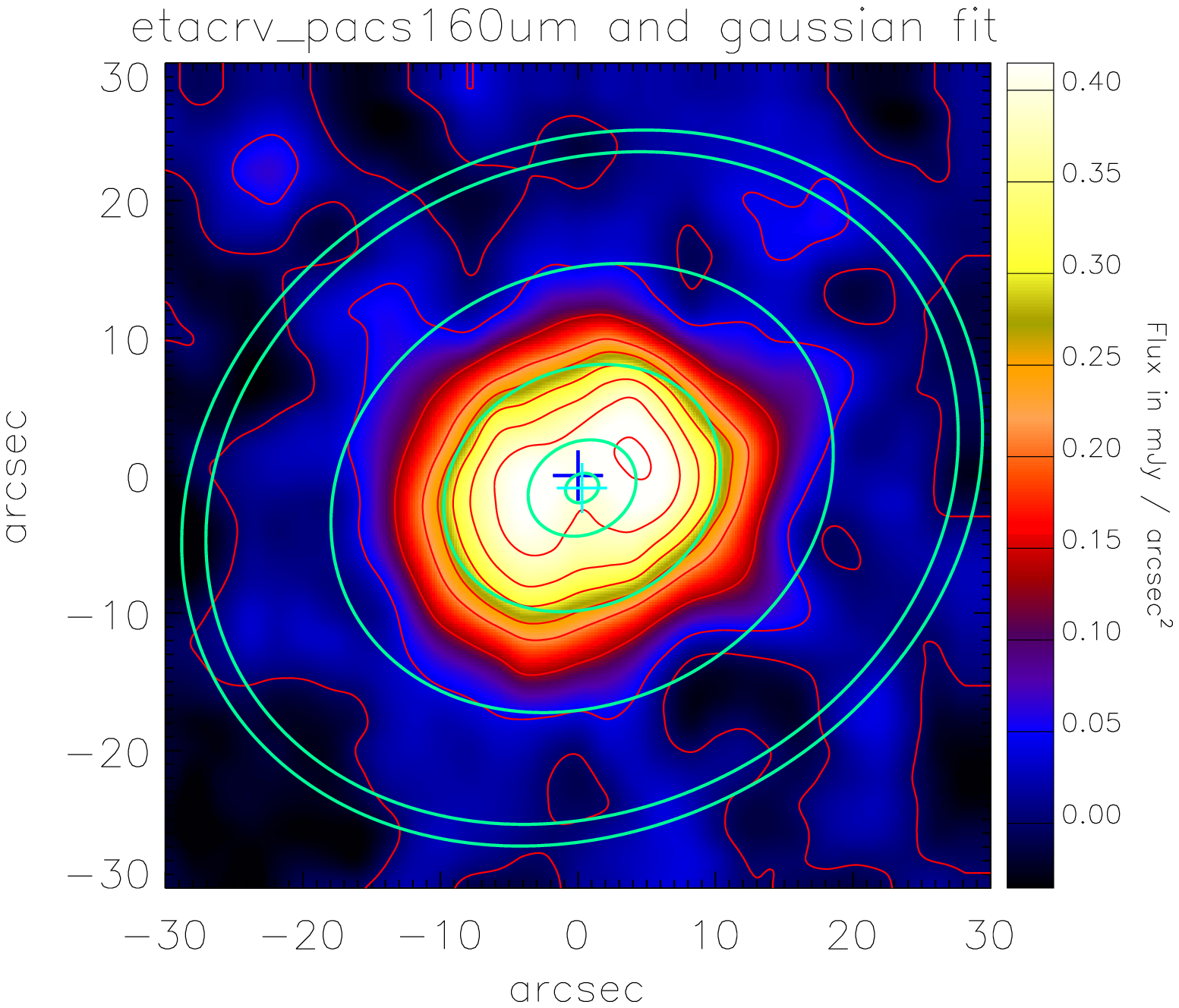}
            \includegraphics[angle=0,width=0.49\textwidth,origin=bl]{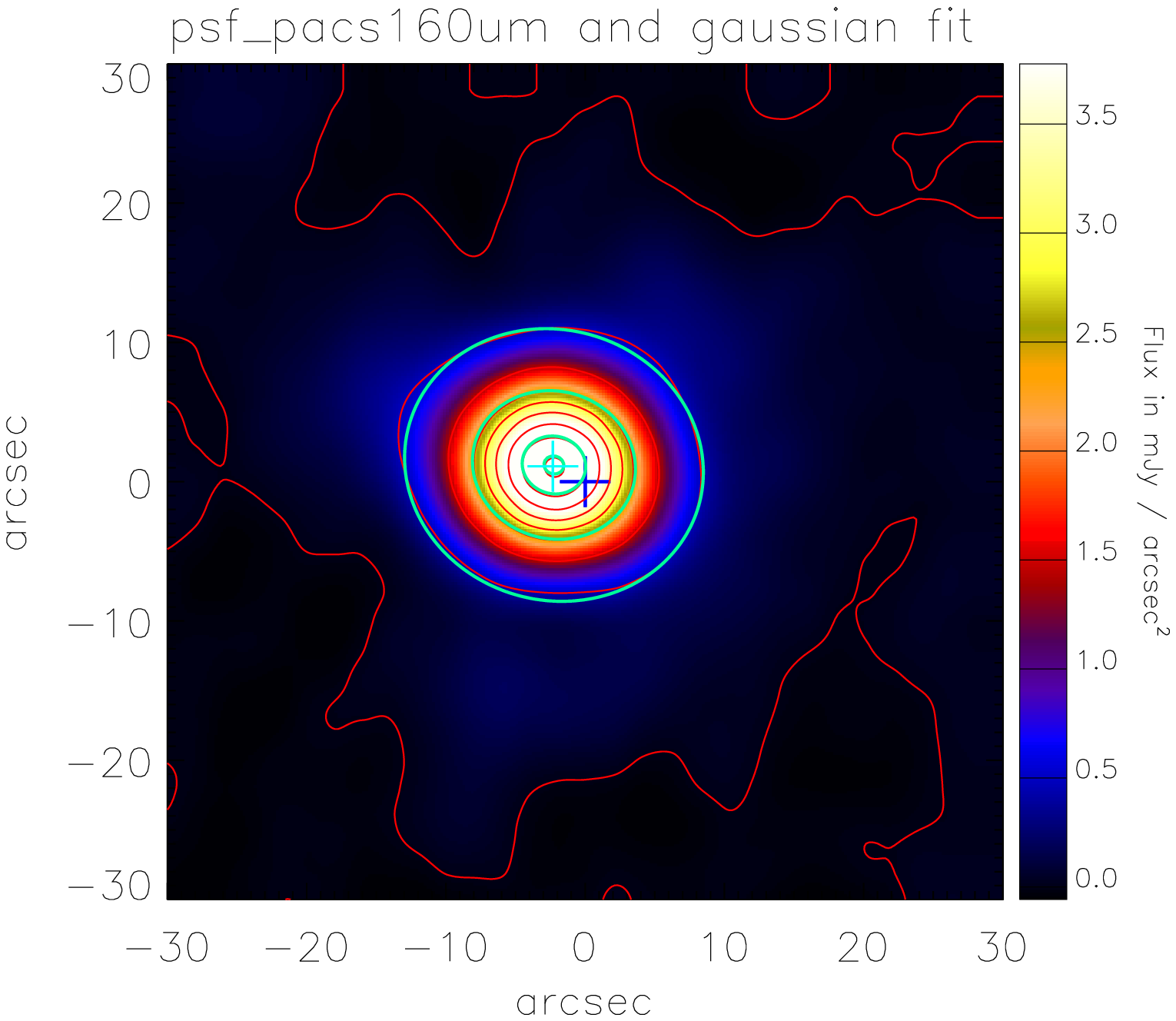}
\caption{From top to bottom: Herschel/PACS 70, 100, 160 {\um} images of $\eta$ Crv (\textbf{left}) and the PSF star (Alf Cet, \textbf{right}). The images are oversampled by a factor 10 wrt to the original resolution (1 $\arcsec$/pixel at 70 and 100 $\mu$m, 2 $\arcsec$/pixel at 160 $\mu$m) and are displayed in the equatorial frame (North is up, East is left). Gaussian fits to the images are shown with contours at levels 0.3\%, 10\%, 50\%, 90\%, 99\% \textit{w.r.t.} the peak of the Gaussian. The dark blue crosses represent the centers of the images. The light blue ones represent the centers of the Gaussians.}\label{fig:herschel_data}
  \end{center}
\end{figure*}

Images were obtained with the Herschel space observatory using the PACS and SPIRE instruments. The data were already presented by \citet{Duchene:2014yu} and consist of resolved images taken with PACS and unresolved images from SPIRE. Here we focus on the PACS images obtained in June 2011 at 70 and 160 \um\ (Obs. ID 1342222622-3) and December 2011 at 100 and 160 \um\ (Obs. ID 1342234385-6).
We perform our own reduction using version 11 of the
Herschel Interactive Processing Environment \citep[HIPE;][]{ott10}
and further develop a dedicated pipeline to extract radial profiles and photometry from the images.
The procedure detailed below is applied to both \etacrv\ and the reference star $\alpha$\,Cet that we use to determine the PACS PSF at all three wavelengths. 

The resolution of the reduced images is chosen to be $1\arcsec/\textrm{pixel}$ at 70 and 100 \um\ and $2\arcsec/\textrm{pixel}$ at 160\um: we refer to these as the original resolution.
We first select a 60$\arcsec$ wide regions centered on the middle of the original images that we rotate by the telescope position angles to align them with the equatorial frame.
The images are then magnified by a factor 10 using a cubic convolution interpolation. We make sure that the results are robust against the value of this factor, finally limiting sampling effects on the derived geometrical parameters such as the disk center.  

The disk geometry is determined by fitting ellipses with a two-dimensional Gaussian profile to the magnified images.
The position angle is consistent within the three images with ${\textrm{PA}=116.2\pm1.1\degr}$. The inclination measurement is more scattered due to the wavelength-dependent apparent size of the disk, with ${\textrm{i}=38.2\pm3.6\degr}$ from face-on. These values as well as the disk FWHM are listed in Table \ref{tab:herschel}. We caution that the uncertainties here only reflect the fitting error, they do not include any measurement error. The latter gives only a limited information on the true physical size of the disk and quantifies how well-resolved it is when compared to the PSF FWHM. The offset between the Gaussian center and the center of the original image is found to be smaller than one pixel (1 or 2$\arcsec$). We conclude that the disk is not shifted with respect to the star within the measurement accuracy. 
Final PACS images of the disk and the PSF are presented in Fig.\,\ref{fig:herschel_data} overlaid with the Gaussian fits.
The background noise that we measure in an annulus located between 22 and 44 arcsec from the star at 70\um\ (\textit{i.e.} between 3 and 6$\sigma$ from the brightness peak assuming a Gaussian profile) is small compared to the statistical error. 

\begin{figure*}[h!btp]
\begin{center}
  \includegraphics[angle=0,width=0.48\textwidth,origin=bl]{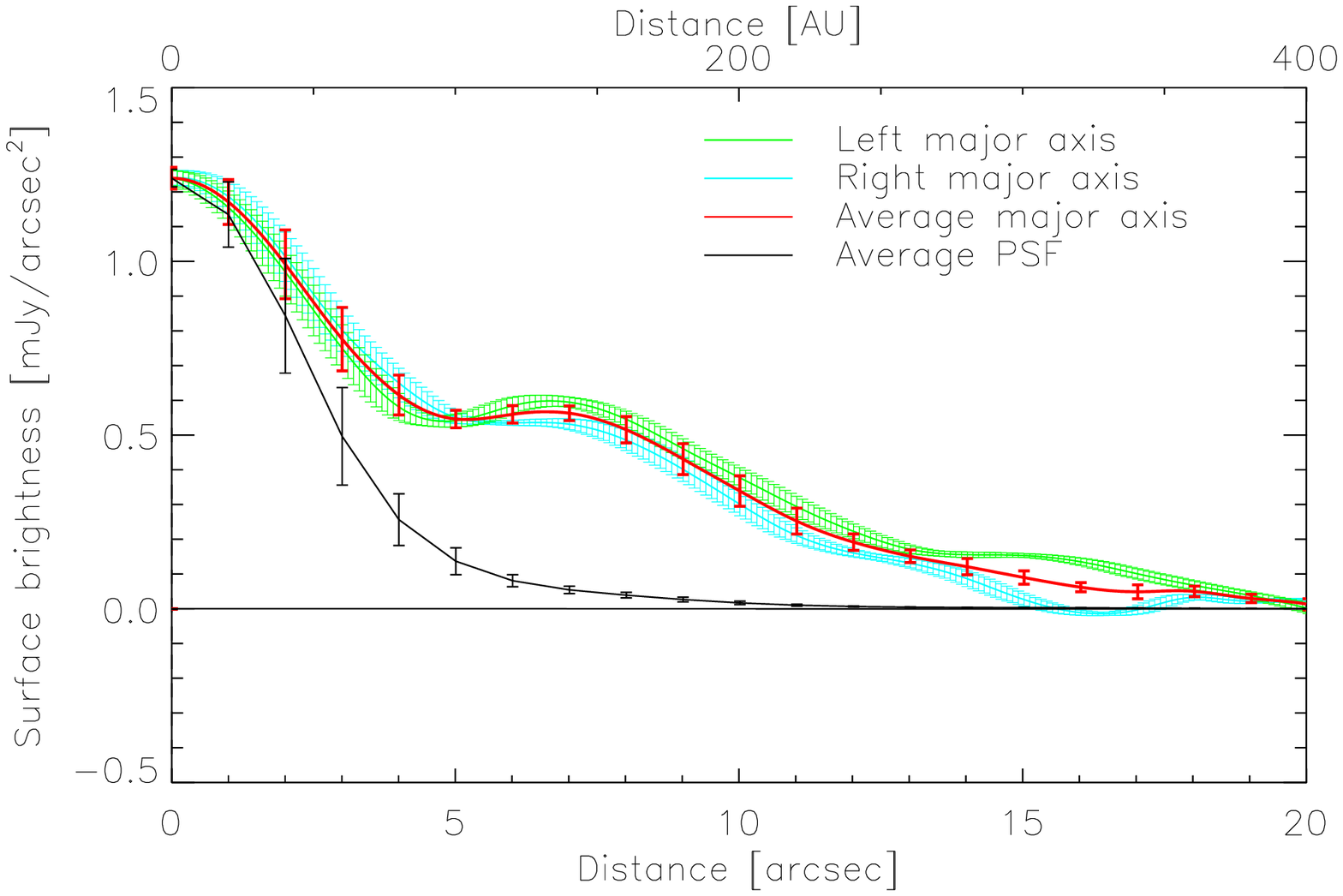}
        \includegraphics[angle=0,width=0.48\textwidth,origin=bl]{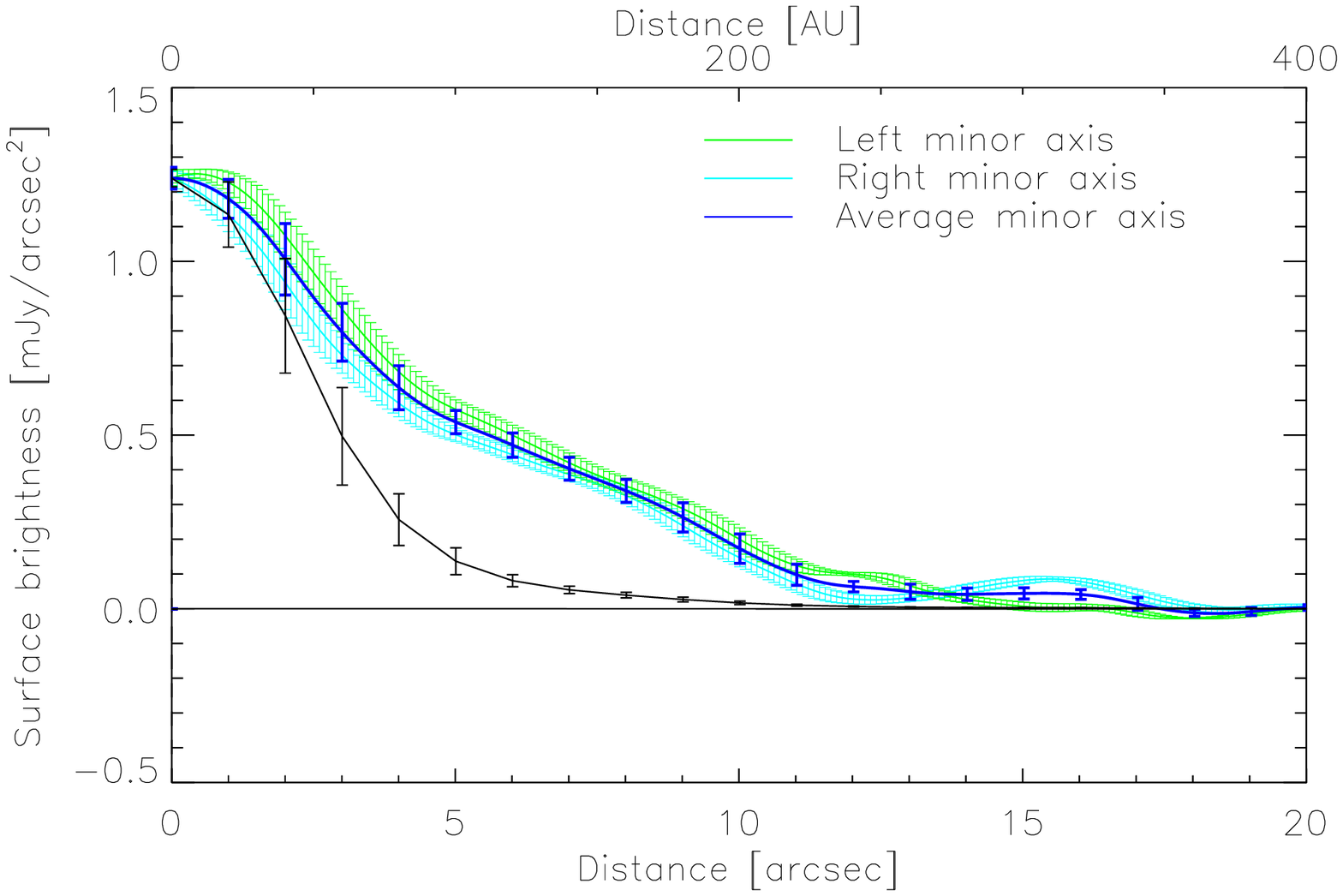}
    \includegraphics[angle=0,width=0.48\textwidth,origin=bl]{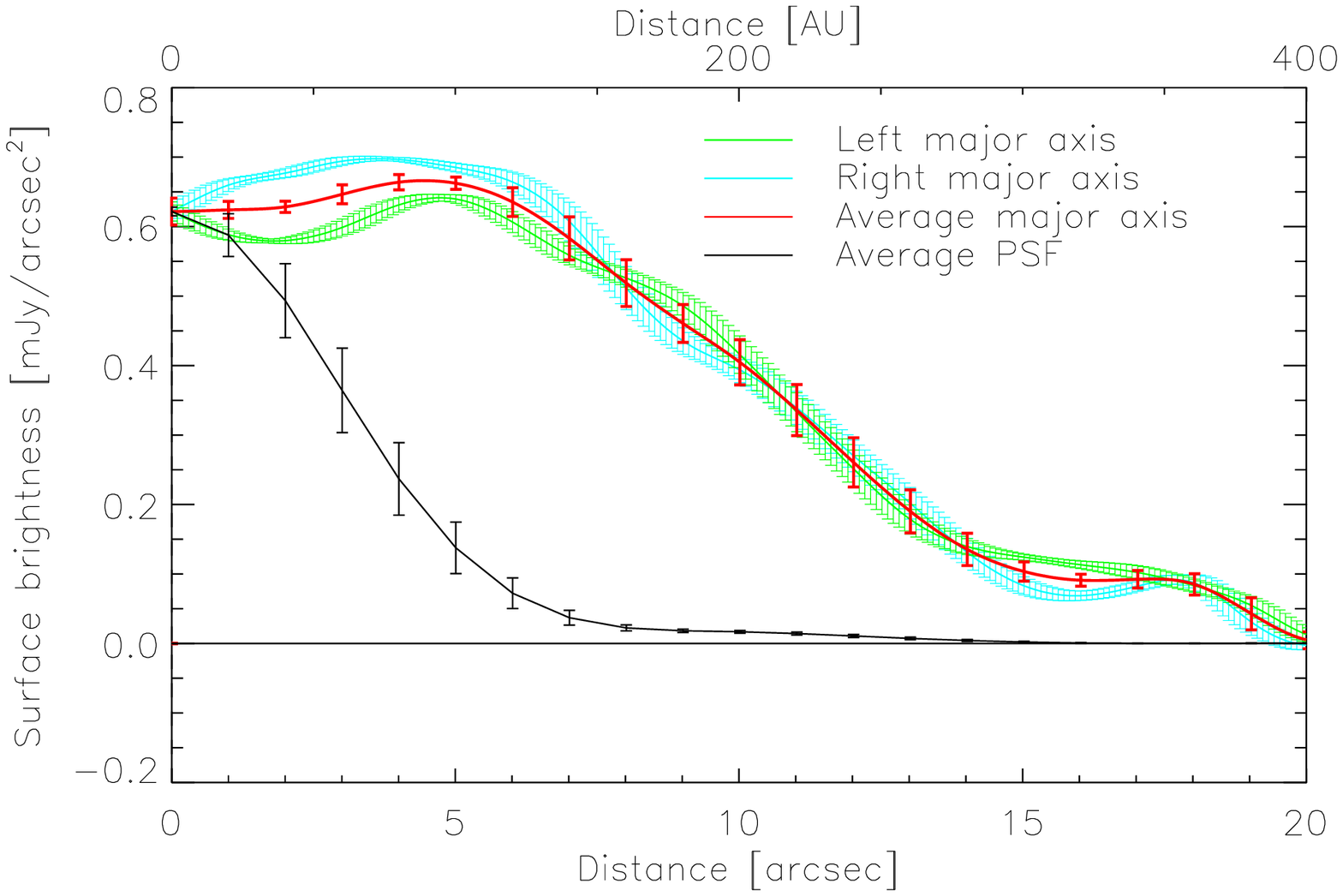}
    \includegraphics[angle=0,width=0.48\textwidth,origin=bl]{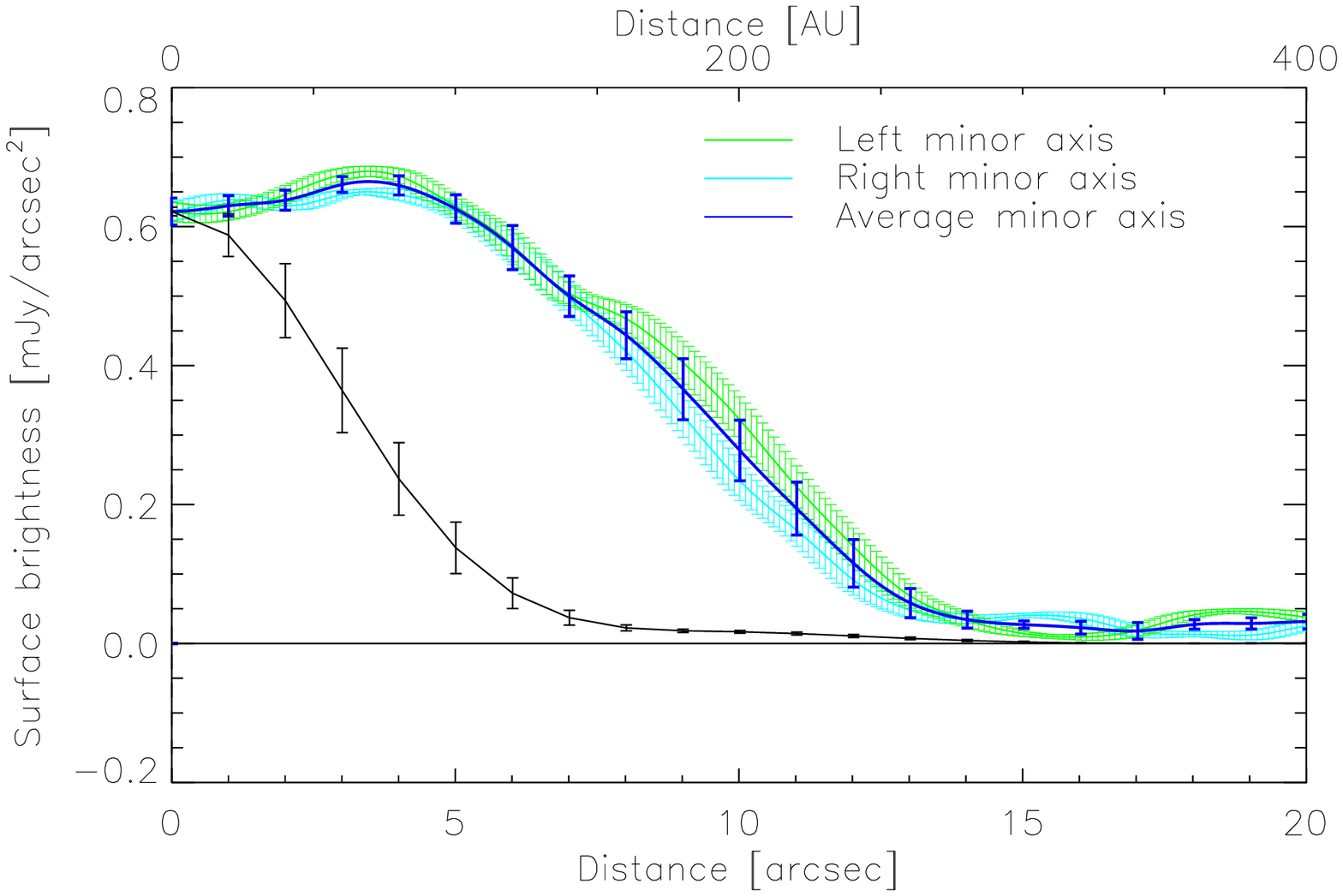}
          \includegraphics[angle=0,width=0.48\textwidth,origin=bl]{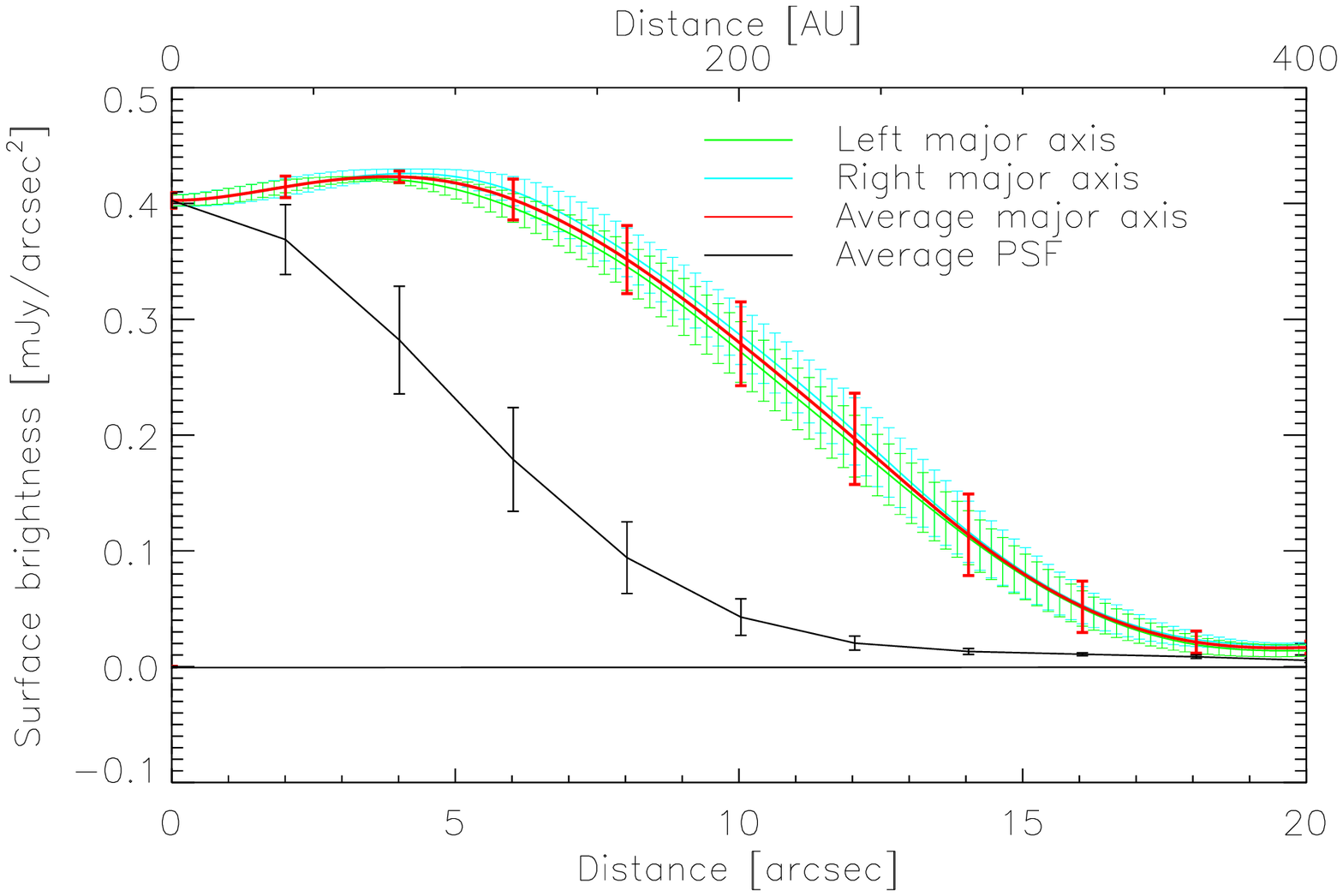}
      \includegraphics[angle=0,width=0.48\textwidth,origin=bl]{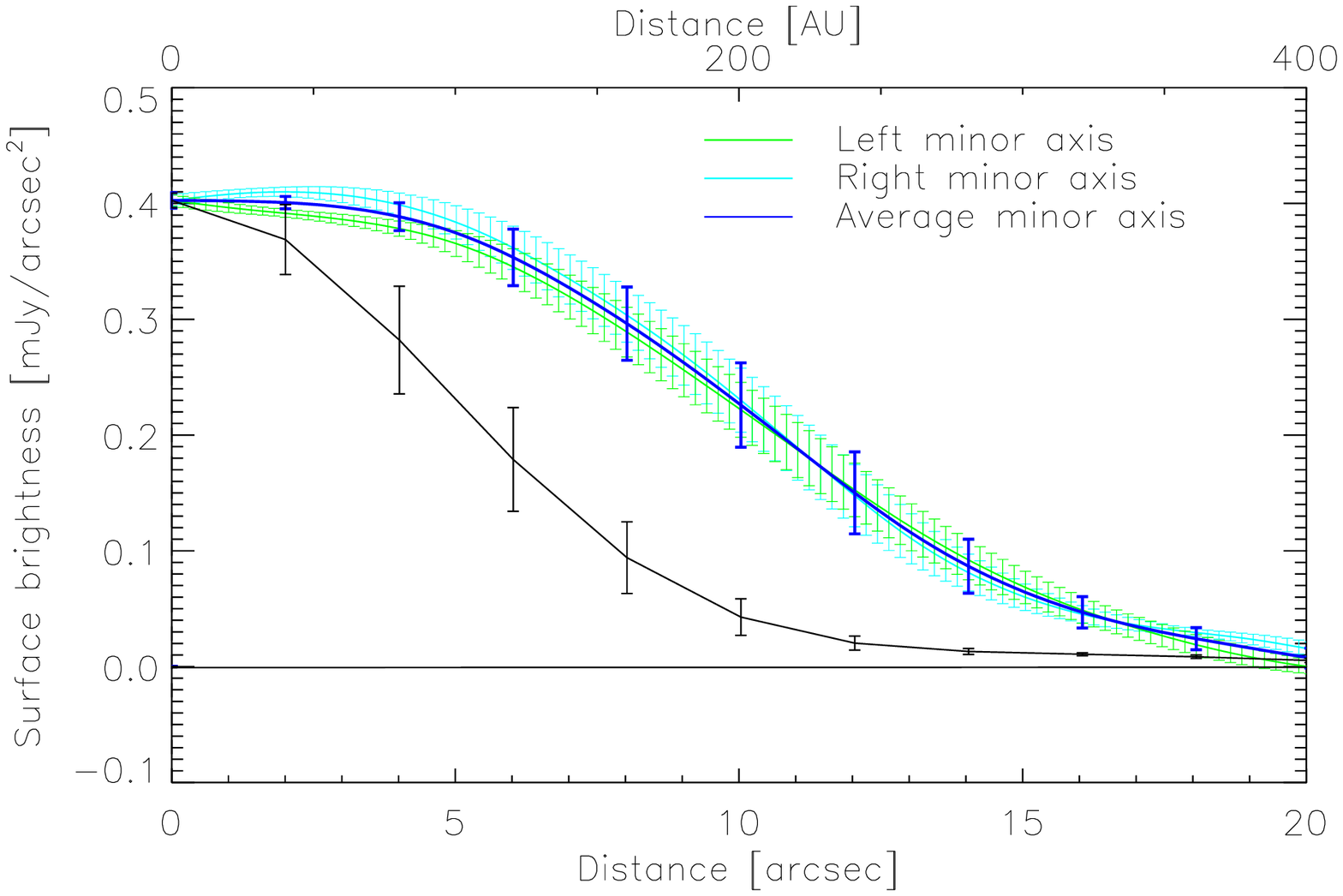}
  \caption{From top to bottom: Herschel/PACS 70, 100 and 160 {\um} radial profiles of $\eta$ Crv along the major axis (\textbf{left}) and the minor axis (\textbf{right}). The light blue curves are taken along each axis at positive separations and the green ones are taken at negative separations in the magnified image. The red and dark blue curves are the average of both brought back to the original image resolution. The PSF profile is the average of the PSF profile along the major and minor axis and it is scaled to the target profile.}\label{fig:herschel_prof}
  \end{center}
\end{figure*}

The next step is to extract radial brightness profiles. We reproduce the above steps with an additional correction for the disk position angle; a value of $38.2\degr$ is used for the three images to ensure consistency between all the data. Radial profiles are extracted along the newly defined major and minor axis, at a resolution matching the original resolution. To achieve this, we measure the mean brightness every 10 pixels in $11\times11$ pixels regions along each axis. The standard deviation in each of these regions defines the statistical error. 
The resulting profiles are shown in Figure\,\ref{fig:herschel_prof}. 

The inner disk is clearly detected at 70\um\ and 100\um.
At all wavelengths, the difference between one side of the minor or major axis and the other is small. 
A closer look at the magnified radial profile shows that at 70\um\ the density profile differs by up to 10\% (at 4$\sigma$) in surface brightness at the locations where the outer belt peaks along the major axis, which is 6.15\arcsec\ on the faintest SE side and 6.35\arcsec\ on the brightest NW side (the positional offset is thus smaller than the pixel size). 

Keeping this in mind, in the rest of the study we consider that the disk surface brightness is essentially symmetric within the uncertainties. We construct combined profiles along the semi-minor and semi-major axis by averaging the measured profiles at positive and negative separations from the center. The final error is the quadratic sum of the statistical error at positive and negative separations, and the background error. These error bars are fairly conservative such that the combined profile is compatible with each pair of measured profiles.
For the PSF case, we additionally average the radial profiles extracted along the disk minor and major axis, and scale the results to match \etacrv\ peak brightness.

Finally PACS fluxes are estimated using aperture photometry. An aperture radius of 20$\arcsec$ is found to be optimal at all wavelengths after accounting for the recommended aperture corrections (0.863, 0.847 and 0.800 respectively)\footnote{PICC-ME-TN-037: \url{http://herschel.esac.esa.int/twiki/pub/Public/PacsCalibrationWeb/pacs_bolo_fluxcal_report_v1.pdf}}. We integrate the flux in ellipctical regions matching the disk PA and inclination. An absolute calibration uncertainty of  2.64\%, 2.75\% and 4.15\% at 70, 100 and 160\um\ respectively is quadratically added to the final error. The total flux is $252.6\pm1.1\pm6.7$ mJy at 70$\mu m$, $297.7\pm1.2\pm8.2$ mJy at 100$\mu m$ and $227.3\pm1.3\pm9.4$ mJy at 160$\mu m$.

We estimate the contribution from the central unresolved component in the 70 and 100 \um\ images using a small aperture that we let vary around a nominal value defined by the PSF HWHM. The combined flux from the star and the inner disk is $\sim72$ mJy at 70$\mu m$ and $\sim63$ mJy at 100$\mu m$. We estimate the uncertainty on these values to be about 5$\%$ and 10$\%$ respectively, representing the dispersion observed as a function on the aperture assumed, and the magnification factor used. These estimates are quite uncertain because the inner disk cannot be strictly separated from the outer one. The 100 \um\ point especially must read as an upper limit. We will use models to revise these values in Section\,\ref{sec:coldmodel}.

The PACS images show no signs for backgrounds contaminants. However the limited resolution of the SPIRE images yield some confusion with background sources located on the North-East side of the disk as observed by \citet{Duchene:2014yu} at 350 and 500\um\ so we use only upper limits for the flux at these wavelengths. 

\begin{table*}[tpb]\caption{Disk parameters from Herschel}\label{tab:herschel}
\begin{center}
\begin{tabular}{ccccccc}
\hline\hline 
wavelength	&	PA* & Major-axis FWHM ($\arcsec$) & Minor-axis FWHM ($\arcsec$) & Inclination ($\degr$) \\
\hline 
70	&	116.5 &  14.3  & 10.7 & 41.5 \\
100 &	115.0 & 16.6  & 12.9 & 38.8 \\
160	&	117.1 &  17.6  & 14.5 & 34.4 \\
Mean$\pm$std &	116.2$\pm$1.1	&		--		&		--	&	38.2$\pm$3.6 \\
\hline
\end{tabular}
\end{center}
\end{table*}

\section{Data selection and modeling strategy}
\subsection{Inspection of the data}\label{sec:data_inspect}
From an observational standpoint we can readily conclude that the \etacrv\ debris disk consists of (at least) two distinct dust populations that we will refer to respectively as the outer ring, and the inner component. This double structure is unambiguously clear in the Herschel images in which radial cuts along the major axis at 70 \um\ (and marginally at 100\um) show a brightness decrease inside of 6\arcsec\ followed by an increase inside of 3\arcsec. This profile makes it possible to separate the inner component from the outer ring and measure their respective fluxes. The inner disk is well detected at 70\um\: with a total flux of 72 mJy, the unresolved component is twice brighter than the star alone. 

The outer ring position angle is $116.2\degr$ and its inclination is 38.2$\degr$ as discussed above and we assume that the inner component shares the same properties. This assumption is reasonable given the fact that the inner disk may originate from the outer one, that they inherited from the same protoplanetary disk, and that any deviation for this orientation is not expected to excess a few degrees as it is observed in the Solar System or in the $\beta$\,Pictoris system.
 The brightness profiles extracted from all PACS images are clearly more extended than the reference PSF, even along the semi-minor axis at 160 \um. At 70 \um\, the width of the belt is marginally resolved and it is clearly separable from the inner component. The peak of the outer disk brightness profile is located at $6.8\pm0.8\arcsec$ at 70\um\ ($123\pm14$\,AU) and $6.4\pm0.4\arcsec$ at 100\um\  ($116\pm7$\,AU) with marginal evidence for side-to-side asymmetry. 
In the subsequent modeling we use a model of a symmetric disk that does not contradict the data given the uncertainties assumed. 
The inner component itself does not appear more extended than the PSF at 70 \um\ implying that it is smaller than the PACS beam. This sets a broad constraint on the inner disk (exozodi) location: most of its flux at 70 \um\ originates from inside of $\sim$40 AU (2.3$\arcsec$).

The analysis of the SED corroborates the separability of the two dust components. The motivation to make the distinction between a warm spectrum and a cold spectrum is illustrated by the blackbody fits in Figure \ref{fig:sed}, which show at least two dust populations with typical temperatures of $350\pm50$\,K and $\sim$40\,K coexist.
The cold component blackbody temperature is uncertain because it depends on how much of the excess comes from the exozodi.
The contribution from the cold component at 34\um\ is negligible, ensuring that the mid-infrared spectrum is not affected by emission from the outer ring.

Hence our strategy to model the outer ring is to adjust the SED in the far-infrared to sub-millimeter domain ($\lambda \geq 54\mu m$) simultaneously with the PACS images independently from the inner disk. The latter is simply accounted for by correcting the stellar spectrum with a point source model that includes both the star and the hot excess. To insure self-consistency the data set for the outer belt is complemented by upper limits at 24 and 33 \um\ obtained after subtracting the hot excess model from the {{\sc{Spitzer}}} data (flux+error).

The first step is thus to derive a model of the inner disk that we can extrapolate to the far-infrared. To achieve this we benefit from a detailed spectrum in the mid-infrared from 9 to 34 \um. The IRS data has already been modelled in detail by \citet{2012ApJ...747...93L} who proposed a thorough spectral decomposition with multiple grain materials. Here we rather focus on pinpointing the disk location while accounting consistently for the dust radiative transfer. This requires a sufficient model of the grain optical properties such as the silicate features, but does not call the need to model the high-resolution spectrum. 
{The spectrum is dominated by broad emission features at 11, 19, 24 and 28 $\mu$m. 
It closely resembles the excess spectrum of the G0V dwarf HD\,69830 at 12.6 pc that \citet{2005ApJ...626.1061B} associated with small crystalline olivine, forsterite and pyroxene dust grains.
Fig.\,\ref{fig:irs_data} suggests that the excess spectrum starts declining at $\sim25\mu m$, with an approximately flat relative excess (excess/star) upon this wavelength. The blackbody temperature for the HD\,69830 excess spectrum was estimated to 400\,K, very similar to \etacrv.}

Assuming a temperature of 300 to 400 K for a 5.06$L\dma{\odot}$ star, the exozodi equilibrium distance is 1 to 2 AU. 
The IRS slit is much larger than this so we can safely conclude that all of the hot excess is accounted for in the {\sc Spitzer} spectrum. Reciprocally, blackbody models show that the contribution from the cold disk is negligible in the IRS range, even more so given that the slit only intercept a small fraction of the outer belt.
However, it is evident that the dust disk does not behave as a blackbody. Furthermore, disk modelling is affected by a well-known degeneracy between grain size and distance, such that the IRS spectrum needs to be complemented by spatial information.

The interferometric data provide the missing constraint on the inner disk geometry. The KIN nulls corroborate the existence of a strong silicate feature at 11 um. Apart from this feature they appear smooth within the measurement errors. 
In Fig.\ref{fig:kin_data} we see that the KIN is sensitive to dust located between 5 to 10 mas (0.1 - 0.2AU) and 0.1 $\arcsec$ (2\,AU) depending on the baseline.
Detailed constructive maps are shown in Figure~\ref{fig:kin_data} overploted with contour levels for the disk. They reveal that if the disk is smaller than ${\sim0.15\arcsec}$ in radius, it does not intercept the destructive fringe pattern. In that case no strong dependence with baseline orientation is expected as we observe here: the nulls show no evidence for azimuthal variations. 
The null excesses become larger with increasing baselines suggesting that the inner disk is compact: the large baseline is more sensitive closer inwards to the first constructive fringes. 

{In summary, the KIN beam only intercepts dust from the inner disk (<2 AU) and cannot be contaminated by the outer disk given the 40 degree inclination of the system to the line of sight. Furthermore, the Spitzer beam only includes contribution from the inner disk because the emission from the outer disk quickly becomes negligible below 40\,$\mu$m and because of the width of the IRS slit, which would capture only a small geometrical fraction of the outer disk. The Herschel beam on the other hand is sensitive to $\lambda>70\mu$m emission from the inner disk owing to the slowly decaying tail of the SED. This effect is easily tackled by incorporating a prior model of the inner disk when studying the outer disk. The spectral spectral decomposition is illustrated in Figure\,\ref{fig:sedcold} that shows how we consistently account for residual signal from each component using upper limits. The spatial arguments are illustrated in Figure\,\ref{fig:radialprofile}.}

\subsection{Modeling strategy}

\begin{table*}[h!tpb]\caption{Model setup}\label{tab:modelsetup}
\begin{center}
\begin{tabular}{lcccc}
\hline\hline 
					&	Inner disk 											& 				Outer disk			\\ [1.2ex]
\hline
Data					& KIN nulls	($n=4\times10$)							& $2\times3$ PACS profiles ($n=2\times(30+30+15)$)	\\
					& IRS (n=34), MIPS, WISE, AKARI (n=6)					& SED points ({\sc Herschel}, SCUBA, MIPS)	\\
					& PACS upper limits (n=2), CHARA upper limit										& {{\sc{Spitzer}} F24 \& F31} upper limits (n=2)				\\[1.2ex]
					\hline
Stellar spectrum		& NextGen												& NextGen + inner disk \\[1.2ex]
\hline
Fixed parameter		& 2D ring, i = 38.2$\degr$, PA = 116.2$\degr$			& 2D ring, i = 38.2$\degr$, PA = 116.2$\degr$				\\
					& $\amax$ = 1 mm											& $\amax$ = 1 mm		\\
					& $r\dma{out}$ = 4\,AU$^*$, $r\dma{in}$ = 0.05\,AU, $\alpha\dma{in} = +10 $ & $r\dma{out}$ = 500\,AU, $r\dma{in}$ = 1\,AU, $\alpha\dma{in} = +10 $ 	\\
\hline
Fitted parameters	& \\ [1.2ex]
\multirow{2}*{\phantom{spacee}Density profile}  					&$\alpha\dma{out} \in \{-3.0, ..., -0.5\}$, n=6, linear 	&	$\alpha\dma{out} \in \{-5.0, ..., 0.0\}$, n=11, linear \\
												&$r\dma{0} \in \{0.05, 4\}$\,AU, n=15, log	&						$r\dma{0}  \in \{10, 300\}$\,AU, n=45, log	\\[1.2ex]
\multirow{2}*{\phantom{space} Grain size}					& 	$a\dma{min} \in \{0.01, ..., 20\}$\um, n=20, log	& $a\dma{min}  \in \{0.05, ..., 50\}$\um, n=45, log	 \\
														& $\kappa  \in \{-5.0, -3.0\}$, n=5, lin  & 
										$\kappa  \in \{-5.0, -2.5\}$, n=6, lin \\[1.2ex]
\multirow{2}*{\phantom{spacee}Scaling}					& \multirow{2}*{mass, \textit{scaled}}										& mass, \textit{scaled}														\\
					& 											&	3 image correction factors								\\[1.2ex]
\hline
\multirow{4}*{Grain composition}	& AstroSil+Forsterite		& AstroSil+ice, n=10 \\
								& Olivine & 						\\
								& Forsterite+Pyroxene & 		AstroSil$\dma{\{1\}}$+C$\dma{\{2\}}$+H$_2$O$\dma{v\dma{Si}}$+porosity$\dma{\mathcal{P}}$ \\
								& Lisse et al. mixture & n = $ 9 \times 15$\\
\hline
\end{tabular}
\end{center}
{{\sc Notes --} $^*$ The model parameters and grain compositions are defined in the text. For the inner disk model, we first test a broader grid extending out to $R\dma{max}=20\,AU$ before restraining the parameter space.}.
\end{table*}

We have at our disposal a wide set of observations that include a spectrum from the optical to the millimeter domain, resolved images of the outer ring and interferometric measurements of the inner ring. To self-consistently interpret the data we need to account for the disk geometry, the dust properties and instrumental models.
These features are implemented in the \gra\ code originally developped by \citet{Augereau1999a}. A detailed mathematical description of the code is presented by \citet{Lebreton:2013uq}. Here we qualitatively introduce our modelling methodology and the motivation for it.
The scheme detailed below is successively applied to the inner component and the outer ring. \\

Debris disks can generally be described as one or several dust belts in thermal equilibrium with a star. They are optically thin and their vertical profiles are generally nearly flat. This is a consequence of the dynamical relaxation that follows the early stages of planet formation in the absence of planetary perturbation \citep{1998AJ....116.2015T,1992Icar...96..107I}. 
For \etacrv\ in particular, we saw that the disk global eccentricity could not exceed $\sim$0.1 providing an order of magnitude for the dispersion of inclinations in the disks. 
Thus for moderately inclined disks, a two-dimensional approach is sufficient unless images at very high spatial resolution are required. 
Furthermore, in the absence of azimuthal asymmetries it is safe to reduce the problem to one dimension.
In the present case both the inner and the outer disk models are described by 1D radial density profiles, that are expanded to 2D and projected at the proper inclination and PA when needed. 

Dust disks originate from collisions in planetesimal belts. Observations and simulations of protoplanetary disks suggest that as leftovers of planetary formation, these belts must be relatively narrow and even in the presence of perturbing planets they have moderate global eccentricities. Once produced, individual grains acquire some eccentricity under the effect of radiation pressure, populating regions outside of the parent belts. In the environment of a massive star drag forces are relatively inefficient such that the regions located inside of the parent-belt are expected to be sparsely fed with dust. 
Thus, an appropriate model for the dust radial density profile is a double power-law centered at a peak position $r_0$, with an outer slope $\alpha\dma{out}$ that serves as a proxy for the grain dynamics. The inner slope is assumed to be very steep ($\alpha\dma{in} = + 10$), such that $r_0$ is closely identified to the inner disk edge.
The power-law expends inwards down to the sublimation distance and outwards up to a large enough distance. 
The scaling of the density profile, $\Sigma_0$, \ie\ the density profile at $r_0$ is calculated at the last stage by least-square scaling each model to the data. \\

Debris disks are by definition collisional systems. The dust grains are the smallest remnants of collisions occurring between kilometer sized planetesimals. The particle size distribution can be approximated by a power-law, the index of which (hereafter $\kappa$) is a proxy of the collisional cascade. The smaller grains have the shorter survival lifetime because they are more sensitive to removal mechanisms (collisions, radiative transfer blowout, drag forces, sublimation). For sufficiently massive stars (K and earlier type stars), there is generally a cut-off size (hereafter \amin) that is theoretically identified to the radiative blowout radius ($a\dma{blow}$).
The upper size corresponds to the larger planetesimals of the collisional cascade and cannot be constrained by observations for extrasolar debris disks. In models we fix its value such that it has no effect on the results ($a\dma{max}\gg\lambda\dma{max}/2\pi$).
Once all parameters of a model are fixed we can calculate the total disk mass $M\dma{dust}$, which is directly proportional to $\Sigma_0$.

\begin{figure}[h!btp]
\begin{center}
  \includegraphics[angle=0,width=0.99\columnwidth,origin=bl]{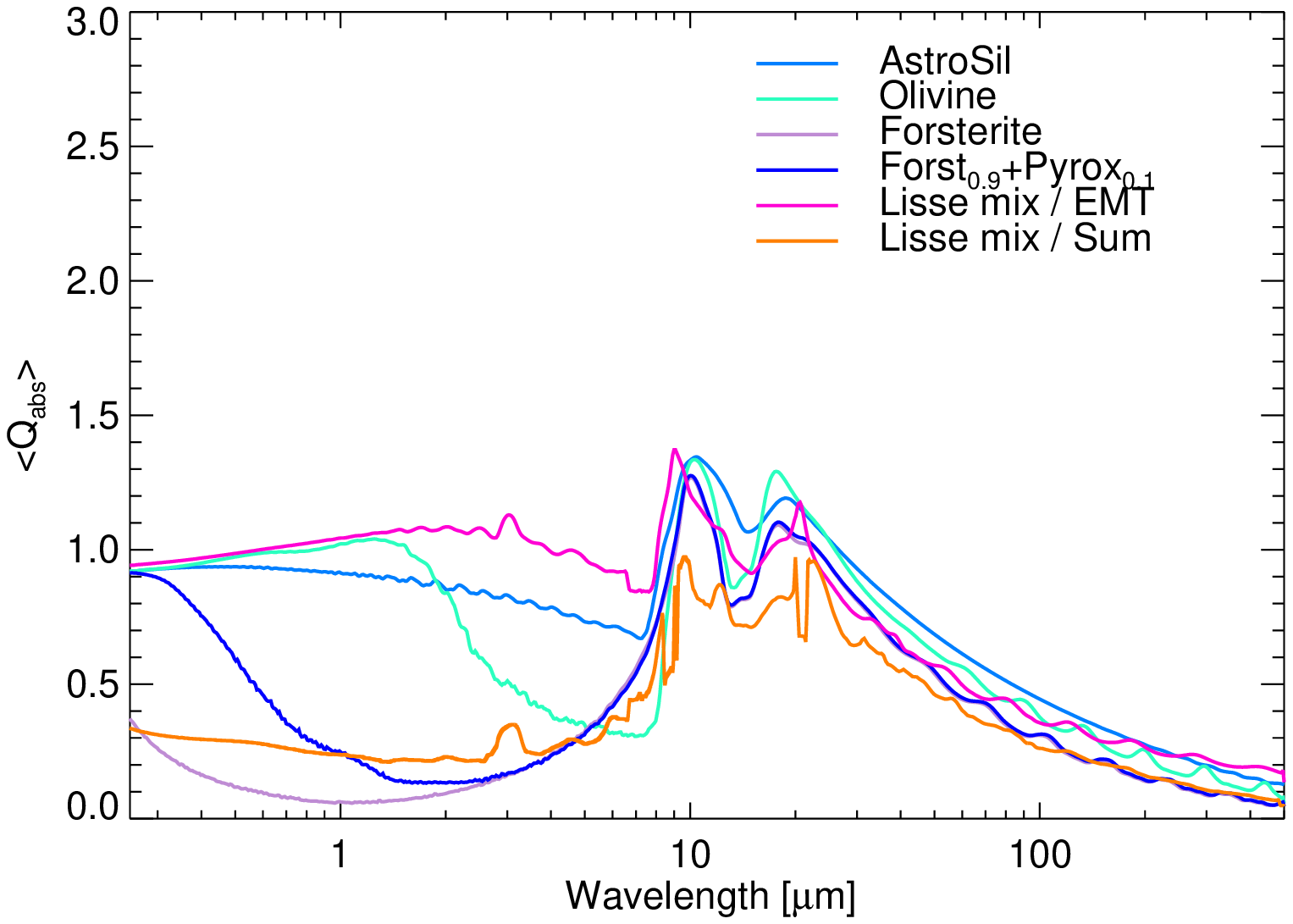}
    \includegraphics[angle=0,width=0.99\columnwidth,origin=bl]{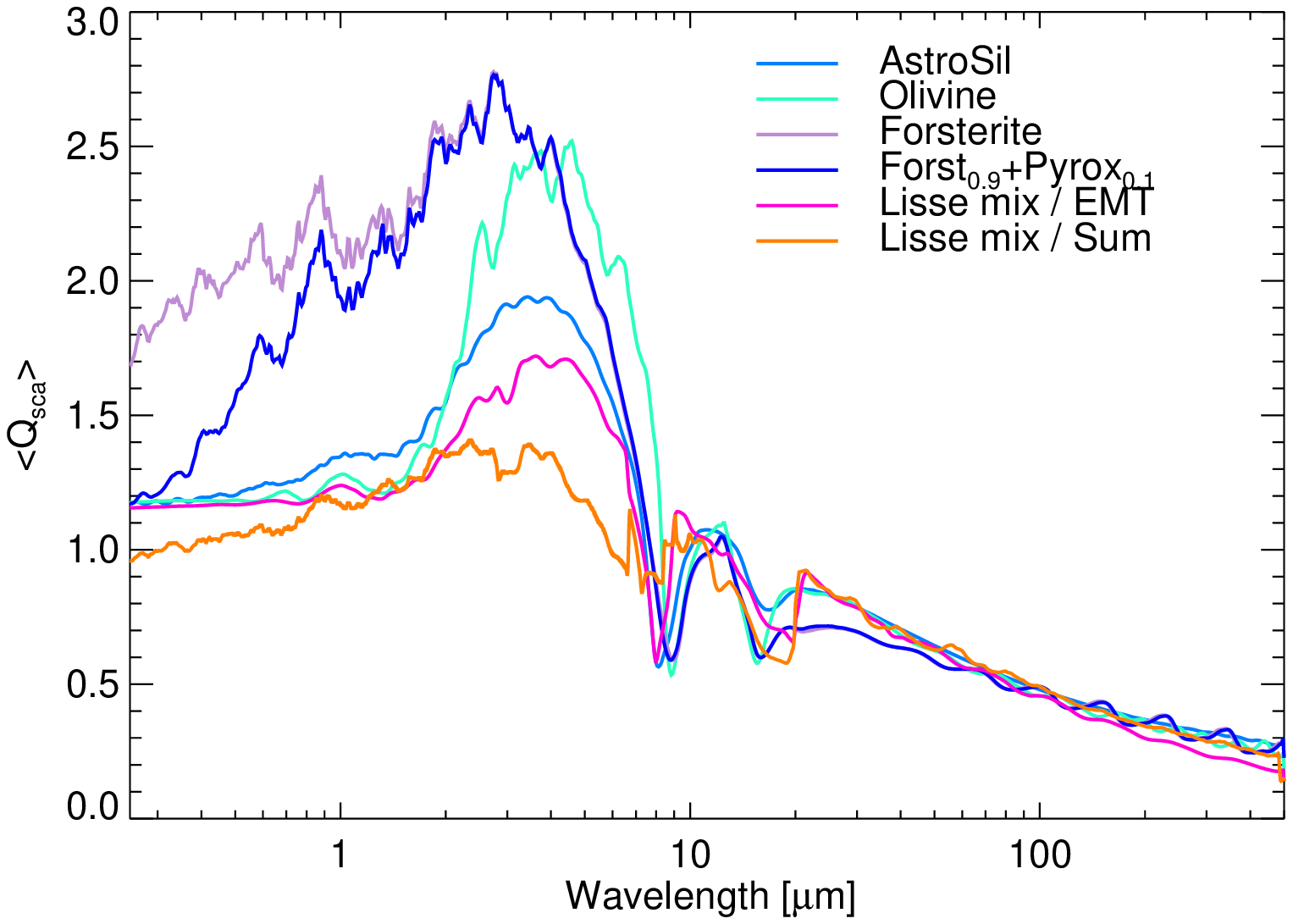}
  \caption{Mean absorption and scattering efficiencies calculated with the Mie theory for the grain compositions used in this study. A power-law size distribution is assumed here with $a\dma{min}=1\,\mu m$, $a\dma{max}=1\,mm$ and $\kappa = -3.5$.}\label{fig:Q}
  \end{center}
\end{figure}

Thus 5 free parameters are used to parameterize a disk density profile and the grain size distribution: $r_0$, $\alpha\dma{out}$, $\kappa$, \amin\ and $M\dma{dust}$. The dust grains scatter starlight and thermally emit depending on their size-, composition- and wavelength-dependent scattering and absorption efficiencies. We compute these assuming the grains are spherical and homogeneous using the Mie theory and a database of optical constants. 
These include dust of the silicate class (astronomical silicate, forsterite, glassy olivine, pyroxene), carbonaceous species, amorphous water ice and vacuum that mimics porosity.
The optical constants were chosen in particular for the availability of measurements over a large wavelength range; when needed the constants were extrapolated. In Table\,\ref{tab:modelsetup} we list the materials explored for both the inner and outer disk. 
In Figure\,\ref{fig:Q} we plot the mean absorption and scattering efficiencies ${\left<Q(\lambda)\right> = \frac{\int_a Q(a,\lambda)\,\mathrm{d}n(a)\,a^2}{\int_a \mathrm{d}n(a)\,a^2}}$ as a function of wavelength, with $a$ the grain size, $\mathrm{d}n(a) \propto a^{-\kappa}\mathrm{d}a$ the size distribution and $Q(a,\lambda)$ the absorption or scattering efficiency. We assumed $a\dma{min}=1\,\mu m$ to $a\dma{max}=1\,mm$ and $\kappa = -3.5$ for this plot.

The optical constants of multiple-material grain are obtained with an effective medium theory (Bruggeman rule, hereafter EMT). The grain composition is parameterized by a volume fraction with respect to the total volume of the grains ($v\dma{x}$). In the following the grain composition is either fixed or parameterized with up to two parameters. At a given wavelength grains with small size parameter ($2\pi a/\lambda \simeq 1$) typically emit more efficiently than blackbodies, significantly affecting the thermal equilibrium distances.

Our algorithm proceeds as follow. First, the optical efficiencies $Q\dma{abs}$ and $Q\dma{sca}$ are computed depending on grain size and composition (independent from the star). Second, equilibrium temperatures are derived for each grain size knowing $Q\dma{abs}$ and the stellar spectrum and they are converted to equilibrium distances. Third, thermal light and scattered light fluxes are computed for each grain around \etacrv. Fourth, the grain fluxes are integrated over the size distribution to produce an image as a function of wavelength. At this stage a large number of models is handled. \\

The models need to be compared to the three types of observations: the SED, the {\sc Herschel} images, and the KIN nulls. 
The SED of the model is obtained by a direct integration of the images. 
Synthetic {\sc Herschel} images are obtained by convolving the models with the properly aligned PACS PSF. Radial brightness profiles are then extracted  from the synthetic images and directly compared with the data. A correction factor is added for each image to insure that the radial profile fitting is not biased by the SED model, adding up 3 parameters to the model \citep{2012A&A...537A.110L}. The $\gra$ code includes a KIN simulator that calculates the nuller constructive and destructive fringe pattern for each epoch and for each wavelength \citep{Mennesson:2013mz}. The images of the model (including the star) are integrated through the KIN transmission map to calculate the interferometric nulls. 

\begin{figure}[h!btp]
\begin{center}
  \includegraphics[angle=0,width=0.99\columnwidth,origin=bl]{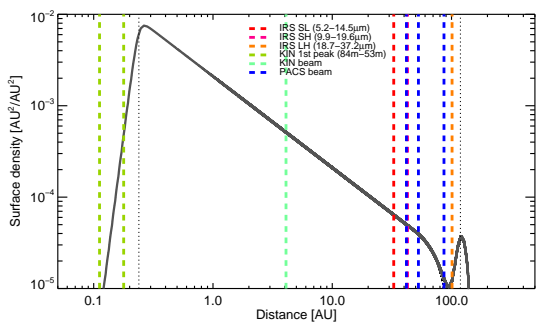}
  \caption{{Radial density of two-component debris disk compared to the PACS and KIN beam FWHM and the width of the IRS slits. Geometrical cross-sections are used rather than optical cross-sections on this figure. The relative importance of the outer belt is visually diminished by the x-axis log-scale. The figure illustrates the separability of the two-components.}}\label{fig:radialprofile}
  \end{center}
\end{figure}

The models are compared to the data using first a least-square optimization quantified by a $\chi^2$ value. Each SED or null measurement is equally weighted when modelling the inner disk. For the outer disk, the SED has an equal weight to one radial profile, the listed $\chi^2$ values are the sum of the $\chi^2$ obtained for the SED and each of the three images.
Then we apply a Bayesian statistical analysis to identify the disk parameters and associated uncertainties. The principle is to associate a probability density to each model as a function of its $\chi^2$ ($P\,\propto\,\exp{(-\chi^2/2)}$), and to integrate it over each parameter, assuming in this case uniform prior probabilities \citep{Lebreton2012a}. 
In table \ref{tab:modelsetup} we summarize the model setup for each of the two disk components, including the parameter range explored. The global SED model is displayed in Fig.\,\ref{fig:sedcold} and a representation of the geometrical surface density profile of the two disk components is shown in Fig.\,\ref{fig:radialprofile}.

\section{Models of the exozodi}
\subsection{Exozodi models: results}\label{sec:zodimodel}

\begin{figure*}[h!btp]
\begin{center}
  \includegraphics[angle=0,width=0.8\textwidth,origin=bl]{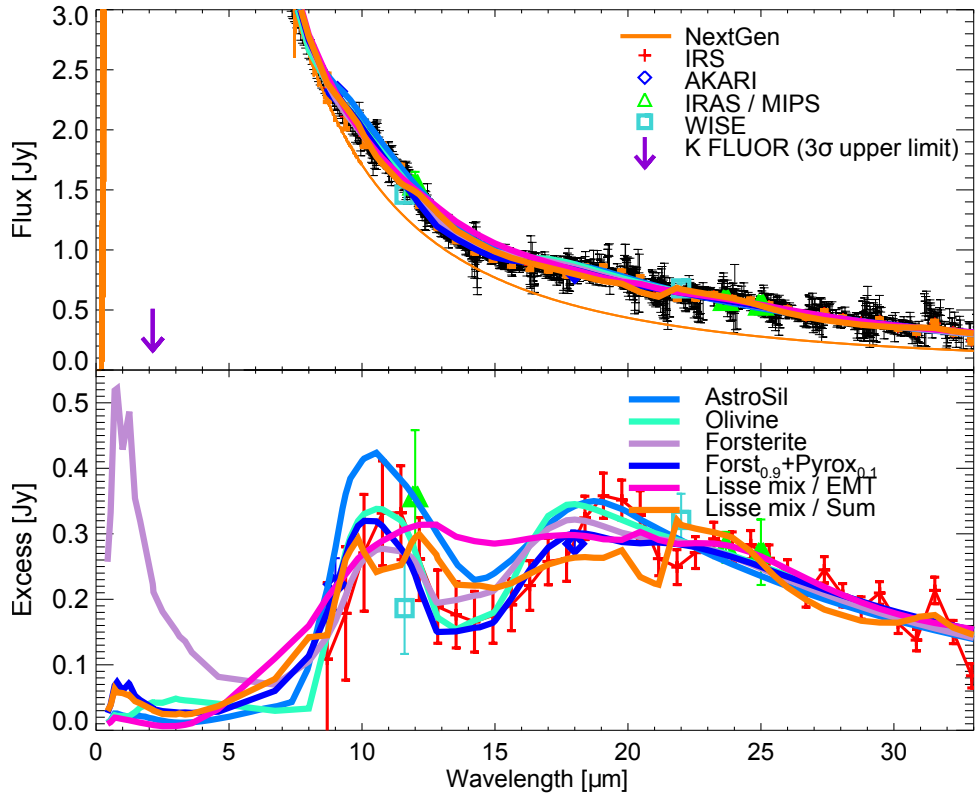}
  \caption{Best fitting exozodi models to the mid-infrared SED from IRS and ancillary photometric measurements. All models presented in Table\,\ref{tab:bestmodels} are shown.} \label{fig:bestfitsed}
  \end{center}
\end{figure*}

\begin{figure}[h!btp]
\begin{center}
  	\advance\leftskip-0.5cm
  	\includegraphics[angle=0,width=0.95\columnwidth,origin=bl]{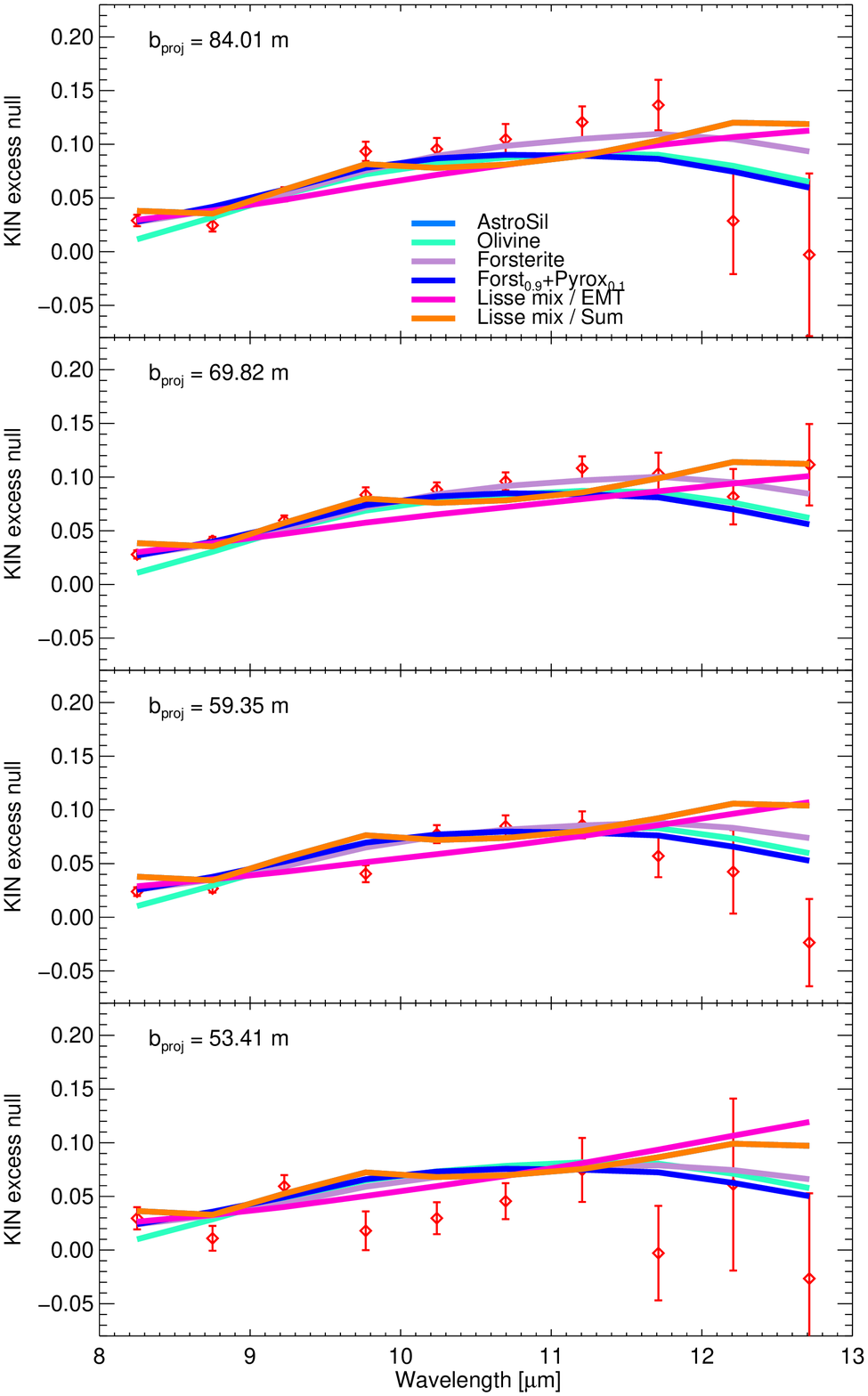}
  \caption{Best fitting exozodi models to the KIN nulls}\label{fig:bestfitnull}
  \end{center}
\end{figure}

Here we attempt to fit the 4 KIN nulls (40 data points) of $\eta$ Crv, its mid-infrared spectrum and upper limits from CHARA and {\sc Herschel} (36+7+3 data points) using a model of the exozodi for several possible compositions with 5 parameters, leading to 81 degrees of freedom (\textit{d.o.f.}).\\

\begin{figure*}[h!btp]
\begin{center}
  	\advance\leftskip-0.5cm
  	  	\includegraphics[angle=0,width=0.95\columnwidth,origin=bl]{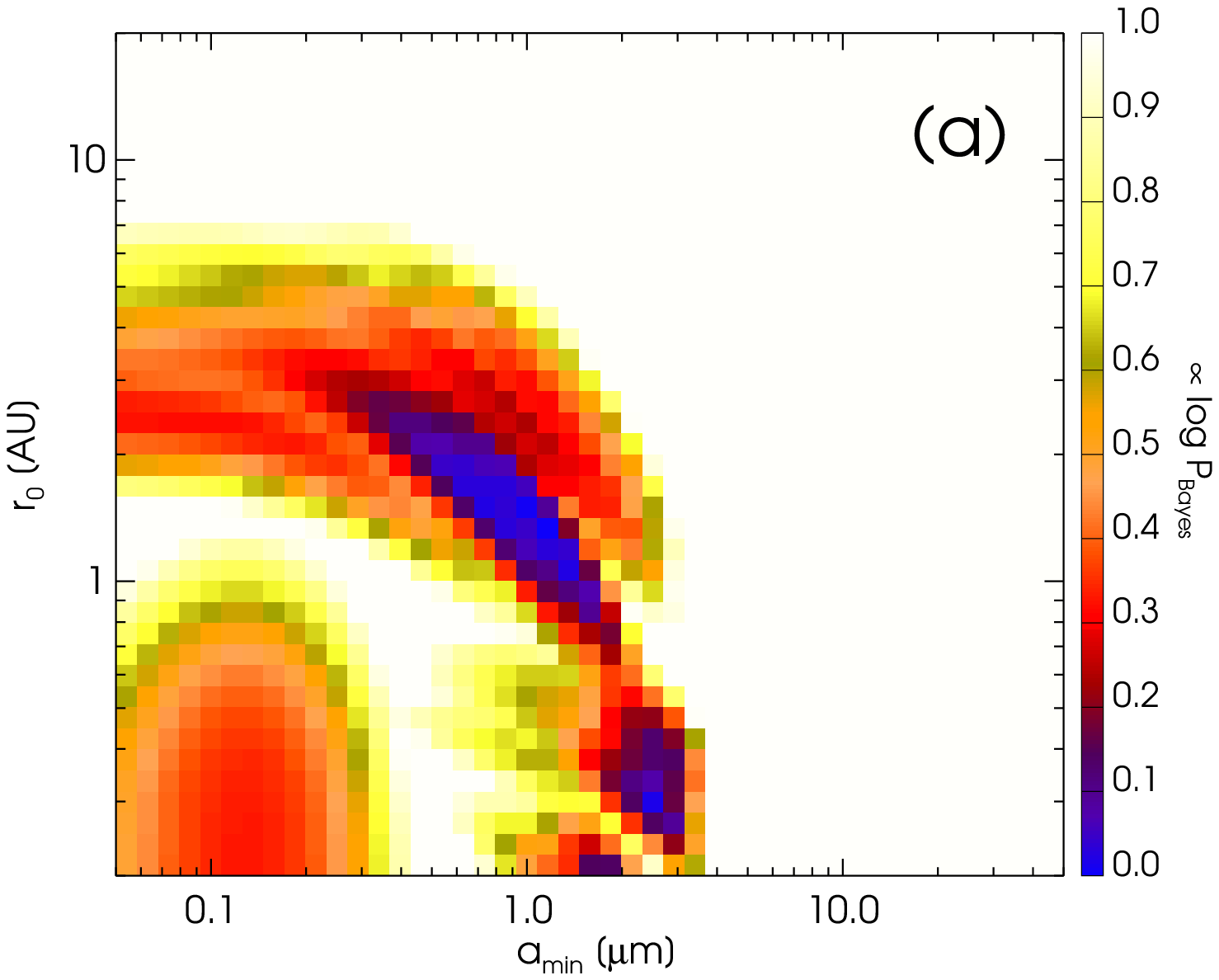}
  	\includegraphics[angle=0,width=0.95\columnwidth,origin=bl]{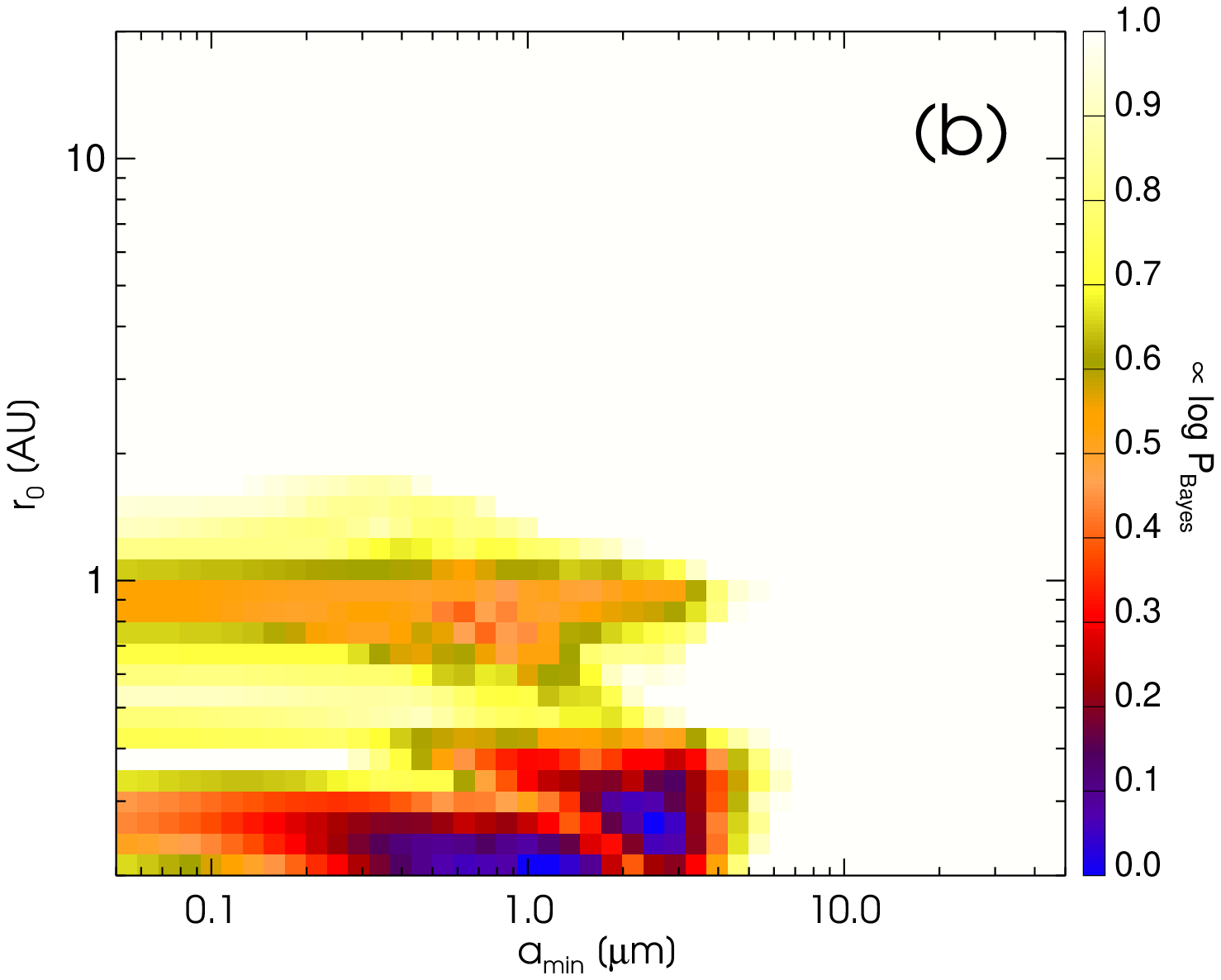}
  \caption{Bayesian probability maps for the $\eta$\,Crv exozodi obtained when comparing the silicate + olivine models with (a) The SED only and (b) the SED together with KIN nulls. The probabilities are represented as a function of minimum grain size and disk peak radius after integrating over the other disk parameters (density slope, grain size slope and \textit{astrosil} to olivine ratio). A logarithmic display stretched between 0 and 1 is used such that the probabilities are homogeneous to a $\chi^2$ and the smallest values are the best models. Two families of solutions at different distances emerge corresponding to the pure olivine case at $\sim0.2$\,AU and the silicate case at $\sim0.9$\,AU.}\label{fig:bayes}
  \end{center}
\end{figure*}

We first use mixtures of astronomical silicates \citep[hereafter \textit{astrosil}, ][]{Draine2003} and pure amorphous forsterite \citep{Jager:2003gf}. \textit{Astrosil} are a well-established reference for debris disk studies. Forsterite (or Fo100) are Mg-rich crystalline silicate and they are expected to reproduce the 11 and 18\,\um\ features with more fidelity. Ten mixing ratios are explored and we look for the best fitting solution in the least-square and Bayesian senses in a space of 6 parameters (Table\,\ref{tab:modelsetup}). 
The best-fitting parameters are listed in Table\,\ref{tab:bestmodels}. 
The pure forsterite model is clearly favored against models that include \textit{astrosil} ($\chi\dma{r}^2 = 1.9$ vs $\chi\dma{r}^2 = 3.0$).
Bayesian probability maps are shown in Figure\,\ref{fig:bayes}. 
They are projected on the two key parameters of the model: the minimum grain size and the peak distance, and they are integrated against all the other parameters ($\alpha\dma{out}$, $\kappa$ and 
$\frac{v\dma{astros.}}{v\dma{astros.}+v\dma{forst.}}$). 
The color scale provides a metric for the probability but it includes a log-transformation that allows to see the less probable solutions.
On the left panel only the spectrum has been fitted. The map shows the degeneracy between grain size and distance. A variety of models are allowed with typically small grains further than 2 AU; these correspond to the astrosil-rich models. At smaller distances, a second family of solutions appears, corresponding to the forsterite-rich models and featuring micrometer grains. Overall, from pure spectral arguments alone, it is not possible to form any conclusions on the disk location.

On the right panel the KIN data have been included and they impose a severe constraint on the models. All models peaking further than 1 AU are excluded and two well-determined solutions are left.
The best-fitting models occur at different distances: 0.8 AU for \textit{astrosil} versus 0.2 AU for forsterite with strong consequence on the inferred dust mass. The surface density profile decreases smoothly ($\alpha\dma{out} = -1.0 \textrm{~or~} -1.5$) as expected for a collisional system.
In both cases the size distribution is relatively steep ($3.7 \leq \kappa \leq 4.7)$, indicating an overabundance of small grains. The minimum grain size is close to 1\um\ which is remarkably coincident with the blowout size ($a\dma{blow} = 1.28\mu m$ and 1.05\um\ respectively).
However, of the two models, the fosterite (Fo100) model yields a vast improvement with respect to the \textit{astrosil} one. 
As can be seen in Figure\,\ref{fig:bestfitsed} this is because the latter fails at reproducing the deep silicate band at 14\um. At colder wavelengths the models converge consistently with the similar slopes derived. 
In terms of reproducing the nulls, there is little variations between the different models. 
The spectral shape of the nulls is again better reproduced by the forsterite model thanks to a slope inversion at 12\um. The modelled nulls become smaller with decreasing baseline but there is little variation between the models at different distances. \\

In sum, the KIN informs us that the exozodi is located at less than 1 AU while the grain properties are mostly probed by the IRS spectrum putting additional constraints on the disk location based on temperature arguments. 
Knowing the star spectrum, the latter is determined by the grain albedo $\omega\dma{\lambda} = 1- \frac{\left<{Q\uma{\lambda}\dma{abs}}\right>}{\left<{Q\uma{\lambda}\dma{abs}+Q\uma{\lambda}\dma{sca}}\right>}$ where the mean is calculated on the grain size distribution. At 0.5\um\ the albedo of the \textit{astrosil} for the best-fit parameters is 61\%. The albedo of forsterites is as high as 99\%: they are a remarkably inefficient absorber justifying the low temperature derived. 

Arguably physical likelihood of the fosterite model could be questioned. Forsterite consists of iron-depleted olivines. \citet{Jager:2003gf} measured the optical indices of pure forsterite in the laboratory. However as shown on their Figure 9, adding the smallest inclusion of iron dramatically increases the absorption efficiency at visible wavelengths. 
For example if we add to the forsterite optical constants inclusions of pyroxene \citep[Mg$_{0.5}$Fe$_{0.5}$Si$_3$]{Dorschner:1995yq} with a volume fraction of 1\% or 10\%, we find that the visible albedo is reduced to 95\% and 73\% respectively. 
Comparable although less good models are produced, with slightly larger distances than the pure forsterite case.
We test an alternative olivine model, the glassy olivine sample from \citet{Dorschner:1995yq} (MgFeSiO$_4$) but the fit is statistically not improved.
Results for pure pyroxene and pure olivine grains are listed in Table \ref{tab:bestmodels} for completeness. \\

The above models do not reproduce the smaller, second-order spectral features.
We take a step further and attempt to test the compositional model proposed by \citet{2012ApJ...747...93L}.
We focus on the 5 dominating species and use the molar fractions listed in their table\,2. 
Our model is an approximation of \citet{2012ApJ...747...93L} mixture using the following materials, with volume fractions and references given within parenthesis: 
forsterites \citep[24.9\%,][]{Jager:2003gf}, amorphous silicas  \citep[\textit{i.e.}\,quartz, 31.4\%, SiO$_2$ at 300 K,][]{1997A&A...327..743H}, metal sulfides \citep[9.7\%, FeS at 300 K,][]{1997A&A...327..743H}, pyroxenes \citep[13.6\%, Mg$_{0.5}$Fe$_{0.5}$SiO$_3$,][]{Dorschner:1995yq}, amorphous carbon \citep[11.6\%,][]{zubko}, water ice \citep[8.8\%][]{LiGr97}. In practice water ice is replaced by porosity because the exozodi is far above the ice sublimation temperature.
The various materials are mixed using the Bruggeman EMT as usual -- \textit{i.e.} assuming grain homogeneity and no hierarchy between the different inclusions (``Lisse mix / EMT''). 
The results are listed in Table\,\ref{tab:bestmodels}, the best model yields a $\chi^2$ of 2.91 with a steep distribution of grains larger than a few microns at $\sim$0.8 AU in a narrow belt.\\

The key difference between the above models lies on how well they reproduce the silicate features at 11 and 18\um. Only the forsterite models yield a satisfying $\chi^2$ because they are the only ones to reproduce the amplitude of the silicate feature. 
Our interpretation is that a consequence of the Mie-EMT approach is to smooth the deep compositional features in contradiction with the data. \citet{2012ApJ...747...93L} rather performed an addition of the spectra of different compositions. 
As a last test, in order to produce results more directly comparable to their work we compute scattering and absorption cross sections for each of the 5 individual materials and we linearly sum them with a weighting factor given by the surface density listed in their table\,2 (``Lisse mix / Sum'').
The results are again listed in Table\,\ref{tab:bestmodels}. Visually, the SED is better reproduced suggesting that the optical model is adequate, but the impact on the least-square statistics is small indicating that this compositional model does not satisfy our new spatial constraints.

\begin{figure*}[h!btp]
\begin{center}
  	\advance\leftskip-0.5cm
  	  	\includegraphics[angle=0,width=0.95\columnwidth,origin=bl]{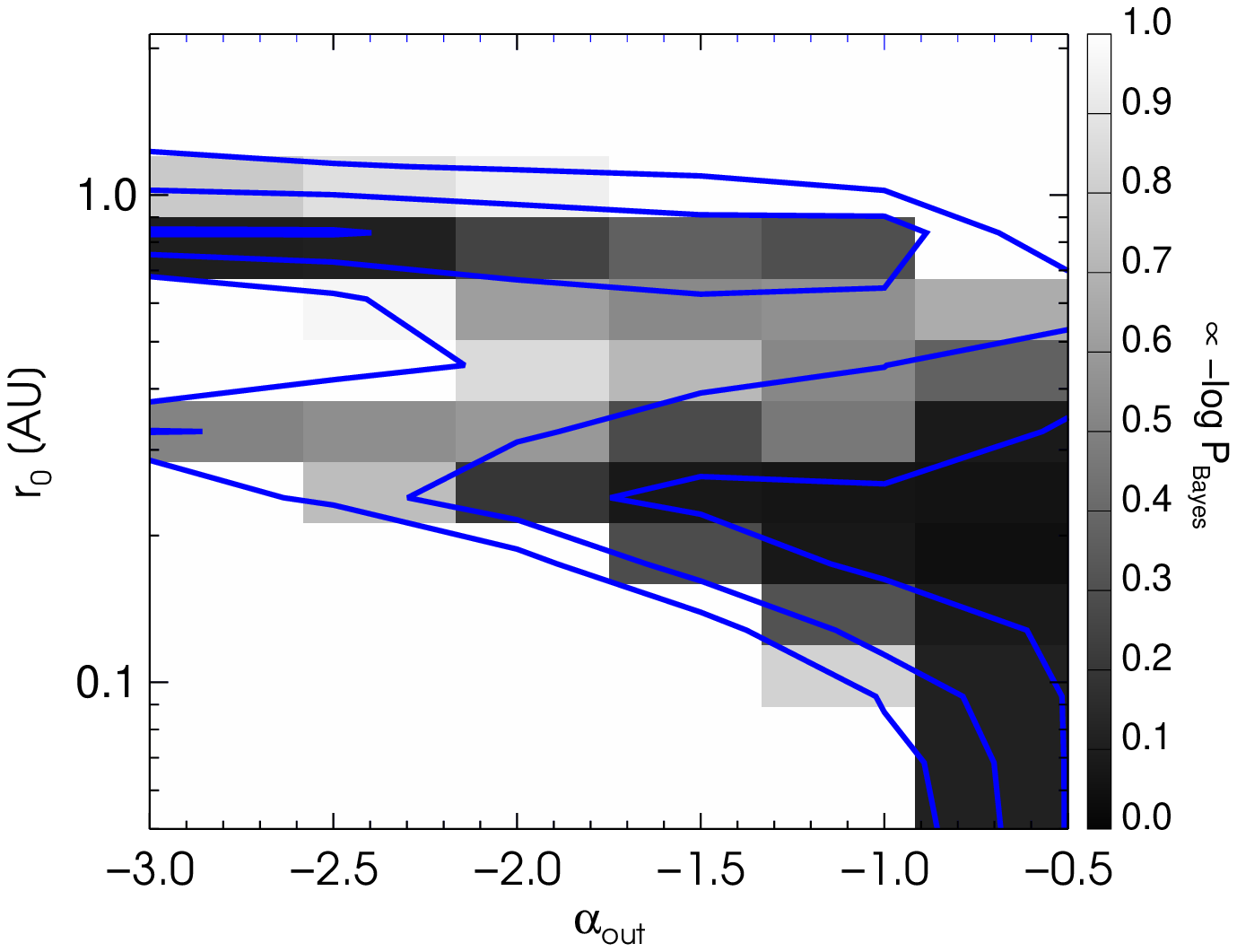}
  	\includegraphics[angle=0,width=0.95\columnwidth,origin=bl]{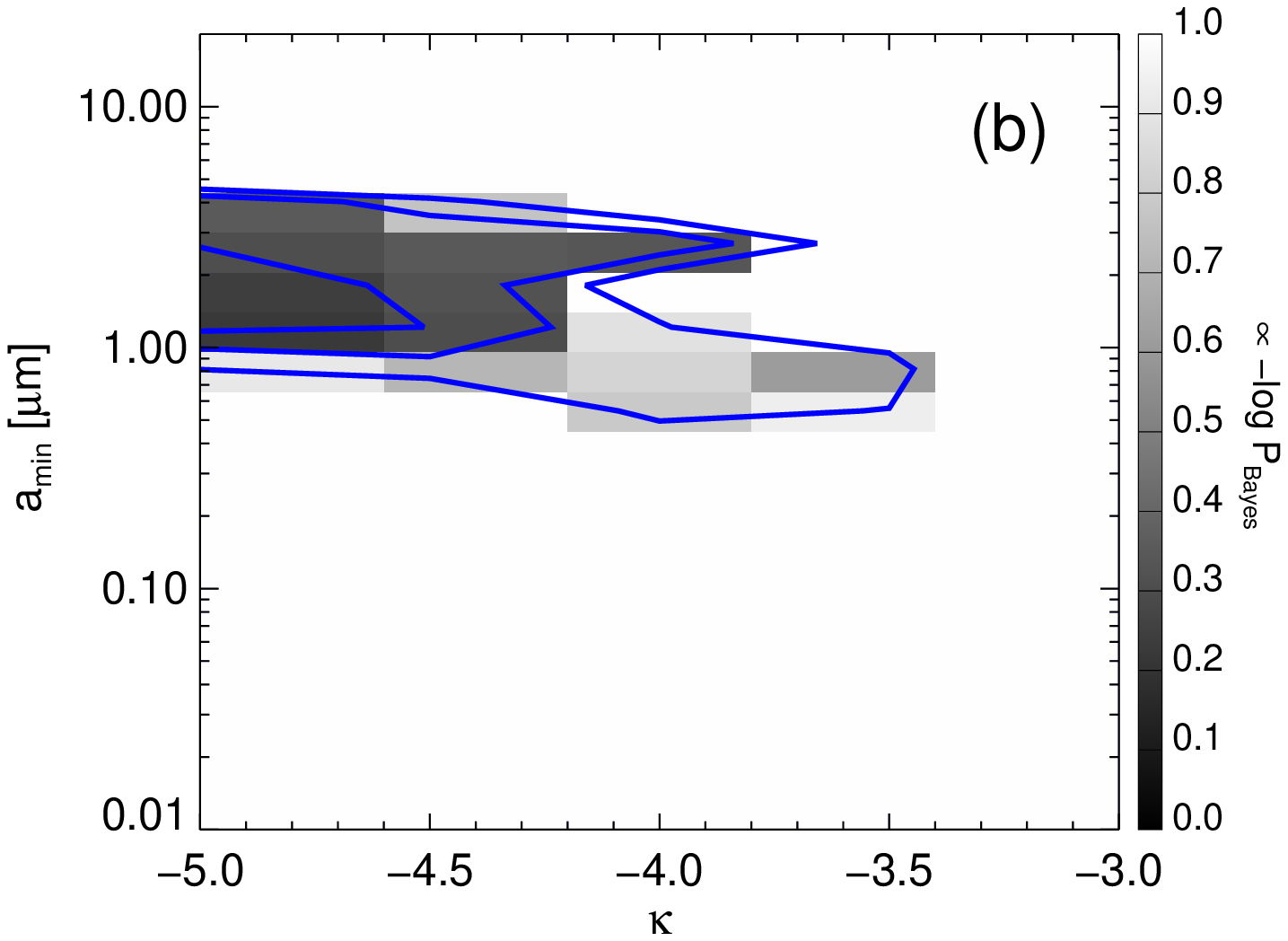}
  \caption{Bayesian probability maps for the $\eta$\,Crv exozodi projected onto (a) the peak of the surface density profile ($r\dma{0}$) and its outer slope ($\alpha\dma{out}$), (b) the minimum grain size ($a\dma{min}$) and the slope of the size distribution ($\kappa$). All models presented in Table\,\ref{tab:modelsetup} were included $\ie$ models with differents compositions were all incorporated though the direct addition of raw probabilities. A logarithmic display stretched between 0 and 1 is used such that the probabilities are homogeneous to a $\chi^2$ and the smallest values are the best models.}\label{fig:proba_sigma_r0}
  \end{center}
\end{figure*}

\begin{table*}[h!tpb]\caption{Best models of the $\eta\,$Crv exozodi}\label{tab:bestmodels}
\begin{center}
\begin{tabular}{lcccccc}
\hline\hline 
Parameter					&	AstroSil	&	Forsterite & Olivine 	& Forst$_{0.9}$+Pyrox$_{0.1}$ & Lisse mix / EMT & Lisse mix / Sum 		\\ [1.2ex]
\hline
$\alpha\dma{\textrm{out}}$ 	&	$-1.0_{-0.4}^{+0.4}$	 &	${-1.5_{-0.1}^{+0.6}}$	&	${-0.5_{-0.2}^{+0.2}}$	& ${-0.5_{-0.2}^{+0.0}}$			& ${-3.0_{-0.0}^{+0.5}}$ & $-1.0_{-0.2}^{+0.2}$	 \\	[1.2ex]
$\kappa$ 					&	$-4.0_{-0.3}^{+0.3}$			&	${-4.5_{-0.2}^{+0.2}}$ &	 ${-5.0_{-0.0}^{+1.5}}$ & ${-5.0_{-0.0}^{+0.2}}$ & ${-5.0_{-0.0}^{+1.2}}$ & ${-5.0_{-0.0}^{+0.4}}$	\\[1.2ex]
$a\dma{\textrm{min}}$ $[\mu m]$&	${0.62_{-0.05}^{+0.12}}$ 		&	${1.2_{-0.1}^{+0.8}}$		&	${0.81_{-0.06}^{+0.98}}$ & ${1.81_{-0.26}^{+0.40}}$ & ${4.04_{-1.80}^{+0.22}}$ & ${2.71_{-0.39}^{+0.58}}$ 	\\[1.2ex]
r$_0$ [AU] 					&	${0.82_{-0.09}^{+0.03}}$ 	&	${0.17_{-0.01}^{+0.08}}$ &	${0.13_{-0.05}^{+0.02}}$  & ${0.17_{-0.03}^{+0.03}}$ & ${0.84_{-0.10}^{+0.13}}$ & ${0.24_{-0.03}^{+0.04}}$	\\[1.2ex]
$M\dma{\textrm{dust}}\uma{\textrm{1mm}}$ $[10\uma{-6}M\dma{\oplus}]$	&	${0.115_{-0.3}^{+0.3}} $  & ${23.5_{-17.2}^{+0.7}}$ & ${20.7_{-12.6}^{+63.0}} $ & ${16.0_{-0.05}^{+0.5}}$ & $0.527_{-0.014}^{+1.113}$ & ${7.67_{0.21}^{+3.56}}$ \\[1.2ex]
  \hline
Density [g/cm$^3$]	& 3.5 & 3.20 & 3.71 & 3.20	& 2.95 & 2.95 \\
Albedo $\lambda=0.5\mu m$	&   0.61							&	0.99		&	0.55  &  0.73 & 0.55 & 0.80\\
Albedo $\lambda=11\mu m$		&   0.32							&	0.39		&   0.47  &  0.47 & 0.18 & 0.62 \\
$g\dma{sca}\uma{1.1\mu m}$ 				&   0.72			& 	0.65		&  	0.92   & 0.75 & 0.85 & 0.79 \\
$T^{\circ}(1\mu m, 1\textrm{AU})$ [K]& 	484					& 	208		&   605 & 396 & 525 & 405  \\
t$\dma{col}$ [years]			&	285							&	2.3		&   30 & 25 & 0.96 & 14\\
$a\dma{blow}$	[\um]	&	1.3	&	1.02 & 1.3 & 1.1 & 1.6 & 0.83\\
${L\dma{disk}}/{L\dma{\star}}$&	$2.98\times 10^{-4}$			&	$1.74\times 10^{-3}$	& $3.47\times 10^{-4}$	 & $4.18\times 10^{-4}$ & 0.00140 & $4.05\times 10^{-4}$	\\[1.2ex]
							\hline
$\min(\chi^2\dma{r})$ 	(\textit{d.o.f.} = 81)			&	2.98	 			&	1.74		& 2.86 & 1.95 & 2.91 & 2.85	\\[1.2ex]
\hline

\hline
\hline
\end{tabular}
\end{center}
{{\sc Notes:} The listed parameters are the ones that give the smallest $\chi^2$. Uncertainties are computed with Bayesian analysis on each grain composition.}
\end{table*}

\begin{figure*}[htpb]\begin{center}
  \includegraphics[angle=0,width=0.8\textwidth,origin=bl]{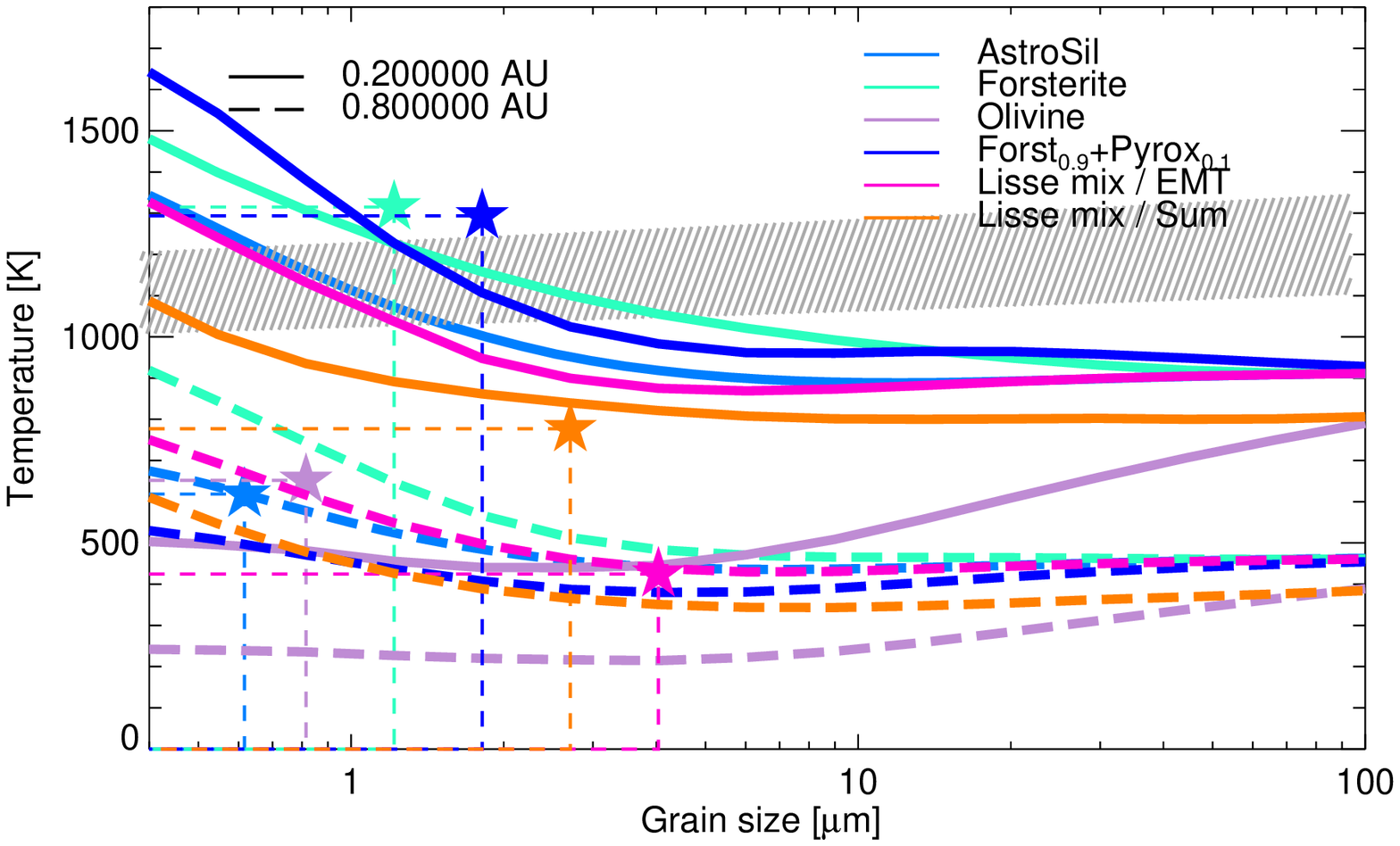}
  \caption{Grain temperature as a function of grain size for a sample of distances and for the 6 compositions explored for the exozodiacal disk. The grey regions is the size-dependent sublimation zone for \textit{astrosil} assuming a dust lifetime of 1 to 100 years. The star symbols denote the best models for each composition.}\label{fig:temperatures}
  \end{center}
\end{figure*}

\subsection{New LBTI results}\label{sec:lbti}
In a companion paper \citep{Defrere:2015lr} we confront our models to new N-band observations from the Large Binocular Telescope Interferometer. {A modified version of our KIN simulator was used to calculate the fringe pattern produced by the LBTI and predict the null expected for our models. Equation\,\ref{eq:nulldef} is revised based on the LBTI setup, which consists of a two-telescope nuller operating at 11.1 $\mu$m with a baseline of 11.4 meters.
the slowly oscillating cosine term is eliminated and the baseline length in the fast oscillating sine term becomes 11.4 meters. 
The LBTI measured a null depth of $4.40\%\pm0.35\%$ over a field-of-view of 140\,mas in radius ($\sim$2.6\,AU at the distance of \etacrv) and shows no significant variation over $35\deg$ of sky rotation. This relatively small null depth suggests that most of the disk emission is suppressed by the LBTI.  
It must therefore be coincident with the central destructive fringe at less than 79 mas (1.4 AU) which corroborates our findings on the disk location.} Our forsterite model at 0.2 AU produces slightly less null than measured, while the astrosilicate one at 0.8 AU largely overestimates the nulls. {Our best model is modestly altered by the inclusion of the new LBTI constraint. The parameters are revised as such for the best-fit model}: $r_0=0.23$\,AU, $a\dma{min}=0.8\,\mu m$, $\kappa=-4.0$, $M\dma{dust}\uma{1mm}=6.4\times 10^{-5}\,M\dma{\oplus}$.
{Thus the new LBTI measurements validate our model of a high-albedo dust belt at $\sim$0.2\,AU as opposed to a belt at $\sim$0.9\,AU. 
They also essentially exclude the presence of additional dust emitting in the 11\um\ range at the 0.1\% level unless a dust clump coincidentally intercepts one of the dark fringes.  
An alternative scenario would indeed be that the disk is not centrosymmetric, \textit{i.e.} that it is perpendicular to the outer disk and aligned with the LBTI fringe pattern. Then if a region of overdensity is coincidentally intercepted by a transmission minimum at a projected separation smaller than  the physical separation, Spitzer models at a few AU could be rehabilitated. However we disfavor this scenario based on the Occam's razor principle.} 

\subsection{Exozodi models: summary}

We presented 6 compositional models that fit simultaneously the $\eta$\,Crv mid-infrared spectrum and interferometric nulls, with various goodnesses of fit. 
Their common trait is that they peak closer inwards than 1 AU and have minimum grain sizes close to 1\um. Most models decline slowly and have steep  size distributions. The dust mass in grains smaller than 1mm is of the order of $10\uma{-5}$ to $10\uma{-7}\,M\dma{\oplus}$ depending on the model.

The key criterion that differentiates the models is their temperature distribution, a function of distance, grain size and albedo. In Figure\,\ref{fig:temperatures} we show the temperature profile of each of the models. 
It is found that there is a variety of dominating temperatures (temperature of the $a\dma{min}$ grains at $r_0$) for different compositional models. We conclude that the mid-infrared spectrum is not a unequivocal probe of the grain temperature in the presence of strong spectral features. The hottest grains of the forsterite models slighty exceed the range of sublimation temperatures calculated for reasonable sublimation timescales, suggesting that the dust may be eliminated at 0.2 AU and rather reside slighly further out. 
In Table\,\ref{tab:bestmodels} we list reference temperatures at 1 AU for 1\um\ grains to allow a direct comparison between the different models.
They demonstrate that the discrepancy in inferred distances for the exozodi largely relies on the variety of equilibrium temperatures for different grain albedos: the KIN measurements are not able to differentiate between models located at 0.2 or 0.8 AU. 
On the other hand, the LBTI nulls complement the spatial constraint and unambiguously favor a 0.2\,AU for the dust at predicted.  

We note that in order to avoid any model-dependent prior from the analysis, we did not remove materials that exceed a certain sublimation temperature (we effectively fixed the sublimation temperature to large enough values >1700\,K). 
In \citet{Lebreton:2013uq} we showed that the sublimation distance is not only size-dependent but also depends on a grain lifetime.
The grey region in Fig.\ref{fig:temperatures} reveals that the best models are all located outside of the dust sublimation zone. The hottest grains are close to the sublimation limit but remain essentially unaffected. 

Most of the models favor steep size distributions. The slope of the density profile on the other hand is best fit by $-1.0$, $-1.5$ slopes although some models can decline faster. What remains to be determined is whether these findings are statistically significant and/or whether they are really constrained by the data. 
In Figure\,\ref{fig:proba_sigma_r0} we investigate on possible degeneracies between the two parameters of the surface density and the two parameters of the size distribution using projections of the Bayesian probabilities. All of the compositions are incorporated simultaneously through a direct addition of their probabilities.
It is found that the narrowest disk models ($\alpha\dma{out} = -3.0$) are generally the ones peaking the furthest out ($r_{0} = 0.8$AU). The most probable models remain the flattest ones ($\alpha\dma{out} = -1.0$) at $r_{0} = \sim0.2$AU. Models both "flat" and peaking close to the star are excluded. In essence, it means that the nulls need material to be present between 0.2 and 1 AU to fit the observations.

The next panels shows a probability map projected onto  $a\dma{min}$ and $\kappa$. It reveals that there are a variety of probable solutions between $a\dma{min} = 0.6$\um\ and $a\dma{min} = 4$\um\ and between $\kappa = -3.5$ and $\kappa = -5.0$. Yet  having a steep size distribution ($\kappa \ll -3.5$) is statistically the most viable solution, and consequently the minimum grain size has to be of the order of 2\um. We consider that the models can not differentiate between slopes as long as they are enough steep because the larger grains have little impact on the observables. 

In summary, both the mid-infrared interferometric and spectral data of $\eta$\,Corvi can be well fitted with an exozodi model consisting of a dust belt located at 0.2\,AU and that declines slowly. 
The minimum size of the grains is about 1 or 2\um\ \textit{i.e.} very close to the blowout size. Nonetheless the size distribution is significantly steeper than a slope of -3.5 resulting in an overabundance of small grains.
Over the wavelength range explored, the spectrum is dominated by strong silicate features that can only be fitted by few micrometer-sized forsterite grains. At these distances the  equilibrium temperatures of the smallest grains range from 450 to 1200 K. The largest grains do not exceed 900K temperatures. At 0.2 AU the hottest grains are subject to sublimation so their lifetime is short (less than 1 year) but grains larger than a few microns are collision dominated. Assuming we can extrapolate the size distribution up to large grain sizes the total dust mass is in the range $5\times10^{-7}$ to $2\times10^{-5}\,M\dma{\oplus}$.
 
Models involving pyroxenes, carbon, sulfides and silicas yield similar best-fitting parameters but modelling such complex grains with the Mie/EMT theory proves inadequate. At wavelengths longer than 30\um\ the Rayleigh–Jeans tail is reached and all the models converge. 
{The models rely on the idealized assumption that the grains are close to spherical and can be model with the Mie theory. Yet the spectrum is very similar to the one of \citet{Chen06} in the 10 to 20\um\ range with emission of the order of 0.2 to 0.4\,Jy. 
More sophisticated mixtures were used by the authors to fit details of the spectral features: amorphous olivine, crystalline forsterite and enstatite grains with a temperature of 360\,K and a crystalline silicate fraction of 31\% were used, in addition to a 120\,K blackbody continuum. Although our model provides a very good fit to the binned IRS spectrum, additional species are needed to produce the finer spectral features of the high resolution spectrum, including the deep observed at 22\um.
Additional dust material would necessarily reach higher temperatures than pure forsterite if they are co-located. In turn, if the materials are mixed, the forsterite would be heated and lose the benefit of their high albedo. Then it is possible that the other species needed to produce the finer spectral from features are eliminated from the inner forsterite zone, and are only present in the outer tail of the exozodi.}

In Figure\,\ref{fig:sedcold} we see that the exozodi model is in agreement with the upper limits on the flux of the {\sc Herschel}/PACS inner component ($F70 < 76 mJy$) . For our best model, the 70\um\ flux from the exozodi is 21 mJy, the stellar flux is 35 mJy so the expected flux of the {\sc Herschel} inner point source is 56 mJy.

\section{Models of the cold belt}\label{sec:coldmodel}

We now undertake to model the $\eta$\,Crv cold debris belt that is seen on the {\sc Herschel} images at $\sim$120 AU. We model simultaneously the 3 pairs of PACS radial profiles (semi-major and semi-minor axis) and the far-infrared SED from {\sc Spitzer}, {\sc Herschel} and SCUBA. We include an upper limit derived from the residual of the fit to the {\sc Spitzer} spectrum at 31\um (< 0.34 Jy) and 24\um\ (< 0.61 Jy) . The exozodi model is added to the stellar spectrum before calculating the cold excess SED. We effectively fit the global SED and the inner disk with a fixed model for the unresolved component and a parametrical model for the disk. The explored range for the 5 parameters of the disk is given in Table\,\ref{tab:modelsetup}. Once the best models are determined we refine the range of explored distances to obtain an accurate estimate of $r_0$.

In terms of grain materials, we first try icy grains composed of \textit{Astrosil} and amorphous water ice (${v\dma{ice}}/({v\dma{Si}}+{v\dma{ice}})$ from 0.0 to 0.9). We then use mixtures of \textit{astrosil} and carbonaceous material (fix ratio ${v\dma{Si}/v\dma{C} = 1/2}$) and we incorporate amorphous water ice (${v\dma{ice}/(v\dma{C}+v\dma{Si}+v\dma{ice})}$ from 0.0 to 0.9) and porosity ($\mathcal{P} = v\dma{vacuum}/v\dma{solid}$ from 30\% to 95\%.). Such models were found to be adequate to fit {\sc Herschel} debris disks with well-populated SEDs \citep{Lebreton2012a,2013ApJ...772...17D}. The carbon to silicate ratio has no detectable effect on the best SED models, the water ice content determines the shape of the SED at its peak and in the mid/far infrared domain; and porosity impacts the slope at sub-millimeter wavelengths.
Three additional image scaling factors are included, setting the number of free parameters to respectively 9 or 10 and the number of degrees of freedom to 140 or 139.

\begin{figure}[h!btp]
\begin{center}
\includegraphics[angle=0,width=0.98\columnwidth,origin=bl]{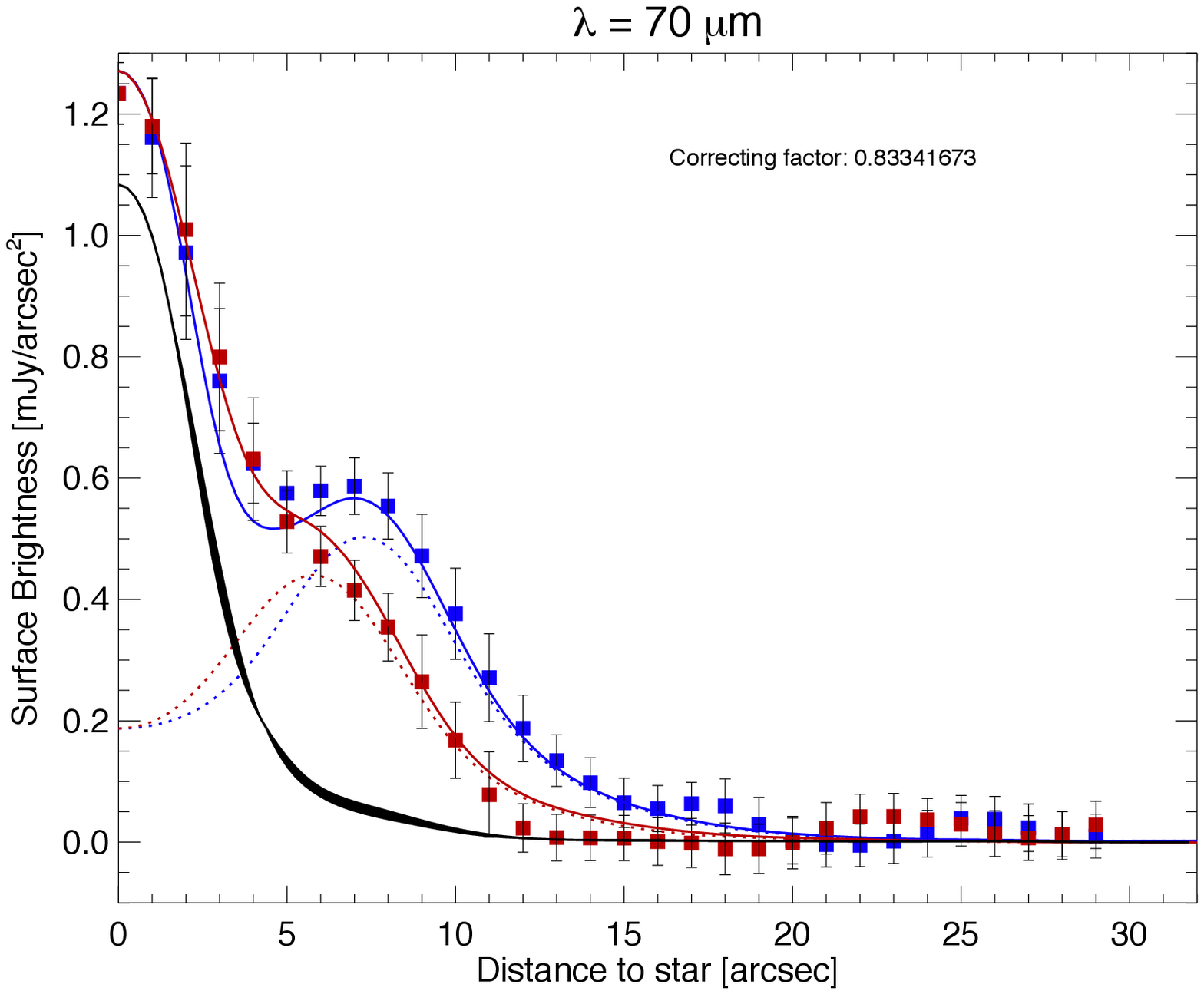}
\includegraphics[angle=0,width=0.98\columnwidth,origin=bl]{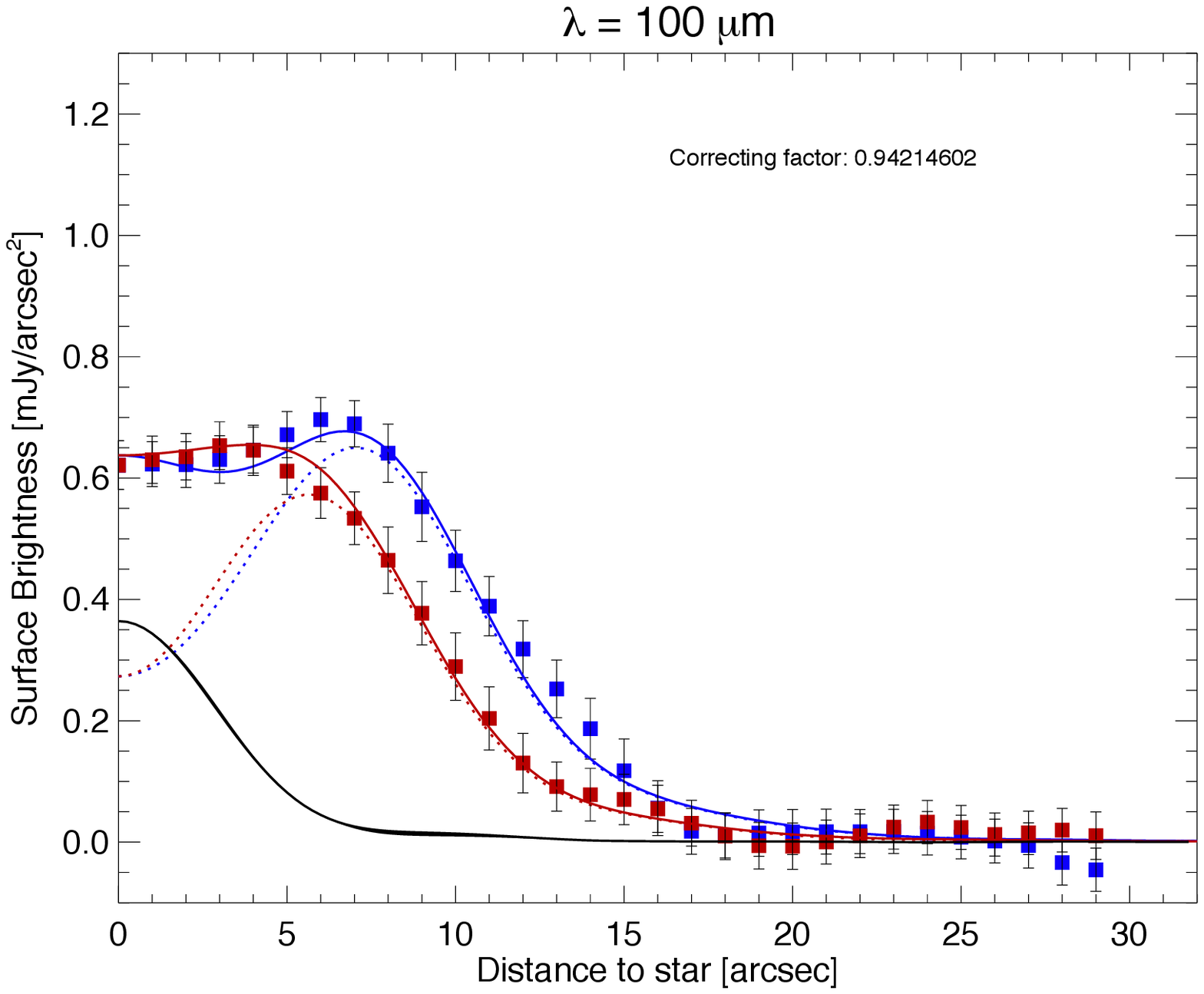}
\includegraphics[angle=0,width=0.98\columnwidth,origin=bl]{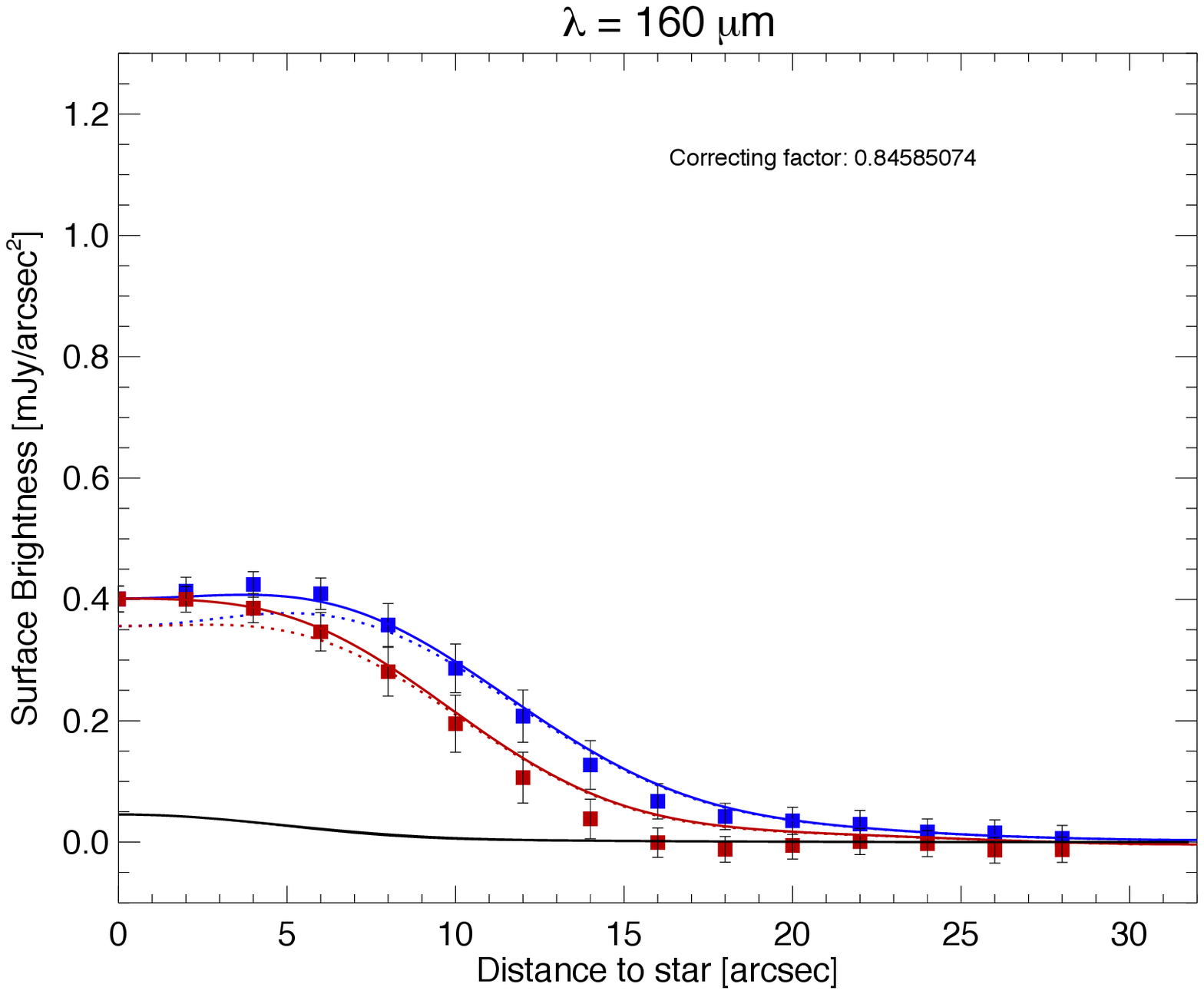}
  \caption{Best fit model of the cold ring radial brightness profile at 70, 100 and 160$\mu m$  along the major axis (blue symbols) and the minor axis (red symbols). The black line is the contribution from the unresolved inner component (star + exozodi), the dotted lines are the disk models, and the solid red and blue lines are the global model. All quantities are convolved with the instrument PSF.}\label{fig:fitimage}
  \end{center}
\end{figure}

\begin{table*}[tpb]\caption{Best models of the cold component}\label{tab:modelcold}
\begin{center}
\begin{tabular}{ccccccc}
\hline\hline 
Parameter					&	AstroSil+ice 	&	Astrosil+C+ice+porosity \\[1.2ex]
\hline
$v\dma{ice}/v\dma{total}$	&$0.1\uma{+0.02}\dma{-0.1}$	&		$0.1\uma{+0.15}\dma{-0.6}$							\\
Porosity 					&	--						&		$68\uma{+4}\dma{-8}\%$			\\
$\alpha\dma{\textrm{out}}$	&$-5.0\uma{+1.1}\dma{-0.0}$	&			$-5.0\uma{+0.7}\dma{-0.0}$				\\	
$\kappa$ 					& $-3.5\uma{+0.2}\dma{-0.2}$	&	$-3.5\uma{+0.2}\dma{-0.2}$			\\
$a\dma{\textrm{min}}$ [\um]	&$4.75\uma{+0.52}\dma{-0.64}$& 		$5.55\uma{+0.58}\dma{-1.97}$			\\
r$_0$ [AU]					&$133.2\uma{+8.5}\dma{-5.6}$&	$133.2\uma{+4.6}\dma{-9.0}$					\\
$M\dma{\textrm{dust,<1mm}}$ [$M\dma{\oplus}$]	
							&   $2.86\dma{-0.02}\uma{+0.47}\times10\uma{-2}$& 	$7.07\dma{-0.89}\uma{0.40}\times10\uma{-3}$	&	 \\[1.2ex]
\hline
Density [g/cm$^3$]	& 3.27 & 0.81 \\
Albedo $\lambda = 0.5\mu m$	&				0.55			&		0.53	\\
$g\dma{sca}\uma{1.1\mu m}$&   0.93 & 0.97 \\
$T^{\circ}(a=a\dma{min},a=1\,mm)$ at $r\dma{0}$ [K]& 		[48, 37]		& [56, 37]\\
t$\dma{col}$ [years]			&  $1.6\times10^6$			&$1.3\times10^6$			 \\
$a\dma{blow}$ [\um]					& 1.4				& 9.3 \\
$\tau\dma{\perp}$			& $2.85\times10^{-4}$ 		&	$3.45\times10^{-4}$	\\
${L\dma{disk}}/{L\dma{\star}}$&	$1.56\times10^{-4}$			&	$1.86\times10^{-4}$	\\[1.2ex]
\hline
min($\chi^2\dma{r})$ (\textit{d.o.f.}) 	&   0.58 (140)		& 0.51 (139)		\\[1.2ex]
\hline
\end{tabular}
\end{center}
\end{table*}

\begin{figure*}[h!btp]
\begin{center}
  \includegraphics[angle=0,width=0.85\textwidth,origin=bl]{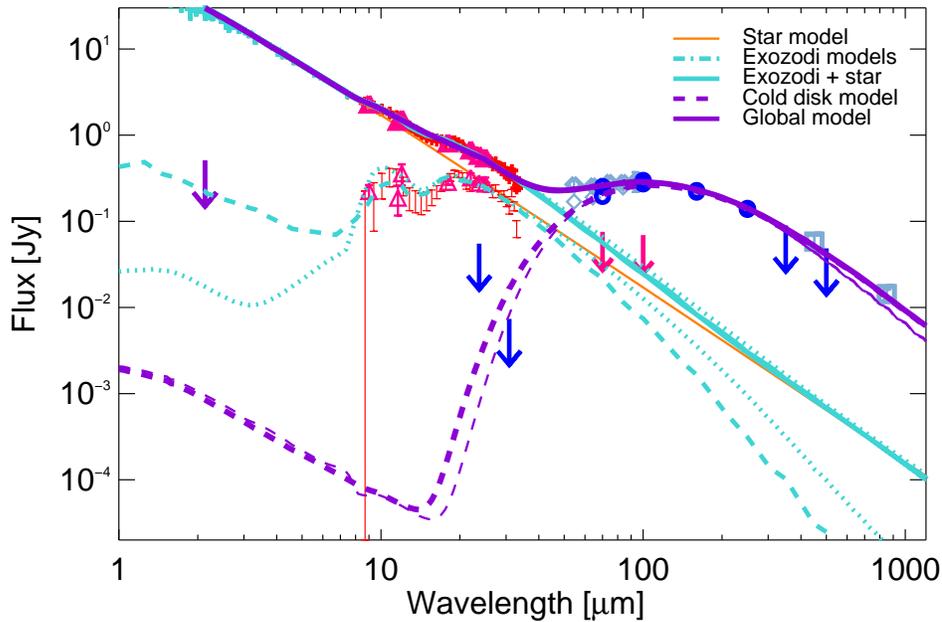}
  \caption{Global SED model of the $\eta$\,Crv debris disk, including the star, the exozodi and the cold belt. Two models are shown for the exozodi but the forsterite one (thick line) is used for the global model. They are based on a fit to the red data points, including upper limits from PACS. The cold disk models are the ones presented in Table\,\ref{tab:modelcold} with either astrosi. with 10\% ice or \textit{astrosil} (thin line) and carbon mixtures with 10\% ice and 68\% porosity (thick line). They are based on a fit to the blue data points, including upper limits from {\sc{Spitzer}}. All models include both thermal emission and scattering.}\label{fig:sedcold}
  \end{center}
\end{figure*}

The best fitting models are given in Table\,\ref{tab:modelcold} and displayed in Figure\,\ref{fig:fitimage} and \ref{fig:sedcold}. The SED and the radial profiles are perfectly fitted using standard disk parameters. 
The pure \textit{astrosil} model are favored against the icy ones: the best fitting models has 10\% of water but the error bars are compatible with 0\% ice. 
The minimum grain size is a few microns ($a\dma{min} = 4\mu m$) and the size distribution has a -3.5 slope. 
{As pointed out by \citet{Pawellek:2014kq}, this disagreement between $a\dma{blow}$ and $a\dma{min}$ is commonly noticed for G- to A-type stars in agreement with collisional models.}
The surface density profile peaks at 133.2 AU and decreases sharply ($\alpha\dma{out} = -5$). 
Model uncertainties on the peak location are larger than the possible 4\,AU (0.2\arcsec) offset between both sides of the major axis. 
The total dust mass is very large with 0.03$M\dma{\oplus}$ in grains smaller than 1 mm. This models yield a reduced $\chi^2$ as small as 0.58. 
We notice that less steep density profiles produce almost as good models as long as the slope $\alpha\dma{out}$ is steeper than -3.5. This tells us that the spatial resolution of {\sc Herschel} does not allow to differentiate between those models because the width of the ring is essentially unresolved. 
This models slightly under-predicts the millimeter and the 50\um\ flux, while it matches exactly the expected 31\um\ residual flux. 

The derived parameters depend on the exact composition assumed. To further improve the fit at the edges of the disk SED and determine more precisely the disk model, we look for the most probable porosity and ice fraction by marginalizing the Bayesian probabilities onto these two parameters. The best fit is again found for ice-free grains although adding up to $\sim$40\% of ice does not alter the fit significantly. A porosity of 45\% is found to improve the fit yielding a smallest $\chi^2\dma{r}$ of 0.51. All the SED measurements are visually perfectly reproduced.

We also test pure forsterite models but they result in worst fit than the \textit{astrosil} ones at the shortest and longest wavelengths ($\textrm{min}(\chi^2) = 1.25$).

At all three wavelengths, the radial profile is very well reproduced by the models along both the semi-minor and semi-major axis. The unconvolved profiles peak slightly further out than the observed surface brightnesses because of a convolution effect.
In Figure\,\ref{fig:fitimage} it appears clearly that light emitted from the 130\,AU ring is distributed towards smaller apparent separations in the convolution process: at 100\um\ about 40\% of the surface brightness in the center actually comes from the ring at 7$\arcsec$, at 70\um\ this number is reduced to 15\%. This explains the discrepancy between the measured and modelled 70\um\ flux previously mentioned: we revise the 70\um\ flux from the inner component (star+exozodi) to 61 mJy.
We note that properly determining the level of unresolved emission was a critical step: our preliminary attempts to fit the radial profiles with either more or less exozodiacal emission lead to very poor results. 

In summary, the SED and the resolved PACS images of $\eta$\,Crv can be very satisfyingly reproduced using a two-component model featuring an exozodi at <1 AU -- seen as unresolved emission by {\sc Herschel} -- and a narrow dust ring at 133.2\,AU. The cold belt is well fitted with classical astronomical silicates. There is no evidence for ice but incorporating carbonaceous material and porosity improves the fit to the SED.

\section{Discussion}
\subsection{Summary and comments}
We presented a two-component model of the $\eta$\,Crv that successfully reproduces all available data from the mid- to the far-infrared, including both spatial and spectral constrains. An image of the final model is shown in Figure\,\ref{fig:modelimage}.

The properties of the cold debris belt are archetypal of a debris belt in collisional equilibrium with a minimum grain size very close to the blowout limit and -3.5 power-law size distribution. The disk peaks at 133.2 AU and it narrow, with a fast decreasing profile towards large distances ($\alpha\dma{out} < -3.5$). We assumed it has a sharp inner edge which proves compatible with the observations. 
{The cold disk is very well modelled by silicate-dominated models, possibly containing carbon, ice and porosity; forsterite models on the other hand are disfavored.}
Despite limited spectral constrains, we conclude that the $\eta$\,Crv dust is most likely ice-free. For a high mass MS star it is expected that ices have been photo-dissociated for the age of $\eta$\,Crv. 

The surprising feature is of course the huge dust mass of the belt ($10^{-2}\,M\dma{\oplus}$). For example, it is almost as massive as the juvenile debris disk of HD\,181327. 
The collision timescale $t\dma{col}$ is of a few Myr: the dust mass is thus not compatible with the natural decay expected from a collisional evolution. Therefore, the system must have preserved its outer planetesimal reservoir (unlike the Kuiper-Belt that was depleted during the LHB) and to be be encountering a transient resurgence of collisional activity. 

The dust from the inner disk does not resemble the one from the outer reservoir. It likely has a complex mineralogical composition, dominated by forsterites. The blowout size is also reflected in the exozodi, but the size distribution is very likely steeper than the canonical -3.5 power law. In other terms, an excess of small grains is observed. This is suggestive of a recent huge collisional event, during which large amounts of dust have been released. This event should be more recent than a few $t\dma{col}$, \textit{i.e.} a few years to a few decades at less than 1 AU, unless the high collision rate is in steady-state. An alternative explanation could be that the dust grains are thermally processed and fragmented when they are smaller than a few microns.

We find that the exozodi is located closer to the star than suggested by previous authors based on pure spectral arguments and models. This is an unambiguous consequence of the Keck interferometric measurements and we are able to reproduce the entire spectrum while respecting this constraint. The blackbody temperature is indeed not an appropriate proxy to estimate the disk location given that there is a distribution of temperatures for different grain sizes and locations. The surface density declines slowly and it is compatible with the theoretical -1.5 slope expected from a purely collisional disk. According to our model the inner disk falls as $r\uma{-0.5}$ to $r\uma{-2.0}$. 

Overall our model of the debris disk compares well with the one proposed by \citet{Duchene:2014yu}, with the exception that we place the bulk of the belt closer than they do. They find {$R\dma{in}\,=\,153.3$\,AU}, {$R\dma{out}\,=\,175.3$\,AU} and {$\alpha\dma{out}\,=\,-2.5$} while we find {$R\dma{in}\,=\,138$\,AU}, {$\alpha\dma{out}\,<\,-3.5$} and we fix {$R\dma{out}$} to an arbitrary large value of 300 AU. We argue that there is no physical mechanism expected to cut-off the density profile at large distances since grains produced in the parent belt with high eccentricities will populate arbitrarily large distances (blowout grains). \citet{Duchene:2014yu} developed an interesting discussion on the discrepancy between the blowout size and the minimum grain size, the latter being $\sim$3 times larger than the former for silicate grains. We find that considering more complex models and decreasing the grain density though the addition of ice and porosity increases the blowout size significantly. We consider that this solves the discrepancy between $a\dma{min}$ and $a\dma{blow}$ as we discussed in \citet{Lebreton2012a}. Furthermore the interaction of the grains with the stellar radiation field can depend on their exact shape. 
The blowout size is a function of the radiation pressure efficiency $Q\dma{RP}\,=\,Q\dma{abs} + (1-g\dma{sca})\,Q\dma{sca}$ and $Q\dma{sca}$ can be modified if the grains are aggregates instead of spheres. 
The anisotropy parameter $g\dma{sca}$ is known to be poorly reproduced by the Mie theory; for example the modeled values of $g\dma{sca}$ listed in Table\,\ref{tab:modelcold} are very high while disk imaged with the HST are consistent with anisotropy coefficients typically smaller than 0.5 \citep[\textit{e.g.}][]{Schneider:2014zl}.

Despite its high mass and because of the large stellocentric distance, the belt fractional luminosity is relatively small ${L\dma{disk}}/{L\dma{\star}} \lesssim 2\times10^{-4}$ and its maximum optical depth is 100 times smaller than the bright HD\,181327, explaining \citet{Duchene:2014yu} failure to detect the disk with HST. In the visible the disk is $2\times10^4$ fainter than the star making it a hard target for scattered light imaging.
There is marginal evidence for a side to side asymmetry in the outer belt, with a surface brightness peak possibly located 1.5$\arcsec$ further in the NW side of the disk and reaching slightly higher values. This conclusion is tempered by the limited resolution and sensitivity of {\sc Herschel} and should be confirmed with high-resolution imaging. 

We observe a inconsistency between the 70-100\,$\mu$m belt and the 160\,$\mu$m one.
At 160\um\ the models slightly overestimate the surface brightness between 13 and 20$\arcsec$. This may be an indication that the density profile is also a function of grain size thus validating the theoretical prediction that the largest grains are more confined than the small ones due to the effect of radiation pressure.

\subsection{Near-infrared spectrum and the possibility of additional dust}\label{sec:moredust}
{We recall that the KIN is dominantly sensitive to dust located between 0.2 and 2\,AU, and that its field of view is at maximum 4\,AU in radius. Our models fall off faster than $r^{-1/2}$ so most of the emission is produced close to $r_0 = 0.2 - 0.9$\,AU. Thus emission from $<1-2$\,AU regions suffices to be consistent with the Spitzer excess so there is no need for additional dust given the measurement accuracy. 
Our models arbitrarily extended out to $R\dma{max} =4\,$AU.
To check that this assumption does not affect the result we compute silicate models with $R\dma{max} =20$\,AU. The best-fit model to the SED and the nulls is unimpacted except for an expected (but limited) increase of the total mass. In sum, the surface density clearly decreases with increasing distances but the exact outer edge of the dust ring is unconstrained.}

We can ask the question of how much supplementary dust could be present at intermediate scales between the exozodi and the cold disk. {\citet{Chen06} suggested the warm disk could have two distinct radial locations detected in the IRS spectrum and \citet{Smith:2008fr} tested this possibility using mid-infrared surface brightness profiles from MICHELLE and VISIR. By fitting PSF subtraction residuals using conservative assumption for the thermal equilibrium distances, their models favor at the 2.6$\sigma$ level a single 320\,K dust population rather than one composed of 360K and additional 120K dust at $\sim$12AU. 
They thus rule out a two-disk component for the inner disk and point out that mid-infrared interferometry is needed to validate this conclusion.}

Any possible missing dust should produce negligible thermal emission as the residual from our model is consistent with 0. Yet it could scatter a significant amount of light without violating the available spectra.  {Indeed in our study, we dismissed spectra shortwards of 9\um\ because we considered the calibration of the photosphere yields uncertainties too large to trust the absolute excess spectrum and our modelling approach requires accurate calibration.}

\begin{figure}[h!tpb]
\begin{center}
  \includegraphics[angle=0,width=0.99\columnwidth,origin=bl]{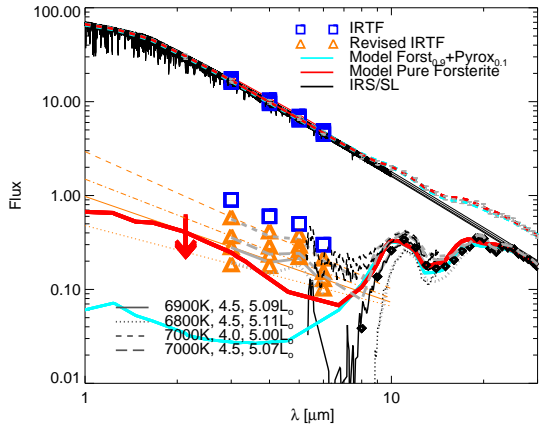}
  \caption{ {Possible near-infrared excess spectrum of $\eta$\,Crv considering various photophere models. The models are a sample of the ones fitted to various subsets of visible and near-infrared magnitudes (bottom-left legend: $T\dma{eff}$, log\,g, luminosity), the solid lines depicts our favored one (see Sec.\,\ref{sec:photosphere}). Four excess values are roughly extracted from the \citet{2012ApJ...747...93L} IRTF spectrum (blue squares) and they are then corrected for the updated stellar models (orange triangles). The orange lines are a linear interpolation of the resulting excesses. The CHARA upper limit (red arrow) sets the range of acceptable spectra. The IRS/SL excess spectrum from 5 to 14$\mu$m is shown as a black line depending on the four different photosphere models (error bars are omitted for clarity). Finally two models of the exozodi spectrum are shown.} }\label{fig:nearir_spectrum}
  \end{center}
\end{figure}

{We now inspect the claimed near-infrared spectrum and perform tests on the level of its absolute calibration. 
According to the IRS/SL and IRTF/SPeX composite spectrum from \citet{2012ApJ...747...93L}, there is a strong upturn in flux shortward of 6\um. In Fig.\,\ref{fig:nearir_spectrum} we show several possible IRS/SL excess spectra depending on the adopted photosphere model. 
The corrected IRTF spectrum suggests that the reported excess lies in the 1.5\% to 3\% range at 3 to 6 \um. 
Assuming a power-law extrapolation to the shorter wavelengths, this revised value proves compatible with the upper limit of 2.0\% found with CHARA at 2.13$\mu$m. The interferometer is not affected by the photosphere level so this measurement is the only one that truly measures the absolute excess level. Fig.\,\ref{fig:nearir_spectrum} also shows that our best photosphere model (6900\,K, log\,g=4.5, $L=5.09\,L\dma{\odot}$) and related excess spectrum (solid lines) are remarkably consistent with the forsterite model in the near-infrared. In the 5-8\um\ range a hotter stellar model ($\sim7950$\,K) would increase the level of the observed excess spectrum at a level compatible with the model.}
The near-infrared excess spectrum decreases linearly from 0.25 Jy to 0.1 Jy between 2 and 6 \um\ consistent with scattering of the starlight. {The spectrum is close to our pure forsterite model, confirming that it requires very high albedo grains. We note that the presence of forsterite could be directly confirmed by the detection of characteristic spectral features in the 2-5\um\ range \citep[see e.g. laboratory spectra from][]{Pitman:2013qy}.}

\citet{2012ApJ...747...93L} argue that the spectrum is consistent with scattering by high-albedo icy dust. They provide a comprehensive discussion about the difficulty of preserving ices so close to the star. 
Even though dusty grains have been found in the environment of comets during Solar System missions, it is harder to explain how sub-thermal dust could populate a circumstellar disk entirely.
The sublimation temperature of icy grains is about 120\,K. 10\um\ grains can survive a temperature of 150K up to a few days. This is much shorter than the expected production timescale that should be comparable to collision timescales. 
Assuming a mixture of half-silicate and half-water ice, these translate into sublimation distances of $\sim$10 and 6\,AU respectively for 10\um\ grains. Due to the size dependence of the temperature, 1\um\ grains sublimate even further (15 to 10 AU). However, small pure icy grains are poor absorbers and their sublimation zone is in fact in the 1.2 to 2.7 AU range for a 150K sublimation temperature.

We complement our grid of models with icy silicate models and pure ice models, assuming a very high sublimation temperature of 150K and attempt to fit the near-infrared spectrum. We find that there is no solution that produces a scattered light spectrum compatible with the data, without producing too much emission in the thermal domain. Pure icy grains, including nanograins models, indeed produce near-infrared excess at the expected level if they are present in huge quantity (for example $10^{-3}\,M\dma{\oplus}$ at 3 AU) but they would then create a large excess in the far-infrared around 40-50\um\ (greater that the PACS measurements). The Spitzer spectrum is not well reproduced by such model either.
Measurements in the 35 to 60\um\ range could help refine this scenario although we consider it is not viable. 
We assumed isotropic scattering but we also note that the Mie models predict scattering anisotropy parameters of 0.65 to 0.93. The nature of the light scattering grains could also be assessed from polarimetric studies. 

\subsection{Interpretation}
Future studies focused on near-infrared spectra of debris disks samples should help understand the nature of hot excesses, but it will be intrinsically hard to subtract the stellar contribution and near- to mid-infrared interferometry remains the best technique to push the knowledge of exozodis further. 
Several solutions have been explored in the literature to explain the unusual warm and hot excesses observed. 
They could be related to peculiar high-albedo grains possibly inspired by Solar System studies as discussed in this paper. 
Another possibility is that the models are facing limitations of the Mie/EMT approach. 
Alternatively, hot excesses could be caused by scattering from dust particles placed on unconventional geometries, \textit{e.g.} an edge-on component \citep{2012A&A...546L...9D} or a dust shell. 
Another appealing mechanism is thermal emission from dust aggregates that suddenly disrupt at the sublimation distance and remain temporarily trapped by gas damping or magnetic fields \citep{Lebreton:2013uq,2013ApJ...763..118S}.

Other mechanism such as stellar winds and mass-loss \citep{2008A&A...487.1041A} or evaporating planet \citep{van-Lieshout:2014xy} have been invoked. 
Some of these explanations have been explored to explain the properties of hot dust for a few individual objects but there is no definitive answer to the mystery of hot exozodis at the time. The origin of the dust-releasing bodies itself is a distinct problem.

{Herschel results from measurements of the 69 \um\ olivine feature have demonstrated the presence of a small but significant fraction (few percent of the total crystalline fraction) of Fo99 to Fo100 in a few young debris disks and protoplanetary disks \citep{2012Natur.490...74D,Sturm:2013vn}. \citet{Su:2015kx} detected crystalline Forsterite in the MIPS-SED spectrum of a 17\,Myr debris disk. 
The object was twice brighter than \etacrv\ at 70\um\ so we cannot conclude on the absence of forsterite, but the SED fitting we performed disfavors a forsterite rich outer disk. Forsterites are thought to form from the condensation from the gas phase at high temperatures thus its presence in the inner disk and absence in the outer disk could suggest a recent high-energy collision.}

\citet{2012ApJ...747...93L} proposed that the entire system is encountering a violent collisional episode comparable to the Late Heavy Bombardment in the Solar System. Strong dynamical perturbations must be invoked in the outer disk to promote a high collisional activity consistent with the dust mass observed as already noted by \citet{Wyatt:2005fj}. 
These perturbations could be induced by a recent dynamical scattering event occurring between giant outer planets. 
The result of this event would be a high collisional activity in the inner disk on secular timescales, either through the transport of planetesimals from the outer disk subsequently colliding at $\sim1$\,AU, or through the \textit{in situ} dynamical stirring of an asteroid belt by a planet placed on an eccentric orbit by giant outer planets. 
The first option may be incompatible with the fact that the grains have different composition in the two populations, unless the dust has been reprocessed through a giant impact as discussed by \citet{2012ApJ...747...93L}.
The second option requires investigations of dynamically perturbed debris disks under the effect of mean-motion resonances and we refer for example to \citet{Beust:2014qf} and \citet{Faramaz:2014qv} for detailed N-body studies applied to the case of Fomalhaut.
The effect of a giant planetesimal collisions on the dust disk itself was tested \textit{e.g.} by \citet{Kral:2014rf}, \citet{Johnson:2012nr} and \citet{2012MNRAS.425..657J}. A steep size distribution is expected for a transient period as well as side to side asymmetries that could yet remain undetected given the limited sensitivity of KIN. 

\subsection{Future observations -- JWST simulations}

\begin{figure*}[h!tpb]
\begin{center}
  \includegraphics[angle=0,width=0.32\textwidth,origin=bl]{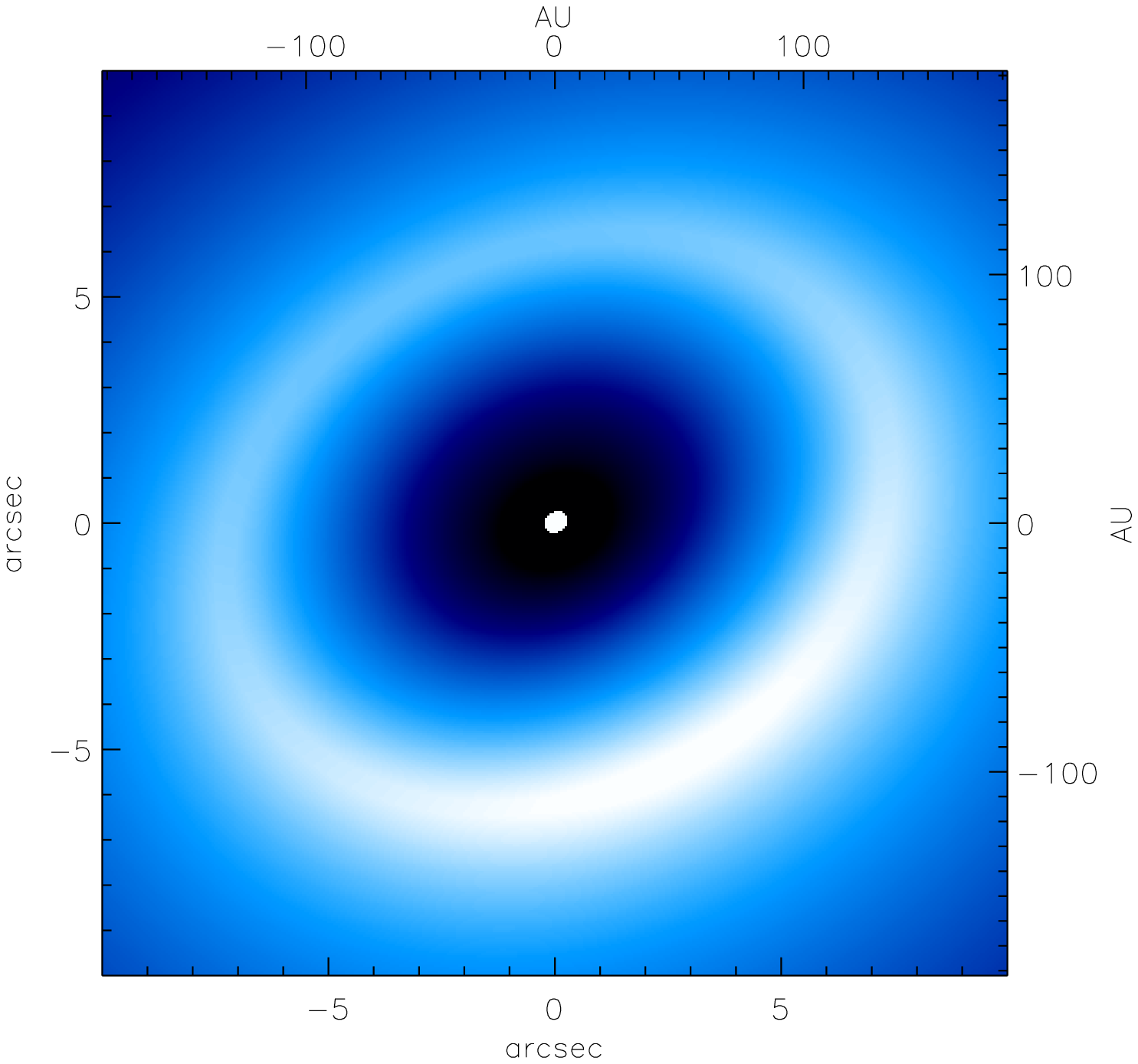}
    \includegraphics[angle=0,width=0.32\textwidth,origin=bl]{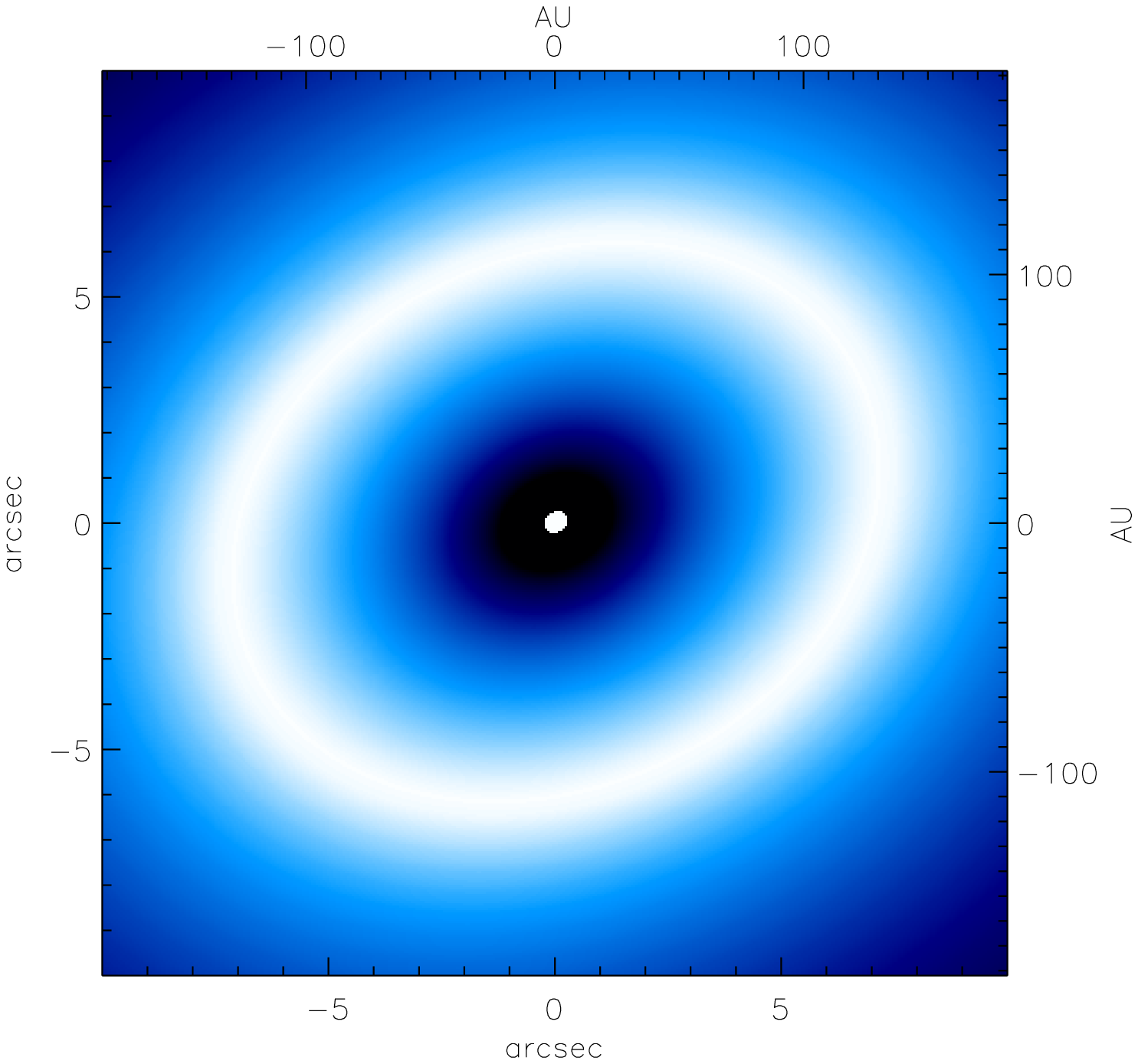}
    \includegraphics[angle=0,width=0.32\textwidth,origin=bl]{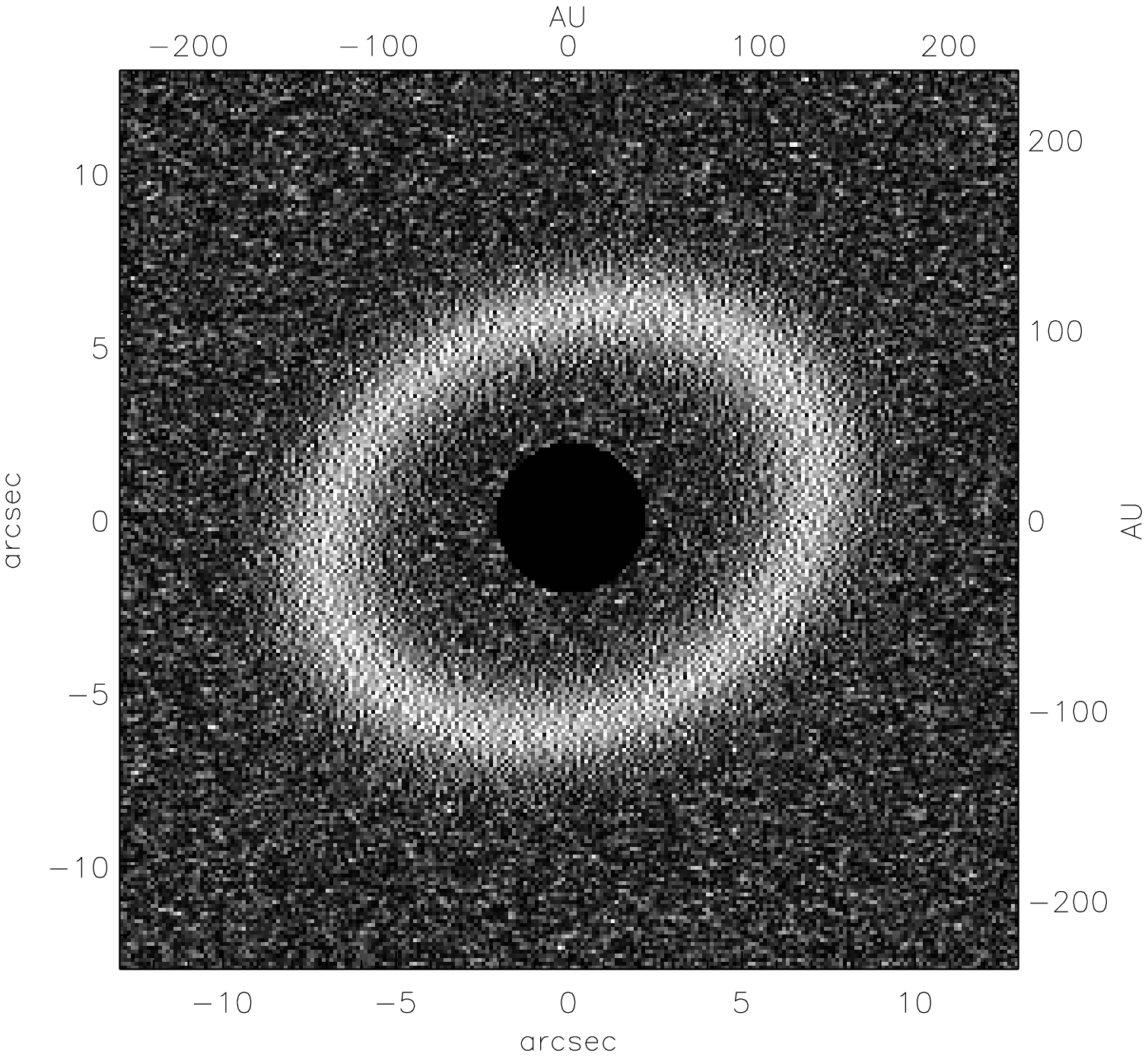}
    \includegraphics[angle=0,width=0.32\textwidth,origin=bl]{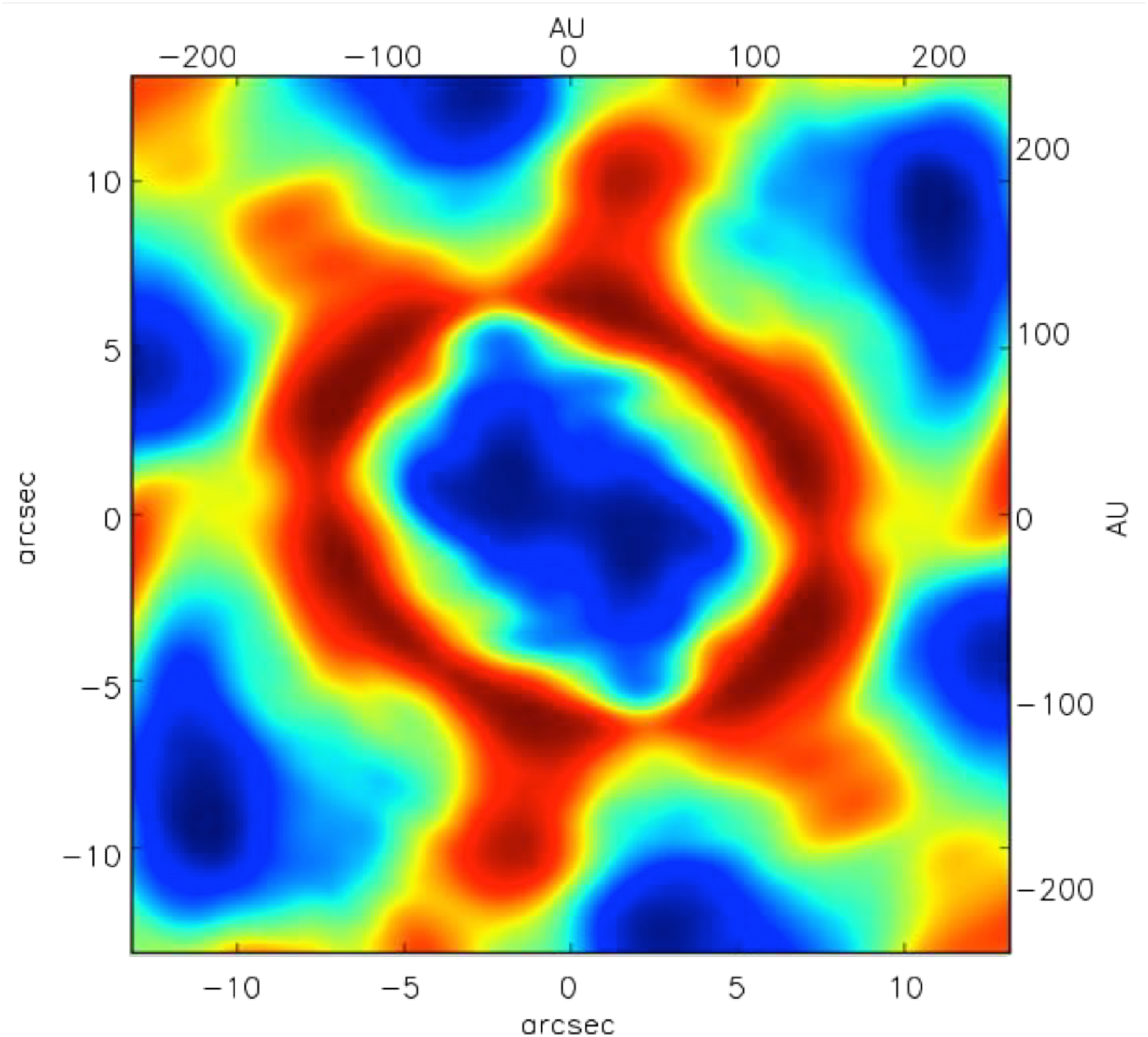}
    \includegraphics[angle=0,width=0.345\textwidth,origin=bl]{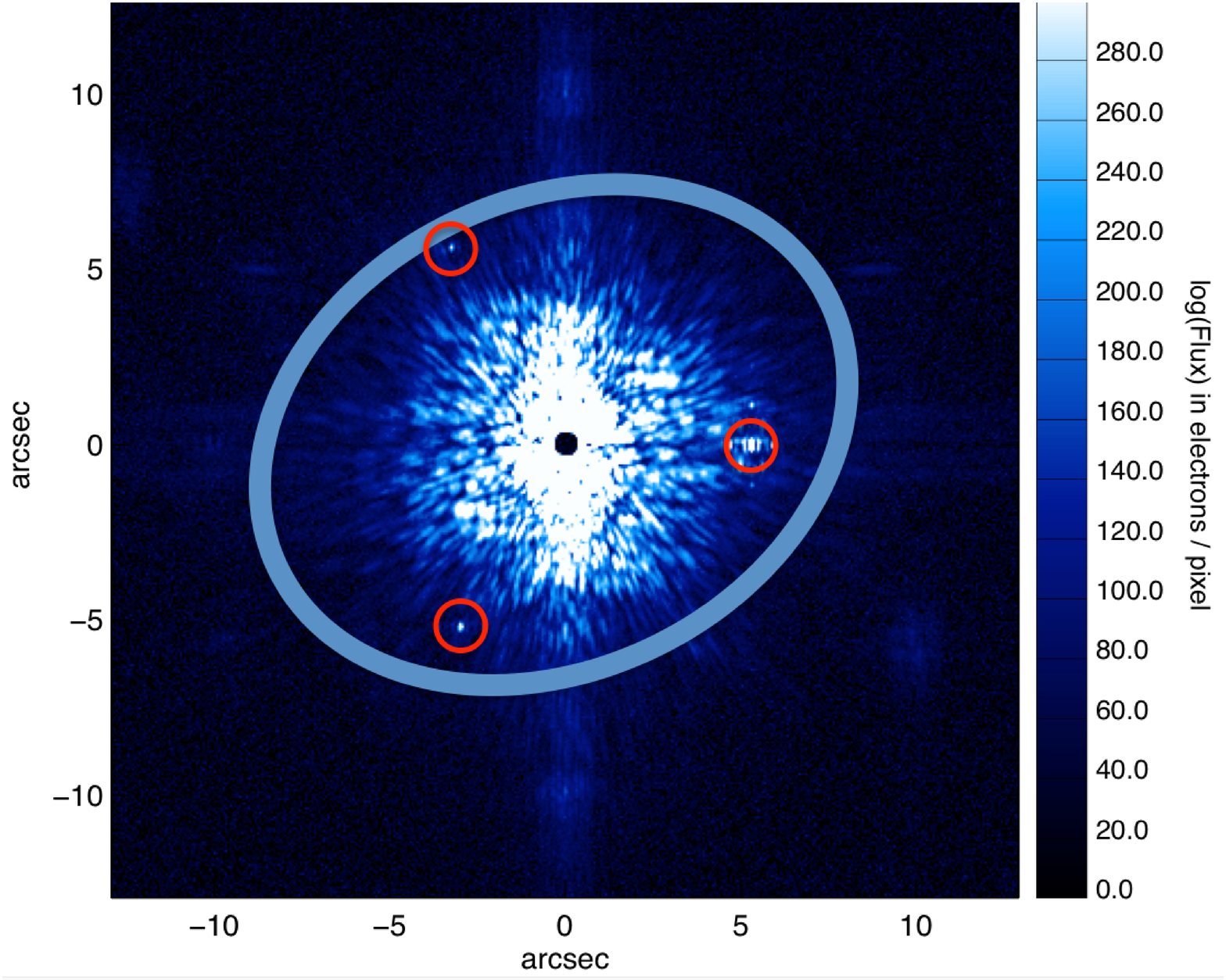}
  \caption{\textbf{Top panel:}  Image of the two-component disk model of $\eta$\,Crv at 11\um\ (\textbf{left}) and 23\um\ (\textbf{middle}) as seen with JWST/MIRI. The star is not included (the inner component is the inner disk). The scale is in magnitude units and includes a cut-off for the brightest 1\% of the pixels in order to highlight the dim cold belt against the bright exozodi. Anisotropic scattering is assumed in this plot (with the arbitrary assumption that the NE side is the closest, $g\dma{sca} = +0.5$). At 11\um\ the outer belt is dominated by scattered light, at 23\um\ by thermal emission. \textbf{Right:} Simulation of the disk as seen by JWST/MIRI with the Lyot coronagraph at 23\um\ after 10 hours of integration in contrast units. The central region is hidden behind the $3\lambda/D$ spot.\\
\textbf{Bottom-left panel:} Simulation of an ALMA image of the belt at 850$\mu$m (353\,GHz, ALMA band~7, bandwidth of 7.5\,Ghz) after 3\,hours of integration with a resolution of 0.22\arcsec\ (East is right). \\
\textbf{Bottom-right panel:} Simulation of potential belt-shaping exoplanets as seen with the JWST/NIRCam coronagraph at 3.6$\mu$m. The planets are 0.5, 1 and 5 Jupiter masses, at 6.2, 6.0 and 5.5$\arcsec$ and they are positioned at the NE, SE and W position angles respectively. 2 hours of integration and direct PSF subtraction are assumed with a wavefront error RMS of 132\,nm and drift of 1\,nm between the target and reference star. The disk is not detected.}\label{fig:modelimage}
  \end{center}
\end{figure*}

{New instruments on large telescopes will make it possible to look in more details at the $\eta$\,Crv disk. The contrast achievable by the Hubble Space Telescope was not sufficient to detect the disk \citep{Duchene:2014yu} impliying the surface brightness in scattered light is low (2.8 to 1.6 $\mu\textrm{Jy}/\textrm{arcsec}^2$ at 0.6 to 0.8\um). Therefore the disk is not considered a good target for James Webb Space Telescope near-infrared imaging. 
Our model predict a peak surface brightness of 19 $\mu\textrm{Jy}/\textrm{arcsec}^2$ at 0.8\um, which exceeds the HST upper limit by an order of magnitude. 
The non-detection suggests either that the belt is relatively broad and therefore more diffuse, or that the scattering phase function translate into a low effective albedo as discussed by \citet{Mulders:2013kx} and \citet{2015ApJ...811...67H}.
We thus correct our synthetic scattered light images to emulate the effect of a low effective albedo.}
The inner disk is very bright in the thermal infrared but it is much too compact for 5\um\ imaging with NIRCam. 
The exozodiacal dust is interesting for infrared spectroscopy with JWST/NIRSpec or MIRI. A continuous spectral coverage from 2 to 5 and 5 to 28 \um\ would make it possible to confirm previous near- and mid-infrared spectra with higher signal-to-noise. Yet we insist that the current limitation in this kind of study is how well we can calibrate the photosphere subtraction.

The flux ratio between the outer disk and the star at 23\,\um\ is about $10^{-2}$ making it a good candidate for coronagraphic imaging with MIRI. At 23\,\um\ the resolution of the JWST will be $0.72\arcsec$ which is 5 times finer than the sharpest Herschel image, and the inner working angle of the Lyot stop at this wavelengths ($3.3\lambda/D = 2.4\arcsec$) makes it possible to comfortably resolve the outer ring.
In Figure\,\ref{fig:modelimage} (top panel), we show a simulation of the outer belt as seen with MIRI at 23\,\um\ obtained using a model of the Lyot coronagraph (see Boccaletti et al. 2015). In the outer belt a contrast of $10^{-5}$ is achieved. The sky background is the dominating noise source. The disk is well detected after 1 hour of integration and subtraction of a reference PSF. Figure\,\ref{fig:modelimage} shows the resulting image after 10 hours that yields a very high signal to noise. At 15.5\,$\mu$m the disk will only be marginally detected.

{ALMA has potentially enough spatial resolution to resolve the outer ring and to assess any structures in it, although the disk will be faint in its wavelength range. Figure\,\ref{fig:modelimage} (bottom-right panel) presents a synthetic image of the \etacrv\ outer belt produced using the ALMA Observation Support Tool \citep{Heywood:2011fk}. We simulated a 3\,hour observation at 870\,$\mu$m (ALMA band~7, 353\,GHz) with a correlator bandwidth of 7.5\,Ghz (optimal for continuum observations) and a spatial resolution of 0.22\,\arcsec. ALMA will be able to resolve the width of the belt, to confirm its offset, to detect possible azimuthal asymmetries and to further constrain the presence of dust at intermediate spatial scales.}

Finally, the belt-like geometry of the disk could be due to the presence of a perturbing planet. Overlapping mean-motion resonances with a planet could produce the gap observed inwards of 130 AU (7$\arcsec$). Based on the \citet{1980AJ.....85.1122W} criterion, the required separation for a 0.5, 1, or 5 Jupiter-Mass planet is 6.2$\arcsec$, 6.0$\arcsec$ and 5.5$\arcsec$ respectively. The temperature of a 10 $M\dma{Jup}$-planet aged 1\,Gyr is 400\,K ($\log{g} = -4.3$). At 4.5\um\, this corresponds to a 15th magnitude object and a planet-to-star contrast of $10^{-6.6}$ that is well within reach of the JWST/NIRCam coronagraph ($10^{-7.2}$ contrast at 5$\arcsec$). 

{In Figure\,\ref{fig:modelimage} (bottom-left panel) we show coronagraphic images obtained using a simulator of the NIRCam coronagraph for debris disk and exoplanet science developed by our team (Lebreton et al., in prep.). 
The simulator makes use of models of the JWST PSF calculated using the WebbPSF package \citep{2012SPIE.8442E..3DP} with a wavefront error RMS of 132 nm, and predictions for the NIRCam flux of gaseous planets obtained with the COND03 models \citep{Baraffe:2003lr,Beichman:2010qy}. 
Disk-to-star flux ratios range from $4.5\times10^{-8}$ to $3.0\times10^{-7}$ for the 3 planets respectively.
We assume two hours of integration in the F360M filter and we emulate PSF subtraction assuming a wavefront error drift of 1\,nm between the target and a reference PSF star.
The dominating noise-source is photon noise rather than PSF subtraction residuals (\textit{i.e.} speckles). We see that the assumed separations are large enough such that all 3 planets are detected even for planets as small as sub-Jupiter masses despite the large star age (1\,Gyr assumed here; \etacrv\ is 1.4\,Gyr-old). 
At smaller separations, the SNR decreases dramatically. At 1.7\arcsec\, \textit{i.e.} 5 times {the NIRCam} 
inner working angle ($IWA\,=\,4\lambda/D\,=\,0.34\arcsec$ at $3.6\mu$m), only the $5\,M\dma{Jup}$ planet is detected (not shown on the image). 
The disk is not detected in scattered light.}

\section{Conclusions}

We studied the $\eta$\,Crv debris disk from its innermost to its outermost regions using complementary data from various instruments. The star was known to host a two-temperature disk from the analysis of its SED, one at $\sim$400\,K and the other one at $\sim$40\,K.
Its {\sc Spitzer}/IRS spectrum in particular reveal that the disk is very bright in the mid-infrared and it is very rich in spectral features. 
We revised this spectrum in order to improve its calibration and measure more accurately the photosphere-subtracted absolute excesses. We carefully propagated the error terms attributable to limited knowledge of the stellar spectrum. 
The spectrum informs us on the dust temperature and mineralogy but lacked sufficient constraints on the geometry of the belt. 
{We obtained null measurements from the Keck interferometer at various baselines that indicate a conservative upper limit on the dust location of 2\,AU.}

{We also revisited recent {\sc Herschel}/PACS resolved images. We performed a precise extraction of the disk radial brightness profiles  with a careful estimate of the uncertainties. 
Two separate components are clearly distinguishable. The inner one is unresolved but it clearly excesses the expected stellar flux. The outer one is consistent with a ring at $\sim$133 AU inclined by 38$\degr$.  
Small asymmetries are detected in the outer ring consisting of a 4\,AU offset along the major axis with a NW side that is up to 10\% brighter and peaks further out.}

Based on this data and ancillary photometric measurements, we were able to construct a detailed model of the two-component debris disk benefiting from both spectral constraints and spatial constraints. 
We applied a radiative transfer approach using various collections of optical constants, and a specific simulator for the KIN nulls and {\sc Herschel} PSF-convolved images. {We proposed two models of the exozodi peaking either at 0.8\,AU or 0.2\,AU depending on grain composition.
New LBTI data validate the second model: the \etacrv\ exozodi is dominated by a dust ring of pure forsterite grains larger than 1$\mu$m.}

{These high albedo grains are needed to reproduce the mid-infrared spectral features from 9 to 33 \um\ while respecting the distances measured by the interferometer and associated dust temperatures. The need for high-albedo grains is also supported by near-infrared spectra, although we demonstrate that limited knowledge of the stellar spectrum makes it impossible to confidently 
calibrate the excess spectrum below 8\,$\mu$m. The inner warm dust is much closer in than predicted by simple ``dark'' dust models and this is of fundamental importance to understand the system and its evolution. The energy balance equation used to calculate equilibrium temperature versus distance must include particle albedo and emissivity.}

We concede that such pure grains are likely hard to find in nature, and that assuming a single population of homogeous grains likely misses some of the complexity observed in the spectra. Yet different grain compositions yield the same conclusions on the basic properties of the dust disk. 
The exozodi is located closer inwards than previously suspected. Most of its dust is located between 0.2 and 0.8 AU and declines shallowly. The grain size distribution seems to be steeper than a canonical collisional equibrium and/or it has an excess of grains slightly smaller than the blowout size in agreement with collisional models. 

On the other hand the outer belt has similar properties as many known debris disks. Its SED is very well fitted using porous ice-free astronomical silicate grains (forsterite produce worst fits). The grain size distribution is that of a collisional equilibrium with a minimum grain size just a few times larger than the blowout size. 

{The far-infrared images show that the geometry of the outer belt is similar to the one of Fomalhaut in terms of size (130 AU), width and inclination. Yet even when comparing absolute fluxes, the Fomalhaut outer belt is about 8 times brighter than the \etacrv\ one at 70\um\ while its age is 3 times smaller ($\eta$\,Crv is 2.5 times further and its absolute luminosity is 3 times smaller).
The Fomalhaut exozodi, with its asteroid belt at 2\,AU, is about 10 times less massive, its spectrum is colder, it is featureless and it is 2 to 3 times fainter at 20\um.}

The origin of the debris disk and especially the massive exozodi is likely attributable to violent dynamical perturbations in a yet undetected population of giant planets that would also shape the outer dust belt. 
{The fosterite/high-albedo grain-scenario adds up to the current list of hypothesis to explain the prevalence of warm and hot excesses around Main Sequence stars.}

Despite the large age of the system, the JWST will be able to achieve the coronagraphic detection of a putative belt-shaping \etacrv\ b planet down to sub-Jupiter massive at a few arcsec.
Future imaging and spectroscopic observations of the disk could provide clues (for example clumps, asymmetries, out of equilibrium grains) on possible formation scenarios of the planetesimal disk such as a collision with a planetary body, a massive asteroid breakup, swarms of exocomets or stochastic scattering events. \\

\newpage
We thank Kate Su who provided the MIPS-SED data. We are grateful to Grant Kennedy and Gaspard Duch\^ene for their comments on  IRS spectrum and photosphere models and to Casey Lisse for his insights on dust properties and models.
We also thank Dimitra Touli for her help with the data analysis. 
Finally we acknowledge Jean-Charles Augereau and Olivier Absil who are the original developpers of the GRaTer code and the KIN simulator respectively.
This paper was based on observations taken with the {\sc Spitzer} Space Telescope and the Keck Interferometer Nuller both funded by NASA. 
{{\sc Herschel} Space Observatory is an ESA space observatory with important participation from NASA.}
This research has made use of the SIMBAD database, operated at CDS, Strasbourg, France.  

\bibliography{./biblio}

\begin{table*}[h!btp]\caption{Photometry and spectrum of $\eta$\,Crv used in this study (inner + outer disk)}\label{tab:photom}
\begin{center}
\begin{tabular}{ccccc}\hline
Wavelength  & Flux & Uncertainty & Excess & Instrument \\
(\um)  & (Jy) & (Jy) & (Jy) &   \\
\hline
2.13           & 28.9         & +0.6               & <2\%         &  CHARA \\
\hline
9.00 & 2.32 & 0.0400 & 0.223 & AKARI \\
11.6 & 1.46 & 0.0700 & 0.187 & WISE \\ 
12.0 & 1.55 & 0.100 & 0.358 & IRAS12 \\ 
18.0 & 0.820 & 0.0200 & 0.285 & AKARI\\ 
23.7 & 0.589 & 0.0236 & 0.279 & MIPS24 \\ 
22.0 & 0.680 & 0.0400 & 0.321 & WISE \\
25.0 & 0.550 & 0.0500 & 0.272 & IRAS25\\ 
\hline
8.69 & 2.35 & 0.116 & 0.109 & IRS\\
9.38 & 2.11 & 0.102 & 0.178 & IRS\\
10.1 & 1.95 & 0.0891 & 0.271 & IRS\\
10.8 & 1.80 & 0.0801 & 0.331 & IRS\\
11.5 & 1.64 & 0.0718 & 0.332 & IRS\\
12.2 & 1.42 & 0.0633 & 0.261 & IRS\\
12.8 & 1.23 & 0.0559 & 0.189 & IRS\\
14.9 & 0.953 & 0.0425 & 0.175 & IRS\\
15.6 & 0.902 & 0.0395 & 0.192 & IRS\\
16.3 & 0.873 & 0.0380 & 0.221 & IRS\\
17.0 & 0.858 & 0.0356 & 0.258 & IRS\\
17.7 & 0.816 & 0.0344 & 0.262 & IRS\\
18.4 & 0.838 & 0.0331 & 0.324 & IRS\\
19.1 & 0.835 & 0.0344 & 0.358 & IRS\\
19.8 & 0.796 & 0.0305 & 0.352 & IRS\\
20.5 & 0.743 & 0.0372 & 0.329 & IRS\\
21.2 & 0.650 & 0.0260 & 0.262 & IRS\\
21.8 & 0.613 & 0.0266 & 0.249 & IRS\\
22.5 & 0.613 & 0.0243 & 0.271 & IRS\\
23.2 & 0.614 & 0.0251 & 0.292 & IRS\\
23.9 & 0.583 & 0.0234 & 0.280 & IRS\\
24.6 & 0.570 & 0.0215 & 0.283 & IRS\\
25.3 & 0.534 & 0.0206 & 0.262 & IRS\\
26.0 & 0.498 & 0.0196 & 0.241 & IRS\\
26.7 & 0.450 & 0.0183 & 0.206 & IRS\\
27.4 & 0.477 & 0.0201 & 0.245 & IRS\\
28.1 & 0.427 & 0.0178 & 0.206 & IRS\\
28.8 & 0.396 & 0.0171 & 0.187 & IRS\\
29.5 & 0.410 & 0.0174 & 0.210 & IRS\\
30.2 & 0.354 & 0.0157 & 0.164 & IRS\\
30.8 & 0.320 & 0.0164 & 0.138 & IRS\\
31.5 & 0.388 & 0.0202 & 0.214 & IRS\\
32.2 & 0.317 & 0.0177 & 0.150 & IRS\\
32.9 & 0.243 & 0.0185 & 0.0833 & IRS\\
\hline
60.0 & 0.263 & 0.0410 & 0.215 & IRAS60 \\
70.0 & 0.259 & 0.00409 & 0.224 & MIPS70 \\
70.0 & 0.253 (0.061) & 0.00680 & 0.217 & PACS \\
100. & < 0.803 & -- & -- & IRAS100 \\
100. & 0.298 (0.036) & 0.00830 & 0.281 & PACS \\
160. & 0.227 & 0.00950 & 0.221 & PACS \\
250. & 0.141 & 0.0127 & 0.138 & SPIRE \\
350. & < 0.0910  & -- & -- & SPIRE \\
500. & < 0.0492 & -- & -- & SPIRE \\
450. & 0.0582 & 0.00980 & 0.0574 & SCUBA \\
850. & 0.0143 & 0.00180 & 0.0141 & SCUBA \\
\hline 
\end{tabular}
\end{center}
{{\sc Notes --} Photometric data used in the paper and introduced in Section\,3. The excess ratio is calculated assuming the photosphere model discussed in Sec.\,\ref{sec:photosphere}. A 4\% error relative to the stellar spectrum is included in the error bars as well as instrumental calibration uncertainty. For CHARA/FLUOR the photometry is the result from the photosphere fitting and the error is a 3$\sigma$ upper limits at the 2\% level. For PACS we list in parenthesis an estimate of the flux of the inner component. 1$\sigma$ upper limits are given for SPIRE and IRAS100 because of confusion with a background source.}
\end{table*}

\end{document}

%% file: commands.tex
\newcommand{\uma}[1]{^{\mathrm{#1}}}
\newcommand{\dma}[1]{_{\mathrm{#1}}}
\newcommand{\amin}{a\dma{min}}
\newcommand{\amax}{a\dma{max}}

\newcommand{\ie}{\textit{i.e.}~}

\def\ie{\textit{i.e.}}
\def\q1{{q1~Eri}}
\def\bp{{$\beta$~Pictoris}}

\def\etacrv{\object{$\eta$ Crv}}

\def\gra{\textsc{GRaTer}}
\def\um{$\mu$m}

\def\aap{Astronomy \& Astrophysics}

\def\amin{$a_{\mathrm{min}}$}

\DeclareMathAlphabet{\mathpzc}{OT1}{pzc}{m}{it}

\DeclareTextSymbol{\degre}{T1}{6}

\def\ie{\textit{i.e.}}

\def\gra{\textsc{GRaTer}}
\def\um{$\mu$m}

\def\aap{Astronomy \& Astrophysics}

\def\degr{\hbox{$^\circ$}}
\def\arcmin{\hbox{$^\prime$}}
\def\arcsec{\hbox{$^{\prime\prime}$}}


%% file: etacrv.bbl
\begin{thebibliography}{62}
\expandafter\ifx\csname natexlab\endcsname\relax\def\natexlab#1{#1}\fi

\bibitem[{{Absil} {et~al.}(2013){Absil}, {Defr{\`e}re}, {Coud{\'e} du Foresto},
  {Di Folco}, {M{\'e}rand}, {Augereau}, {Ertel}, {Hanot}, {Kervella},
  {Mollier}, {Scott}, {Che}, {Monnier}, {Thureau}, {Tuthill}, {ten Brummelaar},
  {McAlister}, {Sturmann}, {Sturmann}, \& {Turner}}]{2013A&A...555A.104A}
{Absil}, O., {Defr{\`e}re}, D., {Coud{\'e} du Foresto}, V., {et~al.} 2013,
  \aap, 555, A104

\bibitem[{{Absil} {et~al.}(2008){Absil}, {di Folco}, {M{\'e}rand}, {Augereau},
  {Coud{\'e} du Foresto}, {Defr{\`e}re}, {Kervella}, {Aufdenberg}, {Desort},
  {Ehrenreich}, {Lagrange}, {Montagnier}, {Olofsson}, {ten Brummelaar},
  {McAlister}, {Sturmann}, {Sturmann}, \& {Turner}}]{2008A&A...487.1041A}
{Absil}, O., {di Folco}, E., {M{\'e}rand}, A., {et~al.} 2008, Astronomy \&
  Astrophysics, 487, 1041

\bibitem[{{Augereau} {et~al.}(1999){Augereau}, {Lagrange}, {Mouillet},
  {Papaloizou}, \& {Grorod}}]{Augereau1999a}
{Augereau}, J.~C., {Lagrange}, A.~M., {Mouillet}, D., {Papaloizou}, J.~C.~B.,
  \& {Grorod}, P.~A. 1999, Astronomy \& Astrophysics, 348, 557

\bibitem[{{Baraffe} {et~al.}(2003){Baraffe}, {Chabrier}, {Barman}, {Allard}, \&
  {Hauschildt}}]{Baraffe:2003lr}
{Baraffe}, I., {Chabrier}, G., {Barman}, T.~S., {Allard}, F., \& {Hauschildt},
  P.~H. 2003, \aap, 402, 701

\bibitem[{{Beichman} {et~al.}(2005){Beichman}, {Bryden}, {Gautier},
  {Stapelfeldt}, {Werner}, {Misselt}, {Rieke}, {Stansberry}, \&
  {Trilling}}]{2005ApJ...626.1061B}
{Beichman}, C.~A., {Bryden}, G., {Gautier}, T.~N., {et~al.} 2005, The
  Astrophysical Journal, 626, 1061

\bibitem[{{Beichman} {et~al.}(2006){Beichman}, {Bryden}, {Stapelfeldt},
  {Gautier}, {Grogan}, {Shao}, {Velusamy}, {Lawler}, {Blaylock}, {Rieke},
  {Lunine}, {Fischer}, {Marcy}, {Greaves}, {Wyatt}, {Holland}, \&
  {Dent}}]{Beichman:2006lr}
{Beichman}, C.~A., {Bryden}, G., {Stapelfeldt}, K.~R., {et~al.} 2006, the
  Astrophysical Journal, 652, 1674

\bibitem[{{Beichman} {et~al.}(2010){Beichman}, {Krist}, {Trauger}, {Greene},
  {Oppenheimer}, {Sivaramakrishnan}, {Doyon}, {Boccaletti}, {Barman}, \&
  {Rieke}}]{Beichman:2010qy}
{Beichman}, C.~A., {Krist}, J., {Trauger}, J.~T., {et~al.} 2010, \pasp, 122,
  162

\bibitem[{{Beust} {et~al.}(2014){Beust}, {Augereau}, {Bonsor}, {Graham},
  {Kalas}, {Lebreton}, {Lagrange}, {Ertel}, {Faramaz}, \&
  {Th{\'e}bault}}]{Beust:2014qf}
{Beust}, H., {Augereau}, J.-C., {Bonsor}, A., {et~al.} 2014, Astronomy \&
  Astrophysics, 561, A43

\bibitem[{{Bryden} {et~al.}(2006){Bryden}, {Beichman}, {Trilling}, {Rieke},
  {Holmes}, {Lawler}, {Stapelfeldt}, {Werner}, {Gautier}, {Blaylock}, {Gordon},
  {Stansberry}, \& {Su}}]{Bryden:2006eu}
{Bryden}, G., {Beichman}, C.~A., {Trilling}, D.~E., {et~al.} 2006, \apj, 636,
  1098

\bibitem[{{Chen} {et~al.}(2006){Chen}, {Sargent}, {Bohac}, {Kim},
  {Leibensperger}, {Jura}, {Najita}, {Forrest}, {Watson}, {Sloan}, \&
  {Keller}}]{Chen06}
{Chen}, C.~H., {Sargent}, B.~A., {Bohac}, C., {et~al.} 2006, The Astrophysical
  Journals, 166, 351

\bibitem[{{Cutri} {et~al.}(2003){Cutri}, {Skrutskie}, {van Dyk}, {Beichman},
  {Carpenter}, {Chester}, {Cambresy}, {Evans}, {Fowler}, {Gizis}, {Howard},
  {Huchra}, {Jarrett}, {Kopan}, {Kirkpatrick}, {Light}, {Marsh}, {McCallon},
  {Schneider}, {Stiening}, {Sykes}, {Weinberg}, {Wheaton}, {Wheelock}, \&
  {Zacarias}}]{Cutri:2003zl}
{Cutri}, R.~M., {Skrutskie}, M.~F., {van Dyk}, S., {et~al.} 2003, {2MASS All
  Sky Catalog of point sources.}

\bibitem[{{de Vries} {et~al.}(2012){de Vries}, {Acke}, {Blommaert}, {Waelkens},
  {Waters}, {Vandenbussche}, {Min}, {Olofsson}, {Dominik}, {Decin}, {Barlow},
  {Brandeker}, {di Francesco}, {Glauser}, {Greaves}, {Harvey}, {Holland},
  {Ivison}, {Liseau}, {Pantin}, {Pilbratt}, {Royer}, \&
  {Sibthorpe}}]{2012Natur.490...74D}
{de Vries}, B.~L., {Acke}, B., {Blommaert}, J.~A.~D.~L., {et~al.} 2012, \nat,
  490, 74

\bibitem[{{Defr{\`e}re} {et~al.}(2011){Defr{\`e}re}, {Absil}, {Augereau}, {di
  Folco}, {Berger}, {Coud{\'e} Du Foresto}, {Kervella}, {Le Bouquin},
  {Lebreton}, {Millan-Gabet}, {Monnier}, {Olofsson}, \&
  {Traub}}]{2011A&A...534A...5D}
{Defr{\`e}re}, D., {Absil}, O., {Augereau}, J.-C., {et~al.} 2011, Astronomy \&
  Astrophysics, 534, A5

\bibitem[{{Defr{\`e}re} {et~al.}(2015){Defr{\`e}re}, {Hinz}, {Skemer},
  {Kennedy}, {Bailey}, {Hoffmann}, {Mennesson}, {Millan-Gabet}, {Danchi},
  {Absil}, {Arbo}, {Beichman}, {Brusa}, {Bryden}, {Downey}, {Durney},
  {Esposito}, {Gaspar}, {Grenz}, {Haniff}, {Hill}, {Lebreton}, {Leisenring},
  {Males}, {Marion}, {McMahon}, {Montoya}, {Morzinski}, {Pinna}, {Puglisi},
  {Rieke}, {Roberge}, {Serabyn}, {Sosa}, {Stapeldfeldt}, {Su}, {Vaitheeswaran},
  {Vaz}, {Weinberger}, \& {Wyatt}}]{Defrere:2015lr}
{Defr{\`e}re}, D., {Hinz}, P.~M., {Skemer}, A.~J., {et~al.} 2015, \apj, 799, 42

\bibitem[{{Defr{\`e}re} {et~al.}(2012){Defr{\`e}re}, {Lebreton}, {Le Bouquin},
  {Lagrange}, {Absil}, {Augereau}, {Berger}, {di Folco}, {Ertel}, {Kluska},
  {Montagnier}, {Millan-Gabet}, {Traub}, \& {Zins}}]{2012A&A...546L...9D}
{Defr{\`e}re}, D., {Lebreton}, J., {Le Bouquin}, J.-B., {et~al.} 2012,
  Astronomy \& Astrophysics, 546, L9

\bibitem[{{Donaldson} {et~al.}(2013){Donaldson}, {Lebreton}, {Roberge},
  {Augereau}, \& {Krivov}}]{2013ApJ...772...17D}
{Donaldson}, J.~K., {Lebreton}, J., {Roberge}, A., {Augereau}, J.-C., \&
  {Krivov}, A.~V. 2013, The Astrophysical Journal, 772, 17

\bibitem[{{Dorschner} {et~al.}(1995){Dorschner}, {Begemann}, {Henning},
  {Jaeger}, \& {Mutschke}}]{Dorschner:1995yq}
{Dorschner}, J., {Begemann}, B., {Henning}, T., {Jaeger}, C., \& {Mutschke}, H.
  1995, \aap, 300, 503

\bibitem[{{Draine}(2003)}]{Draine2003}
{Draine}, B.~T. 2003, Annu. Rev. Astron. Astrophys., 41, 241

\bibitem[{{Duch{\^e}ne} {et~al.}(2014){Duch{\^e}ne}, {Arriaga}, {Wyatt},
  {Kennedy}, {Sibthorpe}, {Lisse}, {Holland}, {Wisniewski}, {Clampin}, {Kalas},
  {Pinte}, {Wilner}, {Booth}, {Horner}, {Matthews}, \&
  {Greaves}}]{Duchene:2014yu}
{Duch{\^e}ne}, G., {Arriaga}, P., {Wyatt}, M., {et~al.} 2014, \apj, 784, 148

\bibitem[{{Faramaz} {et~al.}(2015){Faramaz}, {Beust}, {Augereau}, {Kalas}, \&
  {Graham}}]{Faramaz:2014qv}
{Faramaz}, V., {Beust}, H., {Augereau}, J.-C., {Kalas}, P., \& {Graham}, J.~R.
  2015, \aap, 573, A87

\bibitem[{{Hauschildt} {et~al.}(1999){Hauschildt}, {Allard}, \&
  {Baron}}]{1999ApJ...512..377H}
{Hauschildt}, P.~H., {Allard}, F., \& {Baron}, E. 1999, the Astrophysical
  Journal, 512, 377

\bibitem[{{Hedman} \& {Stark}(2015)}]{2015ApJ...811...67H}
{Hedman}, M.~M. \& {Stark}, C.~C. 2015, \apj, 811, 67

\bibitem[{{Henning} \& {Mutschke}(1997)}]{1997A&A...327..743H}
{Henning}, T. \& {Mutschke}, H. 1997, \aap, 327, 743

\bibitem[{{Heywood} {et~al.}(2011){Heywood}, {Avison}, \&
  {Williams}}]{Heywood:2011fk}
{Heywood}, I., {Avison}, A., \& {Williams}, C.~J. 2011, ArXiv e-prints

\bibitem[{{H{\o}g} {et~al.}(2000){H{\o}g}, {Fabricius}, {Makarov}, {Urban},
  {Corbin}, {Wycoff}, {Bastian}, {Schwekendiek}, \& {Wicenec}}]{Hog:2000qf}
{H{\o}g}, E., {Fabricius}, C., {Makarov}, V.~V., {et~al.} 2000, \aap, 355, L27

\bibitem[{{Holmberg} {et~al.}(2009){Holmberg}, {Nordstr{\"o}m}, \&
  {Andersen}}]{Holmberg09}
{Holmberg}, J., {Nordstr{\"o}m}, B., \& {Andersen}, J. 2009, Astronomy \&
  Astrophysics, 501, 941

\bibitem[{{Ida} \& {Makino}(1992)}]{1992Icar...96..107I}
{Ida}, S. \& {Makino}, J. 1992, \icarus, 96, 107

\bibitem[{{Jackson} \& {Wyatt}(2012)}]{2012MNRAS.425..657J}
{Jackson}, A.~P. \& {Wyatt}, M.~C. 2012, \mnras, 425, 657

\bibitem[{{J{\"a}ger} {et~al.}(2003){J{\"a}ger}, {Dorschner}, {Mutschke},
  {Posch}, \& {Henning}}]{Jager:2003gf}
{J{\"a}ger}, C., {Dorschner}, J., {Mutschke}, H., {Posch}, T., \& {Henning}, T.
  2003, \aap, 408, 193

\bibitem[{{Johnson} {et~al.}(2012){Johnson}, {Lisse}, {Chen}, {Melosh},
  {Wyatt}, {Thebault}, {Henning}, {Gaidos}, {Elkins-Tanton}, {Bridges}, \&
  {Morlok}}]{Johnson:2012nr}
{Johnson}, B.~C., {Lisse}, C.~M., {Chen}, C.~H., {et~al.} 2012, \apj, 761, 45

\bibitem[{{Kral} {et~al.}(2015){Kral}, {Th{\'e}bault}, {Augereau},
  {Boccaletti}, \& {Charnoz}}]{Kral:2014rf}
{Kral}, Q., {Th{\'e}bault}, P., {Augereau}, J.-C., {Boccaletti}, A., \&
  {Charnoz}, S. 2015, \aap, 573, A39

\bibitem[{{Lebreton} {et~al.}(2012){Lebreton}, {Augereau}, {Thi}, {Roberge},
  {Donaldson}, {Schneider}, {Maddison}, {M{\'e}nard}, {Riviere-Marichalar},
  {Mathews}, {Kamp}, {Pinte}, {Dent}, {Barrado}, {Duch{\^e}ne}, {Gonzalez},
  {Grady}, {Meeus}, {Pantin}, {Williams}, \& {Woitke}}]{Lebreton2012a}
{Lebreton}, J., {Augereau}, J.-C., {Thi}, W.-F., {et~al.} 2012, Astronomy \&
  Astrophysics, 539, A17

\bibitem[{{Lebreton} {et~al.}(2013){Lebreton}, {van Lieshout}, {Augereau},
  {Absil}, {Mennesson}, {Kama}, {Dominik}, {Bonsor}, {Vandeportal}, {Beust},
  {Defr{\`e}re}, {Ertel}, {Faramaz}, {Hinz}, {Kral}, {Lagrange}, {Liu}, \&
  {Th{\'e}bault}}]{Lebreton:2013uq}
{Lebreton}, J., {van Lieshout}, R., {Augereau}, J.-C., {et~al.} 2013, Astronomy
  \& Astrophysics, 555, A146

\bibitem[{{Li} \& {Greenberg}(1997)}]{LiGr97}
{Li}, A. \& {Greenberg}, J.~M. 1997, Astronomy \& Astrophysics, 323, 566

\bibitem[{{Lisse} {et~al.}(2012){Lisse}, {Wyatt}, {Chen}, {Morlok}, {Watson},
  {Manoj}, {Sheehan}, {Currie}, {Thebault}, \& {Sitko}}]{2012ApJ...747...93L}
{Lisse}, C.~M., {Wyatt}, M.~C., {Chen}, C.~H., {et~al.} 2012, the Astrophysical
  Journal, 747, 93

\bibitem[{{L{\"o}hne} {et~al.}(2012){L{\"o}hne}, {Augereau}, {Ertel},
  {Marshall}, {Eiroa}, {Mora}, {Absil}, {Stapelfeldt}, {Th{\'e}bault}, {Bayo},
  {Del Burgo}, {Danchi}, {Krivov}, {Lebreton}, {Letawe}, {Magain}, {Maldonado},
  {Montesinos}, {Pilbratt}, {White}, \& {Wolf}}]{2012A&A...537A.110L}
{L{\"o}hne}, T., {Augereau}, J.-C., {Ertel}, S., {et~al.} 2012, \aap, 537, A110

\bibitem[{{L{\"o}hne} {et~al.}(2008){L{\"o}hne}, {Krivov}, \&
  {Rodmann}}]{2008ApJ...673.1123L}
{L{\"o}hne}, T., {Krivov}, A.~V., \& {Rodmann}, J. 2008, The Astrophysical
  Journal, 673, 1123

\bibitem[{{Matthews} {et~al.}(2010){Matthews}, {Sibthorpe}, {Kennedy},
  {Phillips}, {Churcher}, {Duch{\^e}ne}, {Greaves}, {Lestrade}, {Moro-Martin},
  {Wyatt}, {Bastien}, {Biggs}, {Bouvier}, {Butner}, {Dent}, {di Francesco},
  {Eisl{\"o}ffel}, {Graham}, {Harvey}, {Hauschildt}, {Holland}, {Horner},
  {Ibar}, {Ivison}, {Johnstone}, {Kalas}, {Kavelaars}, {Rodriguez}, {Udry},
  {van der Werf}, {Wilner}, \& {Zuckerman}}]{Matthews:2010qy}
{Matthews}, B.~C., {Sibthorpe}, B., {Kennedy}, G., {et~al.} 2010, Astronomy \&
  Astrophysics, 518, L135

\bibitem[{{Mennesson} {et~al.}(2013){Mennesson}, {Absil}, {Lebreton},
  {Augereau}, {Serabyn}, {Colavita}, {Millan-Gabet}, {Liu}, {Hinz}, \&
  {Th{\'e}bault}}]{Mennesson:2013mz}
{Mennesson}, B., {Absil}, O., {Lebreton}, J., {et~al.} 2013, The Astrophysical
  Journal, 763, 119

\bibitem[{{Mennesson} {et~al.}(2014){Mennesson}, {Millan-Gabet}, {Serabyn},
  {Colavita}, {Absil}, {Bryden}, {Wyatt}, {Danchi}, {Defr{\`e}re}, {Dor{\'e}},
  {Hinz}, {Kuchner}, {Ragland}, {Scott}, {Stapelfeldt}, {Traub}, \&
  {Woillez}}]{Mennesson:2014qy}
{Mennesson}, B., {Millan-Gabet}, R., {Serabyn}, E., {et~al.} 2014, \apj, 797,
  119

\bibitem[{{Millan-Gabet} {et~al.}(2011){Millan-Gabet}, {Serabyn}, {Mennesson},
  {Traub}, {Barry}, {Danchi}, {Kuchner}, {Stark}, {Ragland}, {Hrynevych},
  {Woillez}, {Stapelfeldt}, {Bryden}, {Colavita}, \&
  {Booth}}]{2011ApJ...734...67M}
{Millan-Gabet}, R., {Serabyn}, E., {Mennesson}, B., {et~al.} 2011, The
  Astrophysical Journal, 734, 67

\bibitem[{{Mulders} {et~al.}(2013){Mulders}, {Min}, {Dominik}, {Debes}, \&
  {Schneider}}]{Mulders:2013kx}
{Mulders}, G.~D., {Min}, M., {Dominik}, C., {Debes}, J.~H., \& {Schneider}, G.
  2013, \aap, 549, A112

\bibitem[{{Olofsson} {et~al.}(2012){Olofsson}, {Juh{\'a}sz}, {Henning},
  {Mutschke}, {Tamanai}, {Mo{\'o}r}, \& {{\'A}brah{\'a}m}}]{Olofsson:2012vn}
{Olofsson}, J., {Juh{\'a}sz}, A., {Henning}, T., {et~al.} 2012, \aap, 542, A90

\bibitem[{{Ott}(2010)}]{ott10}
{Ott}, S. 2010, in Astronomical Society of the Pacific Conference Series, Vol.
  434, Astronomical Data Analysis Software and Systems XIX, ed. {Y.~Mizumoto,
  K.-I.~Morita, \& M.~Ohishi}, 139

\bibitem[{{Pawellek} {et~al.}(2014){Pawellek}, {Krivov}, {Marshall},
  {Montesinos}, {{\'A}brah{\'a}m}, {Mo{\'o}r}, {Bryden}, \&
  {Eiroa}}]{Pawellek:2014kq}
{Pawellek}, N., {Krivov}, A.~V., {Marshall}, J.~P., {et~al.} 2014, \apj, 792,
  65

\bibitem[{{Perrin} {et~al.}(2012){Perrin}, {Soummer}, {Elliott}, {Lallo}, \&
  {Sivaramakrishnan}}]{2012SPIE.8442E..3DP}
{Perrin}, M.~D., {Soummer}, R., {Elliott}, E.~M., {Lallo}, M.~D., \&
  {Sivaramakrishnan}, A. 2012, in Society of Photo-Optical Instrumentation
  Engineers (SPIE) Conference Series, Vol. 8442, Society of Photo-Optical
  Instrumentation Engineers (SPIE) Conference Series

\bibitem[{{Perryman} {et~al.}(1997){Perryman}, {Lindegren}, {Kovalevsky},
  {Hoeg}, {Bastian}, {Bernacca}, {Cr{\'e}z{\'e}}, {Donati}, {Grenon}, {van
  Leeuwen}, {van der Marel}, {Mignard}, {Murray}, {Le Poole}, {Schrijver},
  {Turon}, {Arenou}, {Froeschl{\'e}}, \& {Petersen}}]{Hipparcos}
{Perryman}, M.~A.~C., {Lindegren}, L., {Kovalevsky}, J., {et~al.} 1997,
  Astronomy \& Astrophysics, 323, L49

\bibitem[{{Pitman} {et~al.}(2013){Pitman}, {Hofmeister}, \&
  {Speck}}]{Pitman:2013qy}
{Pitman}, K.~M., {Hofmeister}, A.~M., \& {Speck}, A.~K. 2013, Earth, Planets,
  and Space, 65, 129

\bibitem[{{Reach} {et~al.}(2003){Reach}, {Morris}, {Boulanger}, \&
  {Okumura}}]{2003Icar..164..384R}
{Reach}, W.~T., {Morris}, P., {Boulanger}, F., \& {Okumura}, K. 2003, Icarus,
  164, 384

\bibitem[{{Schneider} {et~al.}(2014){Schneider}, {Grady}, {Hines}, {Stark},
  {Debes}, {Carson}, {Kuchner}, {Perrin}, {Weinberger}, {Wisniewski},
  {Silverstone}, {Jang-Condell}, {Henning}, {Woodgate}, {Serabyn},
  {Moro-Martin}, {Tamura}, {Hinz}, \& {Rodigas}}]{Schneider:2014zl}
{Schneider}, G., {Grady}, C.~A., {Hines}, D.~C., {et~al.} 2014, \aj, 148, 59

\bibitem[{{Smith} {et~al.}(2008){Smith}, {Wyatt}, \& {Dent}}]{Smith:2008fr}
{Smith}, R., {Wyatt}, M.~C., \& {Dent}, W.~R.~F. 2008, Astronomy \&
  Astrophysics, 485, 897

\bibitem[{{Smith} {et~al.}(2009){Smith}, {Wyatt}, \& {Haniff}}]{Smith:2009zr}
{Smith}, R., {Wyatt}, M.~C., \& {Haniff}, C.~A. 2009, Astronomy \&
  Astrophysics, 503, 265

\bibitem[{{Sturm} {et~al.}(2013){Sturm}, {Bouwman}, {Henning}, {Evans},
  {Waters}, {van Dishoeck}, {Green}, {Olofsson}, {Meeus}, {Maaskant},
  {Dominik}, {Augereau}, {Mulders}, {Acke}, {Merin}, \&
  {Herczeg}}]{Sturm:2013vn}
{Sturm}, B., {Bouwman}, J., {Henning}, T., {et~al.} 2013, Astronomy \&
  Astrophysics, 553, A5

\bibitem[{{Su} {et~al.}(2015){Su}, {Morrison}, {Malhotra}, {Smith}, {Balog}, \&
  {Rieke}}]{Su:2015kx}
{Su}, K.~Y.~L., {Morrison}, S., {Malhotra}, R., {et~al.} 2015, \apj, 799, 146

\bibitem[{{Su} {et~al.}(2013){Su}, {Rieke}, {Malhotra}, {Stapelfeldt},
  {Hughes}, {Bonsor}, {Wilner}, {Balog}, {Watson}, {Werner}, \&
  {Misselt}}]{2013ApJ...763..118S}
{Su}, K.~Y.~L., {Rieke}, G.~H., {Malhotra}, R., {et~al.} 2013, \apj, 763, 118

\bibitem[{{Sylvester} {et~al.}(1996){Sylvester}, {Skinner}, {Barlow}, \&
  {Mannings}}]{Sylvester:1996eu}
{Sylvester}, R.~J., {Skinner}, C.~J., {Barlow}, M.~J., \& {Mannings}, V. 1996,
  \mnras, 279, 915

\bibitem[{{Tremaine}(1998)}]{1998AJ....116.2015T}
{Tremaine}, S. 1998, \aj, 116, 2015

\bibitem[{{van Lieshout} {et~al.}(2014){van Lieshout}, {Min}, \&
  {Dominik}}]{van-Lieshout:2014xy}
{van Lieshout}, R., {Min}, M., \& {Dominik}, C. 2014, \aap, 572, A76

\bibitem[{{Wisdom}(1980)}]{1980AJ.....85.1122W}
{Wisdom}, J. 1980, \aj, 85, 1122

\bibitem[{{Wyatt} {et~al.}(2005){Wyatt}, {Greaves}, {Dent}, \&
  {Coulson}}]{Wyatt:2005fj}
{Wyatt}, M.~C., {Greaves}, J.~S., {Dent}, W.~R.~F., \& {Coulson}, I.~M. 2005,
  the Astrophysical Journal, 620, 492

\bibitem[{{Wyatt} {et~al.}(2007){Wyatt}, {Smith}, {Su}, {Rieke}, {Greaves},
  {Beichman}, \& {Bryden}}]{2007ApJ...663..365W}
{Wyatt}, M.~C., {Smith}, R., {Su}, K.~Y.~L., {et~al.} 2007, \apj, 663, 365

\bibitem[{{Zubko} {et~al.}(1996){Zubko}, {Mennella}, {Colangeli}, \&
  {Bussoletti}}]{zubko}
{Zubko}, V.~G., {Mennella}, V., {Colangeli}, L., \& {Bussoletti}, E. 1996, Mon.
  Not. R. Astron. Soc., 282, 1321

\end{thebibliography}
